\documentclass[reprint,prx,10pt,amssymb, aps,superscriptaddress]{revtex4-2}
\pdfoutput=1
\usepackage{CJK}
\usepackage{xcolor}
\usepackage{physics}
\usepackage{float}
\usepackage{graphicx}
\usepackage{array} 
\usepackage[english]{babel}
\usepackage{subfig}
\usepackage{hhline}
\usepackage{hyperref}

\newcommand{\re}[1]{\mathrm{Re}\left[#1\right]}
\newcommand{\psistab}{\psi_\mathrm{STAB}}

\newtheorem{theorem}{Theorem}
\newtheorem{lemma}{Lemma}
\newtheorem{corollary}{Corollary}

\begin{document}
\begin{CJK*}{UTF8}{} 
\title{Correlation is magic in electronic structure Hamiltonians}

\author{Basie Seibert}
\thanks{These authors contributed equally to this work.
\newline Contact author: bseibert1@unm.edu
\newline Contact author: sam.alterman@tufts.edu}
\affiliation{Department of Physics and Astronomy, Quantum New Mexico Institute, University of New Mexico, Albuquerque, NM 87106, USA}
\author{Sam Alterman}
\thanks{These authors contributed equally to this work.
\newline Contact author: bseibert1@unm.edu
\newline Contact author: sam.alterman@tufts.edu}
\affiliation{Department of Physics and Astronomy, Tufts University, Medford, MA 02155, USA}
\author{Qingfeng Wang}
\affiliation{Department of Physics and Astronomy, Tufts University, Medford, MA 02155, USA}
\CJKfamily{gbsn}
\author{Feng Qian (钱丰)}
\affiliation{Department of Physics and Astronomy, Tufts University, Medford, MA 02155, USA}
\author{Akimasa Miyake}
\affiliation{Department of Physics and Astronomy, Quantum New Mexico Institute, University of New Mexico, Albuquerque, NM 87106, USA}
\author{Peter J. Love}
\affiliation{Department of Physics and Astronomy, Tufts University, Medford, MA 02155, USA}
\affiliation{Department of Computer Science, University of Toronto, Toronto, ON M5S 3H6, Canada}
\affiliation{Department of Physics, University of Toronto, Toronto, ON M5S 1A7, Canada}
\affiliation{QMatter, Inc., Office 109, 254 Chapman Rd, Suite 101-B, Newark, DE 19702, USA}
\affiliation{Computational Science Initiative, Brookhaven National Laboratory, Upton, NY 11973, USA}

\date{June 30, 2026}
\begin{abstract}
    The gate and qubit requirements of quantum computations of electronic structure have been extensively studied. However, the quantum resources present in electronic ground states, as measured by entanglement and magic, remain less well understood. We study the relationship between correlation in electronic structure Hamiltonians and magic as measured by the $2$-stabilizer R\'enyi entropy ($2$-SRE). Perturbative calculations show that the $2$-SRE of a given state is proportional to its overlap with a reference stabilizer state. In the context of quantum chemistry, this links the magic of electronic structure ground states to their Hartree-Fock weight, an established measure of electronic correlation.  We then show that the $2$-SRE of post-Hartree-Fock ground states is proportional to the correlation energy they recover. We explore this connection through the contextual subspace (CS) method. We present a theoretical framework showing that the CS method can be used to monotonically vary the magic of approximate CS ground states, and we prove that the correlation energy recovered by the CS ground states is proportional to the magic present in the approximate ground state. We present simulation results using 190 molecular species under Jordan-Wigner encoding at a range of bond lengths. The linear relationships between magic and correlation are robust across the Hamiltonians in our dataset, but break down at bond lengths beyond the Coulson-Fischer point, where Hartree-Fock fails to capture key physical features of the true ground state wavefunction. By establishing linear relationships for both correlation energy and Hartree-Fock reference weight with the $2$-SRE, we conclude that for weakly- and moderately-correlated electronic structure Hamiltonians, the correlation is directly represented by $2$-SRE, and thus by the magic.
\end{abstract}
\maketitle
\end{CJK*}

\section{Introduction}
The electronic structure problem is the determination of the electronic energy given the position of the nuclei. Classical Hartree-Fock (HF) methods approximately solve the electronic structure problem by restricting the ground state to a single Slater determinant. The full configuration interaction (FCI) method represents the ground state in a linear combination of Slater determinants, but has historically been challenging to implement on classical computers for large molecules of interest~\cite{lewars2003computational,ModernQuantumChemistry,helgaker2013molecular}. The difference between the FCI energy and the HF energy  is the {\em correlation energy}. Recovery of the correlation energy is the target of all post-Hartree-Fock methods ~\cite{lwdin1954quantum-ebe,lowdincorrelation,lewars2003computational,ModernQuantumChemistry,helgaker2013molecular}. 

Electronic structure calculations have been of interest as applications of quantum computers for almost two decades~\cite{doi:10.1126/science.1113479, lanyon2010towards, whitfield2011simulation, kassal2011simulating,McArdle2020,schleich2025cracking}. Small numbers of qubits can represent FCI problems whose dimension puts them out of reach of direct simulation on classical supercomputers. However, quantum algorithms must exploit problem structure to, for example, prepare ground states. This problem structure can also be exploited by classical algorithms, leading to a competition between quantum and classical heuristics that is closer than naive estimates would suggest~\cite{lee2023evaluating}. This motivates the development of algorithms for problems in chemistry beyond FCI electronic structure~\cite{bauer2020quantum,chen2025framework}. It also motivates study of the quantum resources present in chemical ground states, so that the quantum and classical cost of various heuristics can be stated in terms of these resources. This is the topic of the present paper.  

Entanglement has long been a characteristic of uniquely quantum phenomena, and is the prototypical quantum resource~\cite{schrodinger1935discussion}. However, it has long been understood that entanglement alone is insufficient to explain the power of quantum computation. For example, stabilizer states are highly entangled, yet the Gottesman-Knill Theorem tells us that Clifford operations acting on stabilizer states can be efficiently simulated on classical computers~\cite{MikeNIke}. This motivates the study of resource theory that includes both entanglement and non-stabilizerness. For example, Bravyi and Gosset gave an algorithm for the simulation of circuits over the Clifford$+T$ gate set in which the simulation cost rises exponentially with the number of $T$ gates, but only polynomially with the number of qubits~\cite{bravyi2016improved,bravyi2019simulation}.  The property of non-stabilizerness, also known as magic, has previously been linked with other notions of non-classicality in quantum information theory including contextuality \cite{Howard2014,BermejoVega2017} and Wigner negativity \cite{spekkens2008negativity,kocia2017discrete}. The resource theory of magic has been developed significantly in recent years \cite{Veitch2014,BermejoVega2017}, with stabilizer R\'{e}nyi entropy (SRE) emerging as a convenient measure for the amount of magic in any given quantum state \cite{LeoneSREPhysRevLett.128.050402,LeoneMagicMonotonePhysRevA.110.L040403, bittel2025operational}. Previous work has connected the amount of magic in a given quantum state to the number of non-Clifford gates required to prepare it \cite{haug2025probing-d19,Oliviero2022measuringmagic,beverland2020lower}. There is also a growing body of literature which considers the quantum computational resources themselves as measures related to physical structures \cite{PhysRevB.103.075145, Robin2024,Chernyshev2025,PhysRevB.111.L081102,Cao2024,White2024,Gargalionis2025,Aoude2025,tarabunga2023many,Oliviero2022,turkeshi2023pauli-14f,CERNMagic}.

In this paper we connect the resources which distinguish quantum from classical computation to the physics of electronic structure Hamiltonians. We examine the SRE of a given quantum state as a distance measure relative to a reference stabilizer state  \cite{bittel2025operational}. We show that, when the state has large overlap with the reference stabilizer state, there is a linear relationship between the SRE and the overlap. For chemical systems, this is equivalent to saying the magic is proportional to the Hartree-Fock reference weight, an established measure of electronic correlation \cite{xu2023all,helgaker2013molecular,lee1989diagnostic-bca,izsak2023measuring}. The requirement that the overlap with a reference stabilizer state is large is equivalent to stating that the state is weakly- or moderately-correlated in electronic structure Hamiltonians. We then show that, within the weakly- and moderately-correlated regime, the correlation energy recovered by a post-Hartree-Fock state is linearly proportional to its magic.

Drawing on the established connection between magic and contextuality~\cite{Howard2014,BermejoVega2017,spekkens2008negativity,kocia2017discrete}, we utilize the contextual subspace (CS) method to probe the role of magic in physical systems. First introduced as part of the contextual subspace variational quantum eigensolver \cite{kirby2021contextual,Ralli_UP_CS-VQE,weaving2023stabilizer}, the CS method partitions the Pauli Hamiltonian $H$ into two portions: a \textit{non-contextual} $H_{nc}$, and a \textit{contextual} correction $H_{c}$. Given a partition that maximizes the number of Pauli terms in $H_{nc}$, one can construct a sequence of partitions by moving terms from $H_{nc}$ to $H_{c}$.
This allows us to construct approximate solutions to $H$ that vary from the ground state of $H_{nc}$ to the true ground state of $H$. We refer to these approximate Hamiltonians as the \textit{contextual subspace} Hamiltonians $H_{cs}$.

We provide a theoretical framework for the role of magic in the CS method. We demonstrate that the act of projecting a given quantum state into a contextual subspace monotonically decreases the magic of that state. We also show that, in the regime where magic is linearly related to the overlap with a stabilizer reference state, the CS method constructs a family of approximate ground-states where the magic of each ground-state increases monotonically with increasing contextual subspace size.

We present numerical simulations of a wide range of molecules at multiple bond lengths described in Section \ref{subsec:database_overview}. We find numerical evidence for the linear relationship between magic and stabilizer overlap, both in FCI ground-states and in ground-states of the contextual subspace Hamiltonians. We also validate the monotonic relationship between magic in CS ground-states and contextual subspace size. 

We then present numerical evidence that, for weakly and moderately correlated molecules where the overlap between the FCI and HF ground-states is large, there is a linear relationship between the magic of CS ground-states and the portion of the correlation energy those ground-states recover. This linear relationship is robust across a wide range of molecules and a wide range of bond-lengths. We find that the linear relationship degrades significantly at the Coulson-Fischer point \cite{coulson1949xxxiv} where the molecule takes on a multi-reference, strongly-correlated character.

The paper is structured as follows. Section \ref{sec:Background} provides an introduction to quantum chemistry and correlation energy (Section \ref{subsec:correlation}), contextuality and magic as measures of non-classicality (Section \ref{subsec:contextuality}), stabilizer R\'{e}nyi entropy (Section \ref{subsec:sre}), and the contextual subspace projection method (Section \ref{subsec:csp}). Section \ref{sec:analytical} presents our theoretical framework. In Section \ref{subsec:magic_and_correlation} we show the relationship between SRE and overlap with a reference stabilizer state is linear when that overlap is large, and then apply this result to show the correlation energy recovered by a post-Hartree-Fock state is proportional to its magic. In Section \ref{subsec:cs_magic} we introduce a resource theoretic framework for magic in the CS method, and prove that projection into the contextual subspace monotonically reduces the magic of a state. We then use this result to show the relationship between magic and stabilizer overlap in CS ground-states is monotonic, provided that the stabilizer overlap is large. Section \ref{sec:methodology} establishes the methodology for numerical simulation and details the electronic structure Hamiltonian dataset. Section \ref{sec:results} presents the results of numerical simulations validating the theoretical results of Section \ref{sec:analytical}. Finally, Section \ref{sec:Discussion} provides a discussion of the results and potential future avenues for research. 

\section{Background}\label{sec:Background}
This section introduces the concepts required to understand the setting of this work. In Section \ref{subsec:correlation} we define the electronic structure Hamiltonian and correlation energy. In Sections \ref{subsec:contextuality} and \ref{subsec:sre} we discuss non-stabilizerness and contextuality as ideas of non-classicality and present the measure stabilizer R\'{e}nyi entropy to quantify the non-stabilizerness of a given quantum state. Finally in Section \ref{subsec:csp} we introduce a framework for the contextual subspace (CS) method which we use to vary the amount of contextuality in electronic structure Hamiltonians.

\subsection{Quantum chemistry and correlation energy}\label{subsec:correlation}
The objective of the electronic structure problem is to solve the time-independent Schr\"odinger equation
\begin{equation}\label{eq:schrodinger}
    H_\mathrm{mol}(\vec{r})\psi(\vec{r})=E\psi(\vec{r}),
\end{equation}
 where $H_\mathrm{mol}$ is the molecular electronic structure Hamiltonian, $\psi(\vec{r})$ is the multi-particle wavefunction, and $E$ is the eigenenergy associated with $\psi(\vec{r})$. Under the Born-Oppenheimer approximation, $H_\mathrm{mol}$ can be written in atomic units in second-quantized form as \cite{helgaker2013molecular} 
\begin{equation}\label{eq:molham2q}
    H_\mathrm{mol}=\sum_{p,q}^N h_{pq} a^\dagger_pa_q +\frac{1}{2}\sum_{p,q,r,s}^N g_{pqrs} a^\dagger_p a^\dagger_r a_s a_q + h_\mathrm{nucl},
\end{equation}
where $N$ is the number of orbitals included in the simulation, $a^\dagger_i$ and $a_i$ are creation and annihilation operators respectively acting on occupation number vectors of Slater determinants, $h_{pq}$ and $g_{pqrs}$ are the one- and two-electron integrals respectively, and $h_\mathrm{nucl}$ is the energy resulting from electronic repulsion between the nuclei which under the Born-Oppenheimer approximation constitutes a fixed offset for given nuclear coordinates.

If all basis functions are included, an approach known as \emph{full configuration interaction} or FCI, simulating a molecule with $\eta$ electrons in a basis set consisting of $N$ orbitals requires $\begin{pmatrix}
    N\\\eta
\end{pmatrix}=\frac{N!}{\eta!(N-\eta)!}$ Slater determinants \cite{helgaker2013molecular}. The result of simulation is a ground state $\ket{\psi_{FCI}}$ with associated energy $E_{FCI}$. The factorial dependence on the size of the basis set has historically limited the practicality of FCI simulation on classical computers to only relatively simple molecules, with the current state-of-the-art being a simulation of propane, C$_3$H$_8$, involving approximately 1 trillion Slater determinants \cite{gao2024distributed}. 

The \emph{Hartree-Fock} (HF) method approximates the exact many-body wavefunction using the lowest energy Slater determinant, which can be identified classically using an iterative self-consistent field (SCF) algorithm \cite{hartree1928wave,fock1930naherungsmethode,roothaan1951new,hall1951molecular,helgaker2013molecular}. The primary computational advantage of the HF framework is that it maps the exponentially scaling many-body problem onto a continuous polynomial-time optimization. By iteratively solving the self-consistent field (SCF) equations, HF usually achieves $O(N^4)$ scaling, making classical simulation of large molecules tractable \cite{helgaker2013molecular}.  

The most commonly-used form of Hartree-Fock is restricted Hartree-Fock (RHF) which enforces the requirement that the solution be an eigenstate of the total spin operator $S^2$ by keeping electrons in pairs \cite{helgaker2013molecular}. As a result, RHF solutions become unphysical for molecular systems at long bond lengths where the spin singlet restriction artificially forces the wavefunction to retain incorrect ionic character \cite{helgaker2013molecular,lewars2003computational}. At the \emph{Coulson-Fischer} (CF) point, the ground state takes on strong multi-reference character and is not well characterized by a single Slater determinant \cite{coulson1949xxxiv,helgaker2013molecular,Limacher2013}. Beyond the CF point, the \emph{unrestricted Hartree-Fock} (UHF) approximation which does not impose spin restrictions may provide qualitatively better approximations of the ground-state energy. However, because the resulting states exhibit unphysical spin contamination and are no longer eigenstates of the total spin operator $S^2$, it is difficult to rigorously define a single Hartree-Fock state in this regime.

The mean-field nature of HF means it only accounts for the average electron repulsion. Consequently, it fails to capture the full \emph{correlation energy} arising from the instantaneous Coulomb interactions between electrons \cite{lowdincorrelation,lewars2003computational}, which is always negative due to the variational principle. Because the correlation energy is often larger than the chemical accuracy threshold which simulations must achieve to be useful for many applications (generally defined as 1 kcal/mol or 1.6 mHa \cite{pople2003quantum}), Hartree-Fock is frequently not sufficiently accurate. This has led to the development of a family of classical \emph{post-Hartree-Fock} methods which recover a portion of the correlation energy by re-introducing a portion of the many-body terms ignored by Hartree-Fock. Examples of post-Hartree-Fock methods include M\o ller-Plesset perturbation theory, coupled cluster, and DMRG \cite{lewars2003computational,kohanoff2006electronic}.

The difficulty of obtaining the correlation energy with classical computational methods raises the question of its connection to quantum resources. Previous work has connected correlation energy with entanglement \cite{esquivel2015correlation,wang2007quantum-8bf}; we expand this connection to a broader idea of non-classicality measured by contextuality and non-stabilizerness.

\subsection{Measures of non-classicality}\label{subsec:contextuality}

What distinguishes quantum mechanics from classical theories of physics \cite{Belltheorem1964,EPR,BohrEPR1935, Gottesman1999, pusey2012reality,Mermin1993}? Particularly relevant for this work is whether and how non-classical resources allow quantum advantage in quantum computation and information theory \cite{KwonNonclassPhysRevLett.122.040503, GokhaleMeasCostVQEMolecHamArxiv, BennettTeleportationPhysRevLett.70.1895}. Previous work has established multiple measures of non-classicality, including \textit{contextuality} and \textit{non-stabilizerness} \cite{kenfack2004negativity, Howard2014, kocia2017discrete, Gottesman1999, BermejoVega2017, spekkens2008negativity}. While different ideas, contextuality and non-stabilizerness generally describe the same phenomenon, and in recent years their formulations have been shown to be equivalent in many settings \cite{Howard2014, BermejoVega2017, kocia2017discrete, spekkens2008negativity}. In this section we will describe how contextuality and non-stabilizerness are both examples of non-classicality. 

Non-classicality represented by the concept of contextuality was famously demonstrated in the Bell-Kochen-Specker (BKS) theorem \cite{KS1967, Mermin1993}. Suppose that we have two observables $A$ and $B$ with measured values $v(A)$ and $v(B)$. Under a classical (non-contextual) hidden variable model where the outcomes of all measurements are pre-determined, we would expect that for any arbitrary operation $f$, the system would preserve the equality $v(f(A,B))=f(v(A),v(B))$.
For example, we would have $v(AB)=v(A)v(B)$ and $v(A+B)=v(A)+v(B)$. The BKS theorem proves that for certain sets of quantum observables which we call contexts, it is impossible to assign classical hidden variables for each observable without encountering a contradiction \cite{MerminBKS10.1103/physrevlett.65.3373, Mermin1993}. This contradiction is strictly non-classical and uniquely present in quantum mechanical systems, with the dependence of measurement outcomes on what context the observables are measured in referred to as contextuality.

Non-stabilizerness or \textit{magic} is another notion of non-classicality \cite{Howard2014, spekkens2008negativity, BermejoVega2017}. For the set $\mathcal{P}^{(n)}$ of $n$-qubit Pauli strings, a set of Pauli terms $\{P_i\}\subseteq \mathcal{P}^{(n)}$ \textit{stabilizes} an $n$ qubit state $\ket{\psi}$ if $P_i \ket{\psi} = \ket{\psi}$ for all $\{P_i\}\in \mathcal{P}^{(n)}$. The set of states $\ket{\psi}$ which are stabilized by $\{P_i\}$ make up the $+1$ eigenspace of $\{P_i\}$ and are referred to as \textit{stabilizer states}. The Gottesman-Knill theorem states that while the resource cost of simulating an $n$-qubit Hilbert space on a classical computer grows as $2^n$, stabilizer states can be simulated on a classical computer in polynomial time (efficiently) \cite{Gottesman1999, MikeNIke}. The set of stabilizer states is equivalent to the set of states which can be generated by performing only Clifford operations ($n$-qubit unitaries which normalize the $n$-qubit Pauli group) on the all-zero state; non-stabilizer states require non-Clifford operations to generate from the all-zero state. The degree of non-stabilizerness is directly linked to the resource cost of generating the state \cite{Oliviero2022measuringmagic,haug2025probing-d19}.

It is then relevant to ask how we can quantify the amount of non-classicality in a system, as this will represent the necessary contributions of some quantum resource. To do this we look at measures of non-stabilizerness, particularly the stabilizer R\'{e}nyi entropy described in the next section. 

\subsection{Stabilizer R\'{e}nyi Entropy}\label{subsec:sre}
In this work, we utilize the framework of stabilizer R\'{e}nyi entropy (SRE)  \cite{LeoneSREPhysRevLett.128.050402, HaugSRE10.22331/q-2023-08-28-1092} as a measure of non-stabilizerness, which has the advantage of being computable from the state vector without requiring a minimization procedure. For an $n$-qubit pure state $\ket{\psi}$, the $\alpha$-SRE $M_\alpha$ is given by
\begin{equation}\label{eq:SREdef}
    M_\alpha =-\frac{1}{\alpha-1}\log_2[\zeta_\alpha],\quad \zeta_\alpha=\sum_{P\in\mathcal{P}^{(n)}}\frac{\mel{\psi}{P}{\psi}^{2\alpha}}{2^{n}}
\end{equation}
where $\alpha \in \mathbb{R}^+$ and $\mathcal{P}^{(n)}$ is the set of $n$-qubit Pauli strings \cite{LeoneSREPhysRevLett.128.050402}. We note that some literature defines the $\alpha$-SRE in terms of the natural log instead of log base 2; this is equivalent to representing information in nats rather than shannons. Choices of $\alpha$ are continuous, however there are 2 cases which are of particular note. The limit where $\alpha \rightarrow 1$ recovers Shannon entropy, and describes the probability of a Pauli support on the state. $\alpha =2$ is measurable on quantum processors, and is shown to represent distance from a stabilizer state towards a state which is uniform over all normalized states and symmetric across all unitary transformations (Haar random) \cite{rajabpour2025stabilizershannonrenyiequivalenceexact,bittel2025operational, LeoneSREPhysRevLett.128.050402, Sternberg1994groupsphysics, ZyczkowskiSommers2001haarrandom}.
The $\alpha$-SRE is always zero for stabilizer states and any state that is the result of a free operation on stabilizer states (as defined in \cite{LeoneSREPhysRevLett.128.050402}). When looking at an arbitrary $n$-qubit pure state $\ket{\psi}$, the $\alpha$-SRE of the Clifford orbit $\varepsilon_{\psi}$ defined as the collection of all states possible upon the action of the $n$-qubit Clifford group $C^n$ on $\ket{\psi}$: $\varepsilon_{\psi} = \{C\ket{\psi} | C \in C^n\}$, will remain the same. In \cite{bittel2025operational}, the authors prove that the larger the $\alpha$-SRE for a given quantum state, the more closely the Clifford orbit serves as an approximation for a Haar random state, and the more distinguishable from the set of stabilizer states on $n$ qubits. $M_\alpha$ has been proven to be a monotone for non-stabilizerness for $\alpha\geq2$ \cite{LeoneMagicMonotonePhysRevA.110.L040403}. More information on SRE can be found in Appendix \ref{app:srefornerds}. 

Recent works have examined SRE in the context of simulations of the Ising model \cite{Oliviero2022,rattacaso2023stabilizer}, light-front field theory \cite{Sam2025}, and high energy and nuclear physics \cite{tarabunga2023many,Aoude2025,Gargalionis2025,Chernyshev2025,CERNMagic,White2024,Robin2024,Brokemeier2025}. In the context of quantum chemistry, previous work has observed that the SRE of electronic structure Hamiltonian ground states varies dramatically with bond length, with a peak occurring in the region which is most difficult for classical simulation \cite{gu2024zero,sarkis2025are-602}. Additionally, Ref.~\cite{gu2024zero} found that increasing the number of non-Clifford gates in an optimized ansatz circuit approximating an electronic structure ground state increased the quality of the energy approximation provided by the circuit.

The SRE gives us a method for measuring the non-classicality of a system. Below, we describe how the contextual subspace method allows us to vary the non-classicality of a system in a structured way.

\subsection{The contextual subspace method}\label{subsec:csp}

The contextual subspace (CS) method was first introduced by Kirby \emph{et al.} as a component of the contextual subspace variational quantum eigensolver (CS-VQE) \cite{kirby2021contextual}. The CS method partitions the full electronic structure Hamiltonian into a non-contextual part (classically simulable) and a contextual part (contextual correction) \cite{kirby2021contextual}. It was later reformulated by Weaving et al. using the stabilizer formalism, which introduced the ``non-contextual projection ansatz'' to better define the subspace for NISQ devices \cite{weaving2023stabilizer}.

In the CS method, an electronic structure Hamiltonian is mapped to a Pauli Hamiltonian via standard fermion-to-qubit transformations such as Jordan-Wigner or Bravyi-Kitaev mappings \cite{jordan1928ber-654, bravyi2002fermionic,seeley2012bravyi}. The resulting Pauli Hamiltonian then undergoes qubit tapering, a process which utilizes the $\mathbb{Z}_2$ symmetries present to reduce the total number of qubits required for accurate simulation~\cite{bravyi2017taperingqubitssimulatefermionic,doi:10.1021/acs.jctc.0c00113, ralli2025noncontextual}. 

After qubit tapering, the Hamiltonian is separated into contextual and non-contextual parts
\begin{equation}
    H = H_\text{nc} + H_\text{c},
\end{equation}
where the definition for contextuality is given by the conventions presented in \cite{Kirby2019Contextuality, Ralli_UP_CS-VQE}. The symmetries of $H_\text{nc}$ form a stabilizer group with generators $\mathcal{S} = \{G_1, ..., G_n \}$ where the size of the generator set $|\mathcal{S}| = n$ for an $n$-qubit system \cite{Kirby2020, ralli2025noncontextual}. 

The ground state of $H_{\rm nc}$ is $\ket{\text{nc}}$ with associated non-contextual energy $E_{nc}$. In Appendix~\ref{app:ncPauliHam} we provide a definition for non-contextual Pauli Hamiltonians, the stabilizer generator group $\mathcal{S}$, and discuss how the ground state energy is obtained. The ground state lies within a stabilizer subspace identified by a particular assignment to the stabilizer generator eigenvalues. Restriction of the Hamiltonian to this subspace is ensured by the application of operators $ U_{\mathcal{W}}$ and $Q_{\mathcal{W}}$. The operator 
\begin{equation}
U_{\mathcal{W}} = \prod_{G_k \in \mathcal{W}} V_k(G_k)
\end{equation}
rotates each generator $G_k \in \mathcal{W}$  to single-qubit Pauli-$Z$ operators. These $V_k$ operators are all Clifford, except for one. As presented in \cite{weaving2023stabilizer, ralli2025noncontextual, Ralli_UP_CS-VQE, ZhaoMeasRedux}, there is a generator which is defined as a linear combination of anti-commuting Pauli operators which has been rotated to a single Pauli operator via \textit{unitary partitioning}. We denote this generator as $G_A$. The unitary partitioning rotation, $R$, is applied to the Hamiltonian when the generator associated with the anti-commuting operator's classical hidden variable value assignment is enforced. This results in 
\begin{equation}
    U_\mathcal{W} = \left\{
    \begin{array}{lr}
        \prod_{G_k \in \mathcal{W}}V_k(G_k) & \text{if }G_A \notin \mathcal{W} \\
        R\left(\prod_{G_k \in \mathcal{W}}V_k(G_k)\right) & \text{if }G_A \in \mathcal{W}
    \end{array}
    \right\},
\end{equation} 
For more details on this operator and its subtleties see Section 4.1 before Equation (19) in \cite{weaving2023stabilizer} and Appendix \ref{app:ncPauliHam}. 

The operator $Q_{\mathcal{W}}$ is a projector onto the subspace which is consistent with value assignment of our non-contextual stabilizers $G_k \in \mathcal{W}$. We introduce the fixed state $\ket{\psi_{\text{fixed}}}$:
\begin{equation}
    \ket{\psi_\mathrm{fixed}} = \bigotimes_{G_k \in \mathcal{W}}\ket{i}_k \left\{
    \begin{array}{lr}
        i=0 & \text{if } \langle G_k \rangle = +1\\
        i=1 & \text{if } \langle G_k \rangle = -1
    \end{array}
    \right\},
\end{equation}
where $k$ denotes the qubit index where the stabilizers $G_k$ in the set $\mathcal{W}$ are fixed. The operator $Q_\mathcal{W}$ is then
\begin{equation}
    Q_\mathcal{W} = \ket{\psi_\mathrm{fixed}}\bra{\psi_\mathrm{fixed}} \otimes \mathbb{I}_{n-|\mathcal{W}|}.
\end{equation}

We know that for every stabilizer we fix, we are ensuring the eigenvalue of a single qubit Pauli $Z$ operator, so the effective fixing of the expectation value $\langle G_k \rangle \in \{-1, +1\}$ for $G_k \in \mathcal{W}$ resembles the stabilizer projective measurement on the $k$-th qubit
\begin{align}\label{eq:qkdef}
    Q_\mathcal{W}^k &= \frac{1}{2}(\mathbb{I}_2 + \langle G_k \rangle Z_k).
\end{align}
Then the collective application of projective measurements on all qubits fixed in the stabilizer set $\mathcal{W}$ is:
\begin{equation}\label{eq:qwdef}
    Q_\mathcal{W} = \bigotimes_{k=1}^\mathcal{W}Q_\mathcal{W}^k=\bigotimes_{k=1}^\mathcal{W}\frac{1}{2}(\mathbb{I}_2 + \langle G_k \rangle Z_k).
\end{equation}
The action of $Q_\mathcal{W}$ on a general $n$ qubit state $\ket{\phi}$ takes the form 
\begin{equation}\label{eq:projstate}
    Q_\mathcal{W}\ket{\phi} = \ket{\psi_\mathrm{fixed}}\bra{\psi_\mathrm{fixed}}\phi\rangle_{|\mathcal{W}|} \otimes \ket{\phi}_{n-|\mathcal{W}|}.
\end{equation}
The action of $Q_\mathcal{W}$ can then also be understood through the \emph{stabilizer nullity} $\nu$ defined as
\begin{equation}\label{eq:stabnullitydef}
    \nu(\ket{\phi})=n-\log_2\vert\mathrm{STAB}(\ket{\phi})\vert
\end{equation}
where $\mathrm{STAB}(\ket{\phi})=\{P \in \mathcal{P}^{(n)}\}:\ P\ket{\phi}=\ket{\phi}\}$ is the set of all $n$-qubit Paulis which stabilize $\ket{\phi}$ \cite{beverland2020lower}. In the context of our generator formalism, we can think of the set $\mathcal{W}\subseteq \mathcal{S}$ as generating a set of stabilizers $\mathcal{P}^{\mathcal{W}} =\langle G_1, G_2,...,G_{|\mathcal{W}|} \rangle = \{P_i\}_{i=1}^{2^{|\mathcal{W}|}}$ such that $\mathcal{P}^{\mathcal{W}}\subseteq \text{STAB}(\ket{\phi})$.
By manually requiring $2^{\vert\mathcal{W}\vert}$ Paulis to stabilize $\ket{\phi}$, $Q_\mathcal{W}$ upper-bounds the stabilizer nullity such that
\begin{equation}
    \nu\leq \nu^{\vert\mathcal{W}\vert}=n-\vert\mathcal{W}\vert.
\end{equation}

 Applying the operators $ U_{\mathcal{W}}$ and $Q_{\mathcal{W}}$ to the full Hamiltonian yields the contextual subspace Hamiltonian  
\begin{equation}
    H_{cs} = Q_{\mathcal{W}}^\dagger U_{\mathcal{W}}^\dagger H U_{\mathcal{W} }Q_{\mathcal{W}}.
\end{equation}

In previous literature on the CS method \cite{Ralli_UP_CS-VQE, ralli2025noncontextual} the rotation $R$ is only applied if the clique representative is included in $\mathcal{W}$, however in the code package for the CS method, Symmer \cite{symmer}, the rotation is always initially applied under the $\mathtt{StabilizeFirst}$ method which we use. Motivated by this shift, we apply all rotations to diagonalize the stabilizer generators, and we define the contextual subspace Hamiltonian as 
\begin{equation}
    H_{cs}^\mathcal{W} = Q_{\mathcal{W}}^\dagger U_{\mathcal{S}}^\dagger H U_{\mathcal{S} }Q_{\mathcal{W}}.
\end{equation}
The rotations in $U_{\mathcal{S}}$ could cause a change in sign for projectors in $Q_{\mathcal{W}}$. This will appear for $Q_\mathcal{W}^j$ when there is a unitary $V_k(G_k)$ which does not commute with the stabilizer $G_j$. This sign change effectively switches the classical hidden variable value assignment for the stabilizer $G_j$ from $\pm 1$ to $\mp 1$. This change in sign is classically efficient to track \cite{weaving2023stabilizer, Ralli_UP_CS-VQE}. 

For each stabilizer fixed by $Q_{\mathcal{W}}$, the number of active qubits in the contextual subspace Hamiltonian is reduced by one. When $|\mathcal{W}| = n$, all stabilizers are fixed and there are no active qubits in the contextual subspace Hamiltonian, resulting in a fully non-contextual approximation of the ground-state energy. Conversely, when $|\mathcal{W}| = 0$, the rotation operators are equivalent to the identity, $\mathbb{I}_n $, on the full $2^n$-dimensional Hilbert space. We can construct a family of Hamiltonians with $1\leq|\mathcal{W}|\leq n-1$, where the number of active qubits ranges from $n-1$ to $1$. Hamiltonians in this family have associated Hilbert spaces ranging with $|\mathcal{W}|$ with dimensions $2^{n-|\mathcal{W}|}$. The CS method then represents the trade-off between quantum resource requirements through the amount of active qubits, and accuracy in the simulated Hamiltonian. Intuitively, this trade-off can be thought of as exchanging contextuality in a given Hamiltonian for ease of simulation. For a given Hamiltonian, the family of CS Hamiltonians have associated stabilizer sets $\{\mathcal{W}^i\}_{i=1}^{n-1}$ for $i = \vert\mathcal{W}^i\vert$ such that $\mathcal{W}^1\subset\mathcal{W}^2\subset \dots \subset \mathcal{W}^{n-1}$. The family of CS Hamiltonians can be thought of as a collection of approximate Hamiltonians with varying contextuality. This family of CS Hamiltonians have ground states $\ket{\text{cs}^\mathcal{W}}$ with associated ground state energy $E_{\text{cs}}^\mathcal{W}$. 

A challenge for implementing the CS method is determining which stabilizers should be fixed to yield a contextual subspace Hamiltonian with the desired number of active qubits, $\nu^\mathcal{W}$. As is well known, a stabilizer space on $n$ qubits contains $n$ independent stabilizer generators \cite{MikeNIke, weaving2023stabilizer}. The identification of optimal stabilizers is a combinatorial optimization problem which we solve by brute force, searching all $\sum_{i=1}^n {{n}\choose{i}} = 2^n -1$ possible stabilizer choices for all choices of $1 \leq \nu^\mathcal{W} \leq n$ active qubits to find the set of stabilizers $\mathcal{W}_\mathrm{opt}$ which minimize the energy error. 

For all Hamiltonians in this work, we perform an initial qubit tapering derived in \cite{bravyi2017taperingqubitssimulatefermionic}. The subsequent reduction in qubits by the enforcement of value assignments to stabilizer generators is referred to as \emph{qubit tapering} for the remainder of this work.

While previous work \cite{kirby2021contextual,Ralli_UP_CS-VQE,weaving2023stabilizer,weaving2025contextual} has developed the CS method as a tool for reducing resource costs for NISQ algorithms, in this paper we ask how the magic of CS ground states changes as the Hamiltonian is projected into different contextual subspaces. In Section \ref{sec:analytical}, we present an analytical framework for understanding how the CS method manipulates the magic of physical systems.

\section{Theory}\label{sec:analytical}
In this section, we develop a theoretical framework for the role of magic in the electronic structure problem, which we apply to the CS method described in Section \ref{subsec:csp}. In Section \ref{subsec:magic_and_correlation}, we connect the magic of a state to its overlap with a stabilizer reference state. We then show that the correlation energy recovered by a post-Hartree-Fock state is proportional to its magic. We then give a physical interpretation of this result for electronic structure Hamiltonians directly linking ground state magic with electronic correlation for weakly- and moderately-correlated molecules. In Section \ref{subsec:cs_magic}, we provide a theoretical framework for the role of magic in the CS method and show that it can be understood under the framework for post-Hartree-Fock methods established in Section \ref{subsec:magic_and_correlation}. We then conclude by providing a physical interpretation of the CS method for weakly- and moderately-correlated electronic structure Hamiltonians as a mechanism for constructing approximate Hamiltonians that systemically reduce the electronic correlation.

\subsection{Connecting magic and correlation}\label{subsec:magic_and_correlation}
In this section, we demonstrate that for states which have high overlap with a stabilizer state, the magic of the state is linearly related to the overlap. We then provide a physical interpretation of this result for molecular electronic structure Hamiltonians which relates magic to electronic correlation. 
\begin{theorem}\label{th:overlaptheorem}
    For a state $\ket{\psi}$ which has large overlap with a stabilizer state $\ket{\psistab}$ where $\braket{\psi}{\psi_\textrm{STAB}} \approx 1$, the $\alpha$-SRE of $\ket{\psi}$ is given by
    \begin{equation*}
        M_\alpha(\ket{\psi})\approx\frac{2\alpha}{(\alpha-1)\ln(2)}\left(1-\vert\braket{\psi}{\psistab}\vert^2\right)
    \end{equation*}
for integer $\alpha\geq 2$.
\end{theorem}

The proof is as follows. Without loss of generality, we consider an $n$-qubit state written as 
\begin{equation}\label{eq:psi_stab_expand}
    \ket{\psi}=c_0 \ket{\phi_0}+c_1\ket{\phi_1}+\cdots+c_{d-1}\ket{\phi_{d-1}}
\end{equation}
where $\ket{\phi_0}$ is the stabilizer state which maximizes $\vert\braket{\phi}{\psi}\vert^2$ and $\{\ket{\phi_1},\ldots,\ket{\phi_{d-1}}\}$ are the stabilizer states which together with $\ket{\phi_0}$ form the orthonormal stabilizer basis generated by the stabilizer group of $\ket{\phi_0}$. We assume the states are ordered such that $\vert c_0\vert^2\geq\vert c_1\vert^2\geq\cdots\geq \vert c_{d-1}\vert^2$.

Guided by the idea first suggested in \cite{bittel2025operational} that $\alpha$-SRE can be understood as a distance measure from stabilizer states for $\alpha\geq2$, we define a distance $\epsilon$ from the nearest stabilizer state as
\begin{equation}\label{eq:epsilondef}
    \epsilon=1-\vert c_0\vert^2.
\end{equation}
We also define rescaled auxiliary stabilizer weights
\begin{equation}
    \tilde{c}_{i>0}=\frac{c_i}{\sqrt{\epsilon}} 
\end{equation}
which have the property $\sum_{i=1}^{d-1} \vert\tilde{c}_i\vert^2=1$. The state $\ket{\psi}$ can then be written
\begin{equation}\label{eq:psi_nearstab}
    \ket{\psi}=e^{i\theta}\left(\sqrt{1-\epsilon}\ket{\phi_0}+\sqrt{\epsilon}\sum_{i=1}^{d-1} \tilde{c}_i\ket{\phi_i}\right),
\end{equation}
where $e^{i\theta}$ is an overall phase factor. The expectation value of some arbitrary Pauli string $P\in\mathcal{P}^{(n)}$ with respect to $\ket{\psi}$ is then
\begin{equation}\label{eq:pexp}
\begin{split}
\langle P\rangle =& \left(1-\epsilon\right)\langle P\rangle_\mathrm{\phi_0}+\epsilon\sum_{i=1}^{d-1} \vert \tilde{c}_i\vert^2\langle P\rangle_{\phi_i} \\
&+ 2\sqrt{\epsilon(1-\epsilon)}\sum_{i=1}^{d-1}\re{\tilde{c}_i\mel{\phi_0}{P}{\phi_i}}\\
&+2\epsilon\sum_{i=1,j\neq i}^{d-1} \re{\tilde{c}_j^*\tilde{c}_i\mel{\phi_j}{P}{\phi_i}}
\end{split}
\end{equation}
where $\langle P\rangle_{\phi_0}=\mel{\phi_0}{P}{\phi_0}$ and $\langle P\rangle_{\phi_i}=\mel{\phi_i}{P}{\phi_i}.$ 

In the language of resource theory, the leading coefficient $\vert c_0\vert^2$ will be equivalent to the stabilizer fidelity \cite{bravyi2019simulation}
\begin{equation}\label{eq:stabfideldef}
    \mathcal{F}_\mathrm{STAB}(\ket{\psi})=\max_{\phi\in\mathrm{STAB}}\vert\braket{\phi}{\psi}\vert^2.
\end{equation}
The stabilizer fidelity also defines the min-relative entropy of magic \cite{liu2022many,bravyi2019simulation}
\begin{equation}\label{eq:minreldef}
    D_\mathrm{min}(\ket{\psi})=-\log_2(\mathcal{F}_\mathrm{STAB}(\ket{\psi})).
\end{equation}
The min-relative entropy of magic $D_\mathrm{min}$ (and thus, the stabilizer fidelity $\mathcal{F}_\mathrm{STAB}$) has been previously shown to provide an upper bound for the SRE $M_2$ as \cite{HaugSRE10.22331/q-2023-08-28-1092}
\begin{equation}\label{eq:minrelbound}
    M_\alpha\leq\frac{2\alpha}{\alpha-1} D_\mathrm{min}.
\end{equation}
We will now show that when $\epsilon=1-\vert c_0\vert^2\ll1$, this bound is approximately saturated.

Because the SRE is invariant under Clifford rotations, we can assume without loss of generality that all $\{\phi_0,\phi_1,\ldots,\phi_{d-1}\}$ are computational basis states. We then observe that the first two terms of Equation \eqref{eq:pexp} will only be non-zero for $P\in \mathcal{P}_Z^{(n)}$ while the final two terms of Equation \eqref{eq:pexp} will only be non-zero for $P\notin \mathcal{P}_Z^{(n)}$, where $\mathcal{P}_Z^{(n)}$ is the set of all $n$-qubit Pauli strings consisting solely of $Z$'s and $\mathbb{I}$'s. Thus, working to first order in $\epsilon$, we find that for any Pauli string $P\in\mathcal{P}^{(n)}$
\begin{equation}
\begin{split}
\langle P\rangle^{2\alpha} = &(1-2\alpha\epsilon)\langle P\rangle_{\phi_0}^{2\alpha}+2\alpha\epsilon\langle P\rangle_{\phi_0}^{2\alpha-1}\sum_{i=1}^{d-1} \tilde{c}_i\langle P\rangle_{\phi_i}\\
&+\mathcal{O}(\epsilon^2).
\end{split}\label{eq:appxpexp2a}
\end{equation}

The Pauli spectrum $\zeta=2^{-n}\sum_{P\in\mathcal{P}^{(n)}} \langle P\rangle^{2\alpha}$ is then
\begin{equation}
\begin{split}
\zeta
=&(1-2\alpha\epsilon)\sum_{P\in\mathcal{P}^{(n)}}\frac{\langle P\rangle_{\phi_0}^{2\alpha}}{2^n}\\
&+2\alpha\epsilon\left(\sum_{i=1}^{d-1}\tilde{c}_i\sum_{P\in\mathcal{P}^{(n)}}\frac{\langle P\rangle_{\phi_0}^{2\alpha-1}\langle P\rangle_{\phi_i}}{2^n}\right).
\end{split}
\end{equation}
Once again invoking the fact that all $\{\ket{\phi_i}\}$ are computational basis states, we note that $\langle P\rangle_{\phi_0}=\pm1$ and thus $\langle P\rangle_{\phi_0}^{2\alpha-1}=\langle P\rangle_{\phi_0}$ for integer $\alpha\geq2$. We then see that 
\begin{equation}
    \sum_{P\in\mathcal{P}^{(n)}}\langle P\rangle_{\phi_0}\langle P\rangle_{\phi_i}=2^n\delta_{i,0}
\end{equation}
by the trace property $\sum_{P\in\mathcal{P}^{(n)}}\mathrm{Tr}(A P)\mathrm{Tr}(B P)=2^n\mathrm{Tr}(AB)$. We then find that the $\alpha$-SRE will be given by 
\begin{equation}\label{eq:m2epsilon}
	M_\alpha=\frac{1}{1-\alpha}\log_2(\zeta)=\frac{1}{1-\alpha}\log_2(1-2\alpha\epsilon).
\end{equation}
Again assuming $\epsilon\ll1$, this can be expanded to first order in $\epsilon$ as
\begin{equation}\label{eq:SREdistance}
    M_\alpha=\frac{2\alpha}{(\alpha-1)\ln(2)}\epsilon
\end{equation}
or
\begin{equation*}
    M_\alpha=\frac{2\alpha}{(\alpha-1)\ln(2)}\left(1-\vert\braket{\psi}{\psistab}\vert^2\right).\quad \square
\end{equation*}

While Theorem \ref{th:overlaptheorem} is general to all states which have large overlap with a stabilizer state, we will now interpret it in the context of molecular electronic structure Hamiltonians. As discussed in Section \ref{subsec:correlation}, the Hartree-Fock ground-state is defined as the lowest energy Slater determinant. In weakly- or moderately-correlated molecular systems with single reference character, the overlap between the Hartree-Fock and FCI ground states is generally $\gtrsim0.9$ \cite{helgaker2013molecular,lee1989diagnostic-bca}, and thus the Hartree-Fock state will also be the computational basis state with the largest overlap with the FCI ground-state. In these cases, we can identify $\vert\braket{\psi}{\psistab}\vert^2$ with the squared Hartree-Fock weight $c_0^2$, which has long been used in quantum chemistry as a measure of the strength of electronic correlation \cite{xu2023all,helgaker2013molecular,lee1989diagnostic-bca,izsak2023measuring}. We can thus physically interpret Theorem \ref{th:overlaptheorem} as the statement that for single-reference molecules, the magic $M_2$ of the FCI ground state $\ket{\psi_\mathrm{FCI}}$ is determined by the strength of electronic correlation.

Additionally, Theorem~\ref{th:overlaptheorem} can be extended to the Fubini-Study metric, to which all reasonable statistical distance measures (including the min-relative entropy we are using) between quantum states must be reduced in the second order \cite{bengtsson2017geometryquantstates}. The Fubini-Study distance $s_{\text{FS}}(\psi, \phi)$ is defined for two normalized states $\ket{\psi}=\sum_{i}\sqrt{p_i}e^{i \mu_i}\ket{e_i}$ and $\ket{\phi}=\sum_{i}\sqrt{q_i}e^{i \nu_i}\ket{e_i}$, as
\begin{equation}
    \vert\braket{\phi}{\psi}\vert = \cos(s_\text{FS}).
\end{equation}

When the overlap of two states is sufficiently large, the corresponding Fubini-Study distance is sufficiently small so that it can be denoted as $ds_\text{FS}$. Under this assumption,  
\begin{equation}
    \vert\braket{\phi}{\psi}\vert^2 = \cos^2(ds_\text{FS}) 
    =\frac{1+\cos(2 ds_\text{FS})}{2}\approx 1 - ds^2_\text{FS}.
\end{equation} Thus, we can translate the relation of Theorem~\ref{th:overlaptheorem} to the relation between SRE and the Fubini-Study metric as
\begin{equation}
    M_\alpha\approx\frac{2\alpha}{(\alpha-1)\ln(2)}ds^2_\text{FS}(\psi, \psistab).
\end{equation}

This provides corroboration of \cite{bittel2025operational}, where the $\alpha$-SRE for $\alpha \geq 2$ can be interpreted as a distance measure.

We next demonstrate that the amount of magic in a post-Hartree-Fock ground state is proportional to the correlation energy of that state.

\begin{theorem}\label{th:linearity}
    Let $H$ be an electronic structure Hamiltonian with a non-degenerate ground state. Let $\ket{\mathrm{HF}}$ be the Hartree-Fock state for $H$, and $\ket{\psi}$ be an approximate ground-state constructed using a post-Hartree-Fock method with associated perturbation $V$ to the Hartree-Fock Hamiltonian $H_0$. Then to leading order
  \begin{equation}
        |E_\psi-E_\mathrm{HF}|\propto M_\alpha(\ket{\psi})
    \end{equation}
    for integer $\alpha\geq2$ provided that $\|V\|_\mathrm{op}$ is small relative to the spectral gap of the Hartree-Fock Hamiltonian $H_0$.
\end{theorem}
The proof is as follows. Post-Hartree-Fock methods can be understood in a Hamiltonian formalism as constructing Hamiltonians of the form 
\begin{equation}
    H=H_0+V
\end{equation}
where $H_{0}$ are the on-diagonal terms in Equation \eqref{eq:molham2q} which provide the Hartree-Fock energy $E_\mathrm{HF}$ and $V$ are a subset of the off-diagonal terms in Equation \eqref{eq:molham2q} which are reintroduced by the post-Hartree-Fock method. We assume that the computational basis is the basis of Slater determinants, and thus $H_0$ will be entirely diagonal while $V$ is entirely off-diagonal. We then utilize Rayleigh-Schr\"odinger perturbation theory, which for electronic structure Hamiltonians is also known as the M\o ller-Plesset approach and write the Hamiltonian as
\begin{equation}
    H=H_0 +\delta \tilde{V},
\end{equation}
where 
\begin{equation}
    \delta=\frac{\|V\|_\mathrm{op}}{\|H_0\|_\mathrm{op}},\quad \tilde{V}=\frac{V}{\delta}.
\end{equation}
The leading correction to the energy provided by the post-Hartree-Fock method for small $\delta$ will then be 
\begin{equation}\label{eq:en_corr}
    E^{(2)}=-\delta^2 \sum_{i=1}^{d-1} \frac{\left\vert\mel{\phi_i}{\tilde{V}}{\mathrm{HF}}\right\vert^2}{\Delta_{i0}}
\end{equation}
where $\{\ket{\phi_i}\}$ is the set of computational basis states which together with $\ket{\mathrm{HF}}$ form a complete orthonormal basis with $\ket{\phi_0}=\ket{\text{HF}}$ and 
\begin{equation}
    \Delta_{i0}=\mel{\phi_i}{H_0}{\phi_i}-E_{\mathrm{HF}}.
\end{equation}
The $\mathcal{O}(\delta)$ term vanishes due to $V$ being entirely off-diagonal.

The associated (normalized) first-order corrected ground-state will be
\begin{equation}\label{eq:psi_perturb}
    \ket{\psi^{(1)}}=\frac{1}{\mathcal{N}}\left(\ket{\mathrm{HF}}-\delta \sum_{i=1}^{d-1} \frac{\mel{\phi_i}{\tilde{V}}{\mathrm{HF}}}{\Delta_{i0}} \ket{\phi_i}\right)
\end{equation}
where 
\begin{equation}
\mathcal{N}=\sqrt{1+\delta^2\sum_{i=1}^{d-1}\frac{|\mel{\phi_i}{\tilde{V}}{\mathrm{HF}}|^2}{\Delta_{i0}^2}}
\end{equation}
is an overall normalization factor.

For $\delta\ll1$ we can write $\ket{\psi^{(1)}}$ in the form of Equation \eqref{eq:psi_nearstab} by making the identifications
\begin{equation}\label{eq:eps_ctilde_perturbdef}
    \epsilon=\delta^2\sum_{i=1}^{d-1} \frac{|\mel{\phi_i}{\tilde{V}}{\mathrm{HF}}|^2}{\Delta_{i0}^2},\quad \tilde{c}_i=\frac{\delta}{\sqrt{\epsilon}} \frac{\mel{\phi_i}{\tilde{V}}{\mathrm{HF}}}{\Delta_{i0}}.
\end{equation}

The leading order correlation energy $E^{(2)}$ can then be written as 
\begin{equation}\label{eq:corre_eps}
    E^{(2)}=-\epsilon\sum_{i=1}^{d-1} \left\vert\tilde{c}_i\right\vert^2\Delta_{i0},
\end{equation}
which, utilizing Theorem \ref{th:overlaptheorem}, is equivalent to
\begin{equation}\label{eq:corr_vs_magic}
    \left \vert E_{\psi}-E_\mathrm{HF}\right\vert=\frac{(\alpha-1)\ln(2)}{2\alpha}M_\alpha(\ket{\psi})\sum_{i=1}^{d-1} \vert \tilde{c}_i\vert^2 \Delta_{i0}
\end{equation}
for integer $\alpha \geq2.\quad\square$ 

Because $\sum_{i=1}^{d-1}\vert \tilde{c}_i\vert^2=1$, the constant of proportionality relating the correlation energy to the magic can be understood as a weighted average of the relative energies of excited states which sets the overall energy scale for the Hamiltonian. This result could also be obtained by applying Theorem \ref{th:overlaptheorem} to the widely used Davidson correction \cite{DavidsonCorrection,helgaker2013molecular}; we include the above presentation as it is theoretically grounded while the Davidson correction is empirically justified.

Theorem \ref{th:linearity} can be understood as formalizing and generalizing the results of \cite{gu2024zero}, which found that increased non-Clifford gate counts in quantum circuits improved the correlation energy recovered by those circuits for simulations of H$_2$O and LiH. Because $\mathcal{O}(\log n)$ non-Clifford gates induce a linear increase in magic \cite{Oliviero2022measuringmagic,haug2025probing-d19}, increasing the number of non-Clifford gates in an ansatz increases its magic and thus, by Theorem \ref{th:linearity}, its correlation energy. 

We find in Theorem \ref{th:overlaptheorem} an analytic linear relationship between states with large stabilizer overlap and the amount of magic those states possess. For electronic structure Hamiltonians, this has a natural physical interpretation: the magic is determined by the level of electronic correlation. In Theorem \ref{th:linearity}, we show that this implies the magnitude of the correlation energy of a ground-state is proportional to its magic. While this result is general to any post-Hartree-Fock ground-state, we next look to the CS method introduced in Section \ref{subsec:csp}, where we are able to vary the amount of contextuality in a given Hamiltonian. We explore the CS method and its relationship to magic in Section \ref{subsec:cs_magic} below.

\subsection{Magic in the CS method}\label{subsec:cs_magic}
As discussed in Section \ref{subsec:contextuality}, previous literature has established that magic and contextuality are closely related notions of non-classicality \cite{Howard2014,spekkens2008negativity,kocia2017discrete,BermejoVega2017}. Given this connection, we ask how the CS method described in Section \ref{subsec:csp} changes the magic of a system, and in this section we develop a theoretical framework for the role of magic in the CS method. We first prove that projecting a given state into a contextual subspace monotonically reduces the magic of that state. We then show that in the regime where Theorem \ref{th:overlaptheorem} is valid, the ground-states of contextual subspace Hamiltonians monotonically decrease in magic as more generators are fixed. We then discuss how this allows for a physical interpretation of magic in the CS method for electronic structure Hamiltonians.

\subsubsection{Magic monotonicity of the CS projectors}
\label{subsec:proj_monotone}
In this section, we show the monotonic relation between the size of a Hilbert space governed by the contextual subspace method and the $\alpha$-SRE of a particular state projected into that Hilbert space for $\alpha \geq 2$ by using the results found in \cite{LeoneMagicMonotonePhysRevA.110.L040403}.

\begin{lemma}\label{th:proj_monotone}
    Let $\ket{\psi}$ be a general $n$-qubit quantum state. If the unitary partitioning rotation $R$ associated with the anti-commuting stabilizer generator $G_A$ is applied \textit{initially} to the full Hamiltonian such that $H \mapsto R^\dagger HR$ and $\ket{\psi}\mapsto\ket{\psi'} = R^\dagger\ket{\psi} $, then $\ket{\psi^\mathcal{W}} = Q^\dagger_\mathcal{W}\left(\prod_{G_k \in \mathcal{W}}V_k(G_k)\right)^\dagger\ket{\psi'} $ is the state $\ket{\psi}$ projected into the contextual subspace consistent with $\mathcal{W}$ for all $\mathcal{W}\subseteq \mathcal{S}$, and
    \begin{equation}\label{eq:proj_monotone}
    M_\alpha\left(\ket{\psi^\mathcal{W}}\right)\leq M_\alpha(\ket{\psi'})
    \end{equation}
    for $\alpha\geq2$.
\end{lemma}

The proof follows directly from the definitions of $Q^\dagger_\mathcal{W}\left(\prod_{G_k \in \mathcal{W}}V_k(G_k)\right)^\dagger$ in conjunction with Theorem 1 from Ref.~\cite{LeoneMagicMonotonePhysRevA.110.L040403}. 

We consider the structure of the two operators $\prod_{G_k \in \mathcal{W}}V_k(G_k)$ and $Q_\mathcal{W}$ which enforce the CS method tapering presented in Section~\ref{subsec:csp}. We note that $\prod_{G_k \in \mathcal{W}}V_k(G_k)$ is a sequence of purely Clifford rotations which is used to map the resulting set $\mathcal{W}$ of mutually commuting Pauli strings to single-qubit Pauli-$Z$ operators. Because this consists entirely of free operations within the resource theory of magic, it leaves the $\alpha$-SRE invariant, trivially satisfying the monotonicity condition \cite{LeoneMagicMonotonePhysRevA.110.L040403}.

The second and less trivial operator to examine is the projector $Q_\mathcal{W}$. As mentioned in Section \ref{subsec:csp}, we know $Q_\mathcal{W}$ projects the Hamiltonian and the state it acts on into the subspace consistent with the value assignments of the stabilizers $G_k \in \mathcal{W}$ as
\begin{equation}\label{eq:Qproj}
    Q_\mathcal{W} =\prod_{G_k \in \mathcal{W}} \frac{1}{2}(\mathbb{I}_2 + \langle G_k \rangle Z_k).
\end{equation}
Therefore to fix a single stabilizer $P_k \in \mathcal{W}$ we have 
\begin{equation}
     Q_\mathcal{W}^k = \frac{1}{2}(\mathbb{I}_2 + \langle G_k \rangle Z_k).
\end{equation}

Without loss of generality, we assume that for $Q_\mathcal{W}^k, \text{ }k=0$ and our state is 
\begin{equation}
\ket{\psi} = \sqrt{p}\ket{0}_0\otimes\ket{\phi_1}_{1, ...,n} + \sqrt{1-p}\ket{1}_0\otimes \ket{\phi_2}_{1, ...n}
\end{equation}
for $\ket{\phi_1}, \ket{\phi_2} \in \mathbb{C}^{2}\otimes \mathbb{C}^{2(n-1)}$. Then the act of applying the operator on our state is a projective measurement of $\ket{\psi}$ with the collection of subsequent pure states $\{(p, \phi_1), (1-p, \phi_2)\}$, for which according to Corollary 1 in the supplemental material of Ref.~\cite{LeoneMagicMonotonePhysRevA.110.L040403} we obtain monotonicity as
\begin{equation}
    M_{\alpha}(\psi) \geq \min\{M_{\alpha}(\phi_1),M_{\alpha}(\phi_2)\} \geq M_{\alpha}^{\min}(\varepsilon(\psi)).
\end{equation}
Succinctly, $\alpha$-SRE is monotonic under the measurement operation of $Q_\mathcal{W}^kU_\mathcal{W}^k$ represented by $\varepsilon$ for $\alpha \geq 2$. It is of note that this notion of monotonicity is not \textit{strong} monotonicity, which is obtained in \cite{LeoneMagicMonotonePhysRevA.110.L040403} only by linear stabilizer entropies. 

The authors in \cite{LeoneMagicMonotonePhysRevA.110.L040403} state that if the post measurement state is maintained we know that $M_\alpha(\phi_{i+1}) = M_{\alpha}(\ket{i}\bra{i}\otimes \phi_{i+1})$ for $i=0,1$. Therefore we can say generally that for each fixing of a stabilizer $Q_\mathcal{W}^k$, when modeled as acting as a measurement on the general state $\ket{\psi} = \sqrt{p}\ket{0}_k \otimes \ket{\phi_1} + \sqrt{1-p}\ket{1}_k\otimes \ket{\phi_2}$, we retain monotonicity. $\square$

We note that $U_\mathcal{W}$ can contain a non-Clifford operation $R$ in addition to the Clifford rotations previously described. This originates from the non-contextual Hamiltonian structure previously presented in literature \cite{ralli2025noncontextual, Ralli_UP_CS-VQE, weaving2023stabilizer}. In the non-contextual Hamiltonian, one of our symmetry generators, which we term the anti-commuting stabilizer generator, is actually the result of a linear combination of anti-commuting operators which have been rotated using the unitary partitioning technique into a single Pauli term \cite{ZhaoMeasRedux, Ralli_UP_CS-VQE, weaving2023stabilizer, ralli2025noncontextual}. When this stabilizer is enforced we must apply the unitary partitioning rotation $R$ onto the entire Hamiltonian. This action will increase the magic of the ground state by at most $\mathcal{O}(\log(n))$, shown in Appendix \ref{app:taperingCS}. To avoid losing monotonicity of the CS method projections, we must first apply this rotation to the entire Hamiltonian. The application of this rotation $R$ will increase the number of terms in the Hamiltonian by $\mathcal{O}(\vert H_\text{full}\vert (n^2-1))$ \cite{Ralli_UP_CS-VQE}. In practice, the code package used in all simulations presented in this work applies the rotation $R$ initially to the full Hamiltonian $H$, so the increase in magic due to $R$ is already accounted for in our simulations in Section~\ref{sec:results}.

We also can show that if a state exists within the Hilbert space of a contextual subspace defined by $\mathcal{W}$ and then is projected into a contextual subspace defined by $\mathcal{W}'\supset\mathcal{W}$, this results in a monotonic decrease in magic.
\begin{corollary}\label{th:subset_proj_monotone}
    If $\ket{\psi^\mathcal{W}}$ is a quantum state in the contextual subspace generated by $\mathcal{W}$ and $\ket{\psi^{\mathcal{W'}}}$ is the same state projected into the contextual subspace generated by $\mathcal{W}'\supset\mathcal{W}$, then
    \begin{equation}
        M_\alpha\left(\ket{\psi^{\mathcal{W}'}}\right)\leq M_\alpha\left(\ket{\psi^\mathcal{W}}\right)
    \end{equation}
    for $\alpha\geq2$.
\end{corollary}
The proof follows immediately from Lemma \ref{th:proj_monotone}. If we choose to perform additional truncations by fixing more values in the non-contextual stabilizer set $\mathcal{S}$, we will subsequently be using the projection operation described in Equation \eqref{eq:Qproj}. This is a product of projections which each preserve monotonicity, and therefore we retain a monotonic relation between $\alpha$-SRE and CS tapering through the selective fixing of stabilizers in the non-contextual Hamiltonian. $\square$

\subsubsection{Magic monotonicity of the CS ground-states}\label{subsubsec:gs_monotone}
In this section, we use Lemma \ref{th:proj_monotone} and Corollary \ref{th:subset_proj_monotone} to show that the ground states $\{\ket{\mathrm{cs}^\mathcal{W}}\}_{\mathcal{W}=1}^{n-1}$ of the contextual subspace Hamiltonians $\{H_{cs}^\mathcal{W}\}_{\mathcal{W}=1}^{n-1}$ have magic which monotonically decreases with decreasing system size. Succinctly, the magic $M_2(\ket{\mathrm{cs}^\mathcal{W}})$ decreases with increasing $\vert\mathcal{W}\vert$: $M_2(\ket{\mathrm{cs}^{\mathcal{W'}}})\leq M_2(\ket{\mathrm{cs}^{\mathcal{W}}})$ for $\vert\mathcal{W}\vert \leq \vert\mathcal{W'}\vert$.

To utilize the CS method as a mechanism for constructing approximate solutions to a given Hamiltonian $H$ with variable magic, we want to show that projecting $H$ into decreasing contextual subspaces monotonically decreases the magic. While in Lemma \ref{th:proj_monotone} and Corollary \ref{th:subset_proj_monotone} we showed that the act of projecting a given state $\psi$ into the contextual subspace causes a monotonic decrease in magic, this does not automatically guarantee the monotonicity of magic in CS ground-states. At each level of projections in the CS method, the ground state $\ket{\mathrm{cs}^\mathcal{W}}$ is found via diagonalizing the Hamiltonian $H^\mathcal{W}_{cs}$. While the operations which connect $H$ to $H^\mathcal{W}_{cs}$ cause a monotonic decrease in magic, the process of diagonalization is non-Clifford and cannot be classified as a free operation in SRE as defined in \cite{LeoneMagicMonotonePhysRevA.110.L040403}. Thus, we cannot generally assume that the CS ground-states have monotonically decreasing magic purely based on the monotonicity of the CS projection.

As discussed in Section \ref{subsec:csp}, projection into the contextual subspace defined by $\mathcal{W}$ bounds the stabilizer nullity $\nu$ such that $\nu\leq n-\vert \mathcal{W}\vert$. Because the stabilizer nullity provides an upper bound for $M_2$ \cite{LeoneSREPhysRevLett.128.050402}, it is always true that ground-states in a smaller contextual subspace have a smaller maximal $M_2$ compared to ground-states in a larger contextual subspace. However, because the stabilizer nullity $\nu$ only provides an upper bound, this also does not immediately imply the monotonicity of CS ground-states.

\begin{theorem}
    \label{th:gs_monotonicity}
    Let $H$ be a Hamiltonian partitioned as $H=H_{nc}+H_c$ with the ground state of $H_{nc}$ non-degenerate. Let $H_{cs}^\mathcal{W}$ be the contextual subspace Hamiltonian consistent with the generators $\mathcal{W}$. Let $\ket{\text{cs}^\mathcal{W}}$ be the ground-state of $H_{cs}^\mathcal{W}$. Let $M_2\left(\ket{\text{cs}^\mathcal{W}}\right)$ be the magic of $\ket{\text{cs}^\mathcal{W}}$  measured in the computational basis defined by the ground states of $H_{nc}$ projected into the contextual subspace defined by $\mathcal{W}$. Then $M_2\left(\ket{\text{cs}^\mathcal{W}}\right)$ decreases monotonically with increasing $\vert \mathcal{W}\vert$ provided that $\|H_c^\mathcal{W}\|_\mathrm{op}$ remains small relative to the spectral gap of $H_{nc}^\mathcal{W}$.
\end{theorem}
The proof is as follows. Let $\ket{\text{cs}^\mathcal{W}}$ be the ground-state of $H_{cs}^\mathcal{W}$. Let $\ket{\text{cs}^\mathcal{W'}}$ be the ground-state of $H_{cs}^\mathcal{W'}$ with $\mathcal{W}\subset\mathcal{W'}$. If qubit tapering is not performed, then $H_{nc}^\mathcal{W}=H_{nc}^\mathcal{W'}=H_{nc}^\mathcal{S}=U^\dagger_\mathcal{S}H_{nc}U_\mathcal{S}$ and we have
\begin{equation}
    \begin{split}
        H_{cs}^\mathcal{W}&=U^\dagger_{\mathcal{S}}H_{nc}U_\mathcal{S}+Q^\dagger_\mathcal{W}U^\dagger_\mathcal{S}H_{c}U_\mathcal{S}Q_\mathcal{W}\\
        &=H^\mathcal{S}_{nc}+H_c^\mathcal{W}
    \end{split}
\end{equation}
and similarly for $H_{cs}^\mathcal{W'}$. Note that the application of the rotation operator $U_\mathcal{S}$ diagonalizes the non-contextual Hamiltonian, and rotates the contextual Hamiltonian into the consistent basis. Additionally the projector $Q_\mathcal{W}$ ensures that the contextual Hamiltonian only contains Pauli terms which are compatible with the non-contextual hidden variable assignments, i.e. for the stabilizer set $\mathcal{W}$, $Q_\mathcal{W}$ removes Pauli operators which anti-commute with the selected stabilizer generators $G_k \in \mathcal{W}$. For those interested, see Appendix \ref{app:taperingCS} for more details. We can then see that $H_{cs}^\mathcal{W}$ and $H_{cs}^\mathcal{W'}$ are both perturbations of the same parent Hamiltonian $H_{nc}^\mathcal{S}$, and we can utilize the perturbative approach developed in Theorem \ref{th:linearity} provided $\|H_c^\mathcal{W}\|_\mathrm{op}$ remains small relative to the spectral gap of $H_{nc}^\mathcal{S}$.

Let $\{\ket{\phi_i^\mathcal{S}}\}$ be the eigenstates of the rotated non-contextual Hamiltonian $H_{nc}^\mathcal{S}$. Let $\mathcal{I}(\mathcal{W})$ be the subset of the states $\{\ket{\phi_i^\mathcal{S}}\}$ for which $\mel{\phi_i^\mathcal{S}}{Q_\mathcal{W}}{\phi_i^\mathcal{S}}=1$, excluding $\ket{\text{nc}^\mathcal{S}}$. 

Utilizing Equation \eqref{eq:eps_ctilde_perturbdef}, we then have the distance between $\ket{\text{cs}^\mathcal{W}}$ and the rotated non-contextual reference state $\ket{\text{nc}^\mathcal{S}}=U_\mathcal{S}\ket{\text{nc}}$ will be given by
\begin{equation}
    \epsilon^\mathcal{W}=\sum_{i=1}^{d-1} \frac{\left\vert\mel{\phi_i^\mathcal{S}}{Q^\dagger_\mathcal{W}U^\dagger_\mathcal{S} H_{c}U_\mathcal{S}Q_\mathcal{W}}{\text{nc}^\mathcal{S}} \right\vert^2}{\left(\Delta_{i0}^\mathcal{S}\right)^2}
\end{equation}
where
\begin{equation}\label{eq:deltai0}
    \Delta_{i0}^\mathcal{S}=\mel{\phi_i^\mathcal{S}}{H_{nc}^\mathcal{S}}{\phi_i^\mathcal{S}}-\mel{\text{nc}^\mathcal{S}}{H_{nc}^\mathcal{S}}{\text{nc}^\mathcal{S}}
\end{equation}
and similarly for $\epsilon^\mathcal{W'}$. From Equation \eqref{eq:projstate} we can see that $Q_\mathcal{W}\ket{\text{nc}^\mathcal{S}}=\ket{\text{nc}^\mathcal{S}}$ by construction, while $Q_\mathcal{W}\ket{\phi_i^\mathcal{S}}=0$ unless $\ket{\phi_i^\mathcal{S}}\in \mathcal{I}(\mathcal{W})$. Thus,
\begin{equation}
    \epsilon^\mathcal{W}=\sum_{\phi_i^\mathcal{S}\in \mathcal{I}(\mathcal{W})} \frac{\left\vert\mel{\phi_i^\mathcal{S}}{U^\dagger_\mathcal{S} H_{c}U_\mathcal{S}}{\text{nc}^\mathcal{S}} \right\vert^2}{\left(\Delta_{i0}^\mathcal{S}\right)^2}
\end{equation}
and 
\begin{equation}
    \epsilon^\mathcal{W'}=\sum_{\phi_i^\mathcal{S}\in \mathcal{I}(\mathcal{W'})} \frac{\left\vert\mel{\phi_i^\mathcal{S}}{U^\dagger_\mathcal{S} H_{c}U_\mathcal{S}}{\text{nc}^\mathcal{S}} \right\vert^2}{\left(\Delta_{i0}^\mathcal{S}\right)^2}.
\end{equation}

Because $\mathcal{W}\subset\mathcal{W'}$, $\mathcal{I}(\mathcal{W'})\subseteq \mathcal{I}(\mathcal{W})$, and thus
\begin{equation}\label{eq:epsdiff}
    \epsilon^\mathcal{W}-\epsilon^\mathcal{W'}=\sum_{\phi_i^\mathcal{S}\in \mathcal{I}(\mathcal{W})\setminus \mathcal{I}(\mathcal{W'})}\frac{\left\vert\mel{\phi_i^\mathcal{S}}{U^\dagger_\mathcal{S} H_{c}U_\mathcal{S}}{\text{nc}^\mathcal{S}} \right\vert^2}{\left(\Delta_{i0}^\mathcal{S}\right)^2}\geq0
\end{equation}
We then have
\begin{equation}
    \epsilon^\mathcal{W}\geq\epsilon^\mathcal{W'}
\end{equation}
and the result immediately follows from Theorem \ref{th:overlaptheorem}.$\quad \square$

We demonstrated in Theorem \ref{th:linearity} that the correlation energy of a molecular Hamiltonian is linearly proportional to the magic in the ground state of the Hamiltonian when there is a large overlap of the ground state with the Hartree-Fock state. The Theorem uses the Hartree-Fock state as a reference stabilizer state, however if we are in the framework of the CS method, it is appropriate to instead use the non-contextual ground state $\ket{\text{nc}}$ as the reference stabilizer state for the contextual subspace ground states $\ket{\text{cs}^\mathcal{W}}$. To this end, we consider the projected non-contextual ground state overlap with the CS ground state $\braket{\text{nc}^\mathcal{W}}{\text{cs}^\mathcal{W}}$ for each stabilizer set $\mathcal{W}$ of a molecular Hamiltonian under the CS method. We find directly following from Theorem \ref{th:linearity} and intuition from Theorem \ref{th:gs_monotonicity} that 
\begin{equation}
   \left\vert E_{cs}^\mathcal{W}-E_{nc}\right\vert\propto M_2\left(\ket{\text{cs}^\mathcal{W}}\right).
\end{equation}

In this section, we developed an analytical framework for understanding the connection between states, both their overlap and their associated energy, with the amount of magic those states possess. We found in Theorem \ref{th:overlaptheorem} that states with a large stabilizer overlap will possess a small amount of magic, with an inverse linear relationship between the two quantities. We related this property of large stabilizer overlap to the ground states of electronic structure Hamiltonians.

We then presented the CS method in our analytical framework and proved its connection to manipulating magic in electronic structure Hamiltonians. We showed in Theorem \ref{th:gs_monotonicity} that increasing the size of the contextual subspace provides approximate ground states with monotonically increasing magic. We then concluded that, following Theorem \ref{th:linearity} and Theorem \ref{th:gs_monotonicity}, the amount of correlation energy the approximate Hamiltonians and their ground states recover is linearly related to the amount of magic in their ground states. We next show numerical evidence for our analytics using electronic structure Hamiltonians.

\section{Methodology}\label{sec:methodology}
\begin{figure*}[t!]
    \centering
    \includegraphics[width=\linewidth]{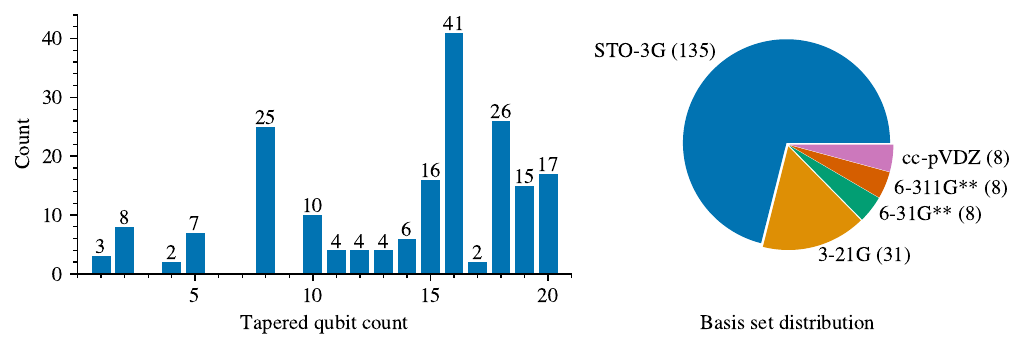}
    \caption{Qubit count and basis set distributions for the 190 species included in the Symmer Hamiltonian dataset used for this work \cite{symmer_hamiltonian_database}. More information on the methodology for generating this dataset can be found in Appendix \ref{app:hamiltonian_database}.}\label{fig:dataset_overview}
\end{figure*}

In this section we provide explanation for how we conducted our numerical simulations. We discuss the development of a new Hamiltonian dataset in Section \ref{subsec:database_overview} which is publicly available. We then describe in Section \ref{subsec:methodology} the implementation of the CS method on our Hamiltonians, first discussed in Section \ref{subsec:csp}, and the calculations for 2-SRE which we use to quantify the non-classicality of our states, described in Section \ref{subsec:sre}.

\subsection{Electronic structure Hamiltonian dataset}\label{subsec:database_overview}
To investigate the relationship between non-stabilizerness and correlation energy in contextual subspace ground-states, we utilize a dataset of Jordan-Wigner encoded electronic structure Hamiltonians representing 190 species developed for this work and available at \cite{symmer_hamiltonian_database}. For the remainder of this paper, we use ``species'' to refer to a particular molecule modeled in a certain point group and basis set choices. For numerical tractability, the majority of species are simulated in the STO-3G minimal basis set; we include simulations in other basis sets when those representations remain under 20 qubits.  To investigate behavior away from equilibrium bond lengths, diatomic and polyatomic species are simulated with the equilibrium geometry uniformly scaled by a factor $\lambda\in\{0.5,0.7,0.8,0.9,1,1.2,1.5,2,2.5,3\}$. This results in a total of 1,594 Hamiltonians in our dataset. Geometry data are sourced from NIST CCCBDB \cite{CCCBDB}, PennyLane molecular datasets \cite{PennyLaneAI_DatasetsSource}, and the Symmer reference collection \cite{symmer}. Demographic information of the 190 species included in our dataset is shown in Figure \ref{fig:dataset_overview}. More information on the Hamiltonian generation process can be found in Appendix \ref{app:hamiltonian_database}.

\subsection{Numerical simulation methodology}\label{subsec:methodology}
To reduce computational complexity, initial qubit tapering was performed on all Hamiltonians as described in Section \ref{subsec:csp} \cite{bravyi2017taperingqubitssimulatefermionic,doi:10.1021/acs.jctc.0c00113}. For simplicity we will refer to the qubit count after initial tapering as the total number of qubits $N$ required for FCI ground state energy simulation and the maximal ground state magic. 

For each molecular Hamiltonian $H$ defined on $n$ qubits, we project into the contextual subspace with $\nu^{\vert\mathcal{W}\vert} \in \{0,1,...,n\}$ active qubits using the $\mathtt{StabilizeFirst}$ method of the Symmer Python package available at \cite{Ralli_UP_CS-VQE}. Each contextual subspace Hamiltonian $H_{cs}^\mathcal{W}$ is then classically diagonalized to find the ground state eigenvector $\ket{\text{cs}^\mathcal{W}}$ and associated eigenenergy $E(\mathcal{W})$. 

In Section \ref{subsec:cs_magic}, we consider a series of nested contextual subspaces, which is also the approach considered by previous CS literature \cite{kirby2021contextual,weaving2023stabilizer,Ralli_UP_CS-VQE}. However, the approach implemented in the Symmer package instead seeks to find the optimal set of $\vert \mathcal{W}\vert$ operators to maximize the recovered correlation energy. Symmer constructs the $\nu$-qubit contextual subspace by doing a greedy search over all possible combinations of $\nu$ operators in $\mathcal{S}$ to find the choice of $\mathcal{W}$ which best preserves the symmetries of the true ground state via a supplied auxiliary operator (usually the unitary coupled cluster operator). In practice, this may then result in a sequence of contextual subspaces which are not in fact concentric. 

However, we can show that the magic of these non-concentric subspaces will still increase monotonically provided the heuristic does in fact maximize the recovered correlation energy and preserve the key qualities of the true ground state. All of the subspaces share a partitioning of the Hamiltonian between contextual and non-contextual terms, and thus share an overall set of operators $\mathcal{S}.$ Thus, they will all be considered as perturbations of the same Hamiltonian $H_{nc}^\mathcal{S}=U^\dagger_\mathcal{S}H_{nc}U_{\mathcal{S}}$ through the framework of Theorem \ref{th:linearity}.  From Equation \eqref{eq:corre_eps} we can see the correlation energy recovered within the contextual subspace defined by a given $\mathcal{W}$ will be
\begin{equation}\label{eq:corr_en}
    \Delta E(\mathcal{W})=-\epsilon^\mathcal{W}\sum_{\phi_i^\mathcal{S}\in \mathcal{I}(\mathcal{W})}|\tilde{c}_i|^2 \Delta_{i0}^\mathcal{S},
\end{equation}
where $\{\ket{\phi_i^\mathcal{S}}\}$ are the eigenstates of $H_{nc}^\mathcal{S}$, $\mathcal{I}(\mathcal{W})$ is the subset of those eigenstates for which $\mel{\phi_{i}^\mathcal{S}}{Q_{\mathcal{W}}}{\phi_{i}^\mathcal{S}}=1$, $\Delta_{i0}^\mathcal{S}$ is given by Equation \eqref{eq:deltai0}. Let $\mathcal{W}_\nu$ be the set of $n-\nu$ generators from $\mathcal{S}$ which maximizes the $\Delta E(\mathcal{W})$ of the $\nu$-qubit CS ground state. We could always construct a candidate $\mathcal{W}_{\nu+1}$ by removing an operator from $\mathcal{W}_\nu$, which would monotonically increase the magnitude of the recovered correlation energy because a contextual subspace includes all states in concentric contextual subspaces. Thus, if the optimal $\mathcal{W}_\nu$ is found, $|\Delta E(\mathcal{W}_\nu)|\leq|\Delta E(\mathcal{W}_{\nu+1})|$. Additionally, because the Symmer approach preserves the structure of the true ground state, the participation ratios $|\tilde{c}_i|^2$ of the different states in $\mathcal{I}(\mathcal{W}_\nu)$ should remain generally consistent as larger contextual subspaces are constructed. We can then see from Equation \ref{eq:corr_en} that if $|\Delta E(\mathcal{W}_\nu)|\leq|\Delta E(\mathcal{W}_{\nu+1})|$, we should have $\epsilon^\mathcal{W_\nu}\leq \epsilon^{\mathcal{W}_{\nu+1}}$.

The contextual subspace method is guaranteed to find an eigenstate with energy less than or equal to the non-contextual solution (i.e., Hartree-Fock or the solution to the non-contextual Hamiltonian from Section \ref{subsec:csp}) when implemented correctly \cite{kirby2021contextual, Ralli_UP_CS-VQE}. Therefore any CS ground state must have $E(\mathcal{W})\leq E_{HF}$. Additionally, because any CS state can be represented in the Hilbert space of the full Hamiltonian $H$ we must also have $E_{FCI}\leq E(\mathcal{W})$ due to the variational principle. We thus require that for all CS ground-states, $E_{FCI}\leq E(\mathcal{W})\leq E_{HF}$. Contextual subspace Hamiltonians which result in an eigenenergy that is either worse than Hartree-Fock or better than the FCI reference are understood as reflective of errors in Symmer's implementation of contextual subspace projection and excluded from our dataset. Subsequently the 2-SRE is calculated for each $H_{cs}^\mathcal{W}$ ground state denoted as $M_{2}(\mathcal{W})$ using an algorithm outlined in Appendix \ref{app:SREcalc} which is similar to the approaches of \cite{huang2026fast-9ec} and \cite{xiao2026exponentially-96a}. All code is available at \cite{magic-and-correlation-code}.

\section{Simulation Results}\label{sec:results}
In this section, we present the result of our simulations performed on the Tufts University High Performance Compute Cluster (HPCC) according to the methodology outlined in Section \ref{sec:methodology}. Out of 1,594 Hamiltonians in our dataset, 1,301 (81.6\%) were successfully simulated while 293 (18.4\%) failed. Of the 293 simulations which failed, 263 (89.8\%) failed due to Symmer errors, 29 (9.9\%) failed due to exceeding the allowed memory for HPCC jobs, and 1 (0.3\%) failed due to exceeding the 48-hour time limit for HPCC jobs. In total, 121 of 190 species (64\%) were successfully simulated for all bond lengths and 159 of 190 species (84\%) were successfully simulated for at least one bond length.

In Section \ref{subsec:overlap_evidence} we present numerical evidence for the linear relationship between FCI magic $M_2(\ket{\psi_\mathrm{FCI}})$ and HF weight $\vert \braket{\mathrm{HF}}{\psi_\mathrm{FCI}}\vert^2$ predicted in Theorem \ref{th:overlaptheorem}. In Section \ref{subsec:monotone_evidence}, we present empirical evidence for the monotonic relationship between the size of a contextual subspace and the magic of ground-states in that contextual subspace predicted by Theorem \ref{th:gs_monotonicity}. In Section \ref{subsec:linear_energy}, we present empirical evidence that the correlation energy recovered by a CS ground-state $\Delta E(\ket{\psi_\mathrm{gs}^\mathcal{W}})$ is linearly proportional to the magic of that ground state $M_2(\ket{\psi_\mathrm{gs}^\mathcal{W}})$ as predicted by Theorem \ref{th:linearity}. All data is available at \cite{magic-and-correlation-data}.

\subsection{Linear relationship between magic and HF weight}\label{subsec:overlap_evidence}

\begin{figure*}[p!]
    \centering
    \hfill
    \subfloat[\label{subfig:fci_equ_high_overlap}]{
    \includegraphics[width=0.49\textwidth]{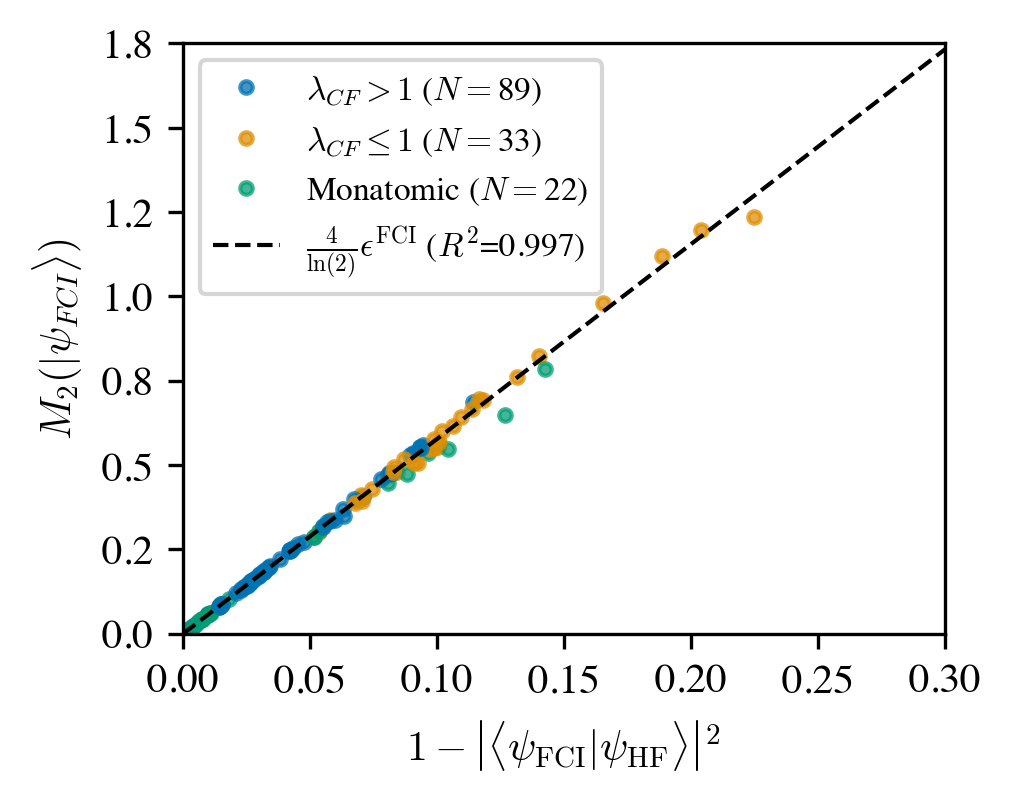}
    }
    \subfloat[\label{subfig:fci_equ_all}]{\includegraphics[width=0.49\textwidth]{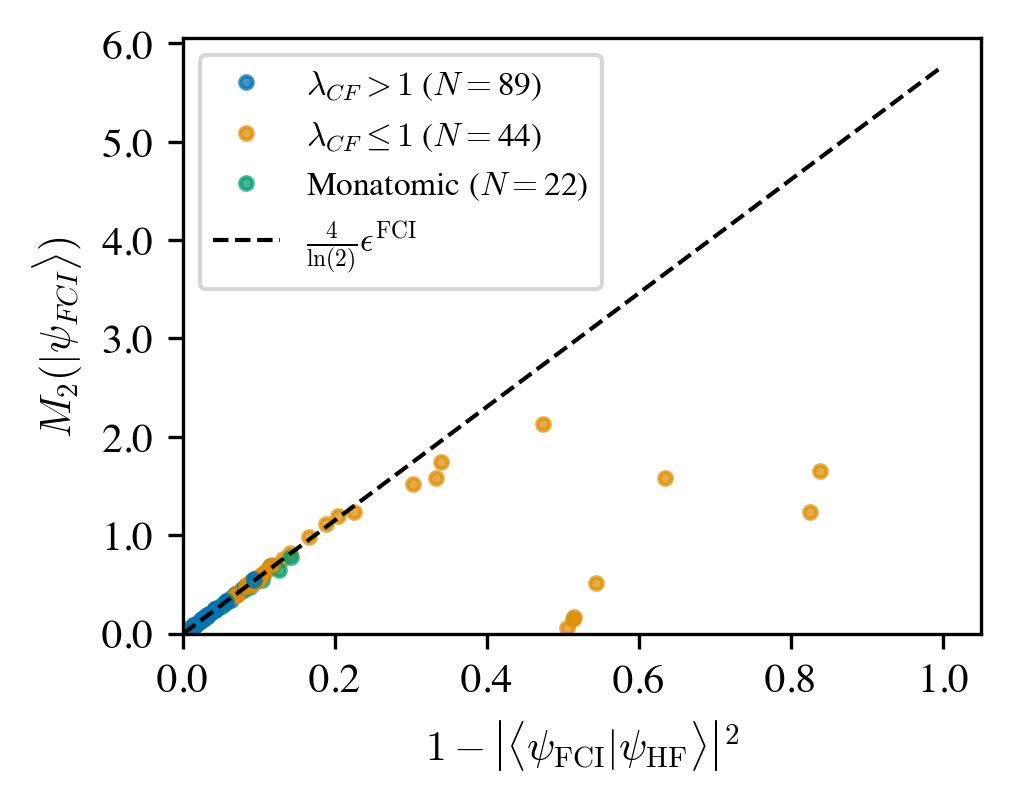}
    }
    \caption{SRE of FCI ground states $M_2(\ket{\psi_\mathrm{FCI}})$ and distance from the Hartree-Fock reference state $\epsilon^\mathrm{FCI}=1-\vert\braket{\mathrm{HF}}{\psi_\mathrm{FCI}}\vert ^2$ (a) for 144 species where $\left\vert \braket{\mathrm{HF}}{\psi_\mathrm{FCI}}\right\vert^2\geq0.7$ at equilibrium bond length and (b) all 155 simulated species. Blue dots are species below their Coulson-Fischer point, orange dots are species above their Coulson-Fischer point, and green dots are monatomic species (which do not have bonds and thus lack a Coulson-Fischer point). The black dashed line is the $M_2(\ket{\psi_\mathrm{FCI}})=\frac{4}{\ln(2)}\epsilon^\mathrm{FCI}$ relationship predicted by Theorem \ref{th:overlaptheorem}.}\label{fig:overlap_proof_eq}

    \hfill
    \subfloat[\label{subfig:fci_belowcf}]{
    \includegraphics[width=0.49\textwidth]{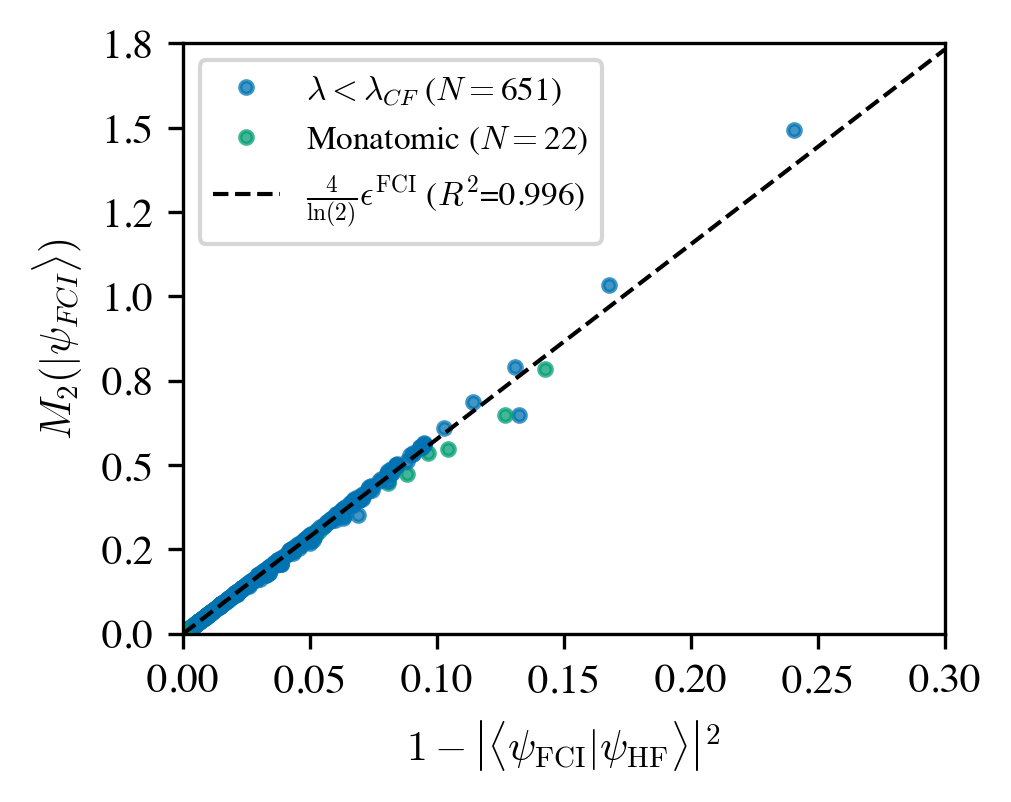}
    }
    \subfloat[\label{subfig:fci_abovecf}]{\includegraphics[width=0.49\textwidth]{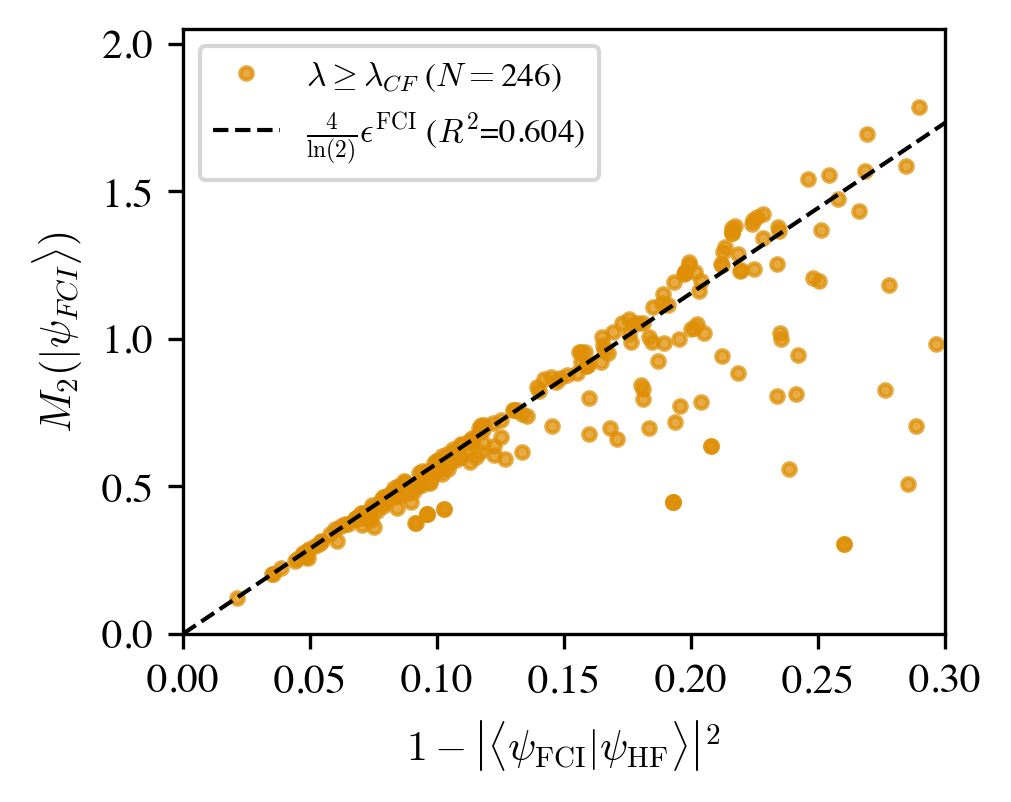}
    }
    \caption{SRE of FCI ground states $M_2(\ket{\psi_\mathrm{FCI}})$ and distance from the Hartree-Fock reference state $\epsilon^\mathrm{FCI}=1-\vert\braket{\mathrm{HF}}{\psi_\mathrm{FCI}}\vert ^2$ for 919 Hamiltonians where $\epsilon^\mathrm{FCI}\leq0.3$. As in Figure \ref{fig:overlap_proof_eq}, blue dots are species below their Coulson-Fischer point, orange dots are species above their Coulson-Fischer point, and green dots are monatomic species (which do not have bonds and thus lack a Coulson-Fischer point). The black dashed line is the $M_2(\ket{\psi_\mathrm{FCI}})=\frac{4}{\ln(2)}\epsilon^\mathrm{FCI}$ relationship predicted by Theorem \ref{th:overlaptheorem}. In (a) we examine species which are either monatomic or simulated below their Coulson-Fischer point, while in (b) we examine species which are simulated beyond their Coulson-Fischer point.}\label{fig:overlap_proof_noneq}
\end{figure*}
\begin{figure*}[p!]
\centering
    \hfill
    \subfloat[\label{subfig:cs_overlap_belowcf}]{
    \includegraphics[width=0.49\textwidth]{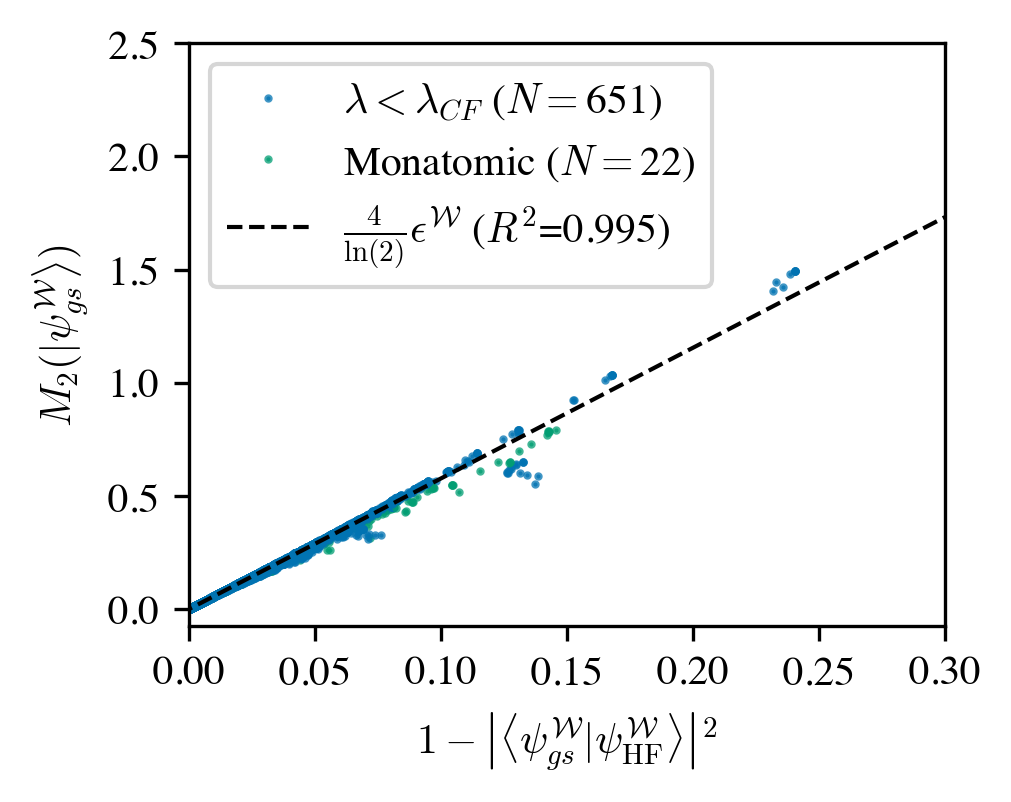}
    }
    \subfloat[\label{subfig:cs_overlap_abovecf}]{\includegraphics[width=0.49\textwidth]{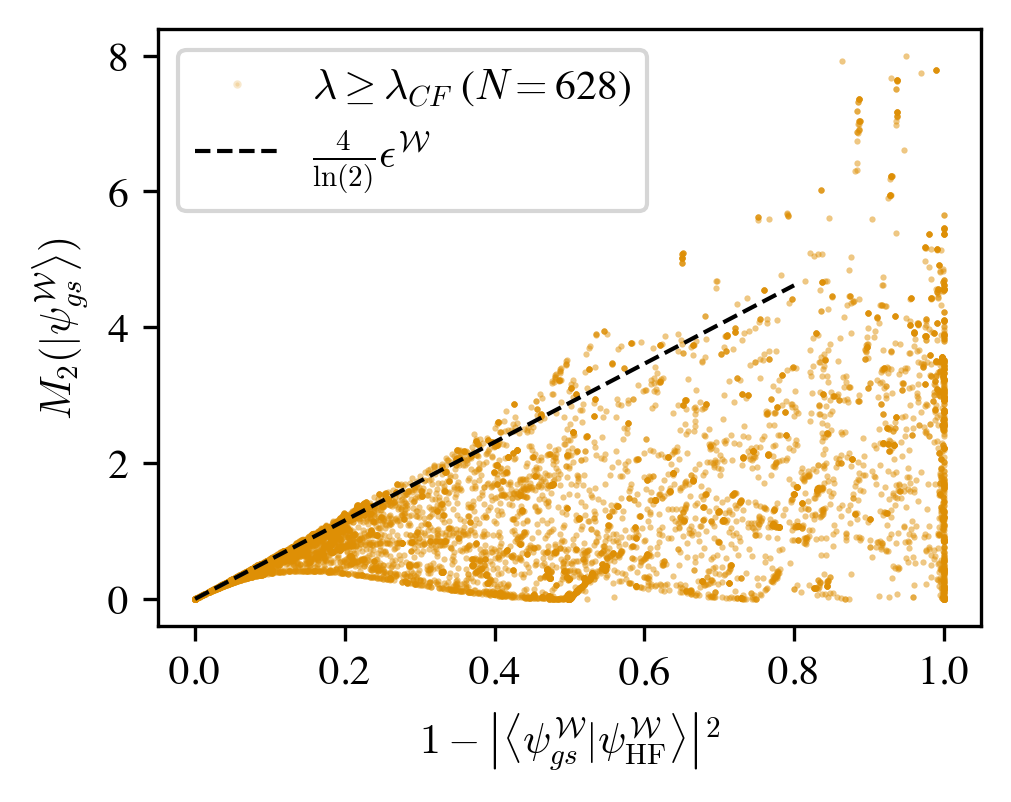}
    }
\caption{SRE of contextual subspace ground states $M_2(\ket{\psi_\mathrm{gs}^\mathcal{W}})$ and distance from the CS-projected Hartree-Fock reference state $\epsilon^\mathcal{W}=1-\vert\braket{\psi^\mathcal{W}_{gs}}{\psi^\mathcal{W}_\mathrm{HF}}\vert ^2$ for 1,301 unique combinations of species and bond length. Blue dots are species simulated below their Coulson-Fischer point, orange dots are species simulated above their Coulson-Fischer point, and green dots are monatomic species (which do not have bonds and thus lack a Coulson-Fischer point). The black dashed line is the $M_2(\ket{\mathrm{cs}^\mathcal{W}})=\frac{4}{\ln(2)}\epsilon^\mathrm{\mathcal{W}}$ relationship predicted by Theorem \ref{th:overlaptheorem}. In (a) we examine species which are either monatomic or simulated below their Coulson-Fischer point, while in (b) we examine species which are simulated beyond their Coulson-Fischer point.}\label{fig:overlap_cs}

\hfill

    \subfloat[\label{subfig:alldiffs}]{
    \includegraphics[width=0.49\textwidth]{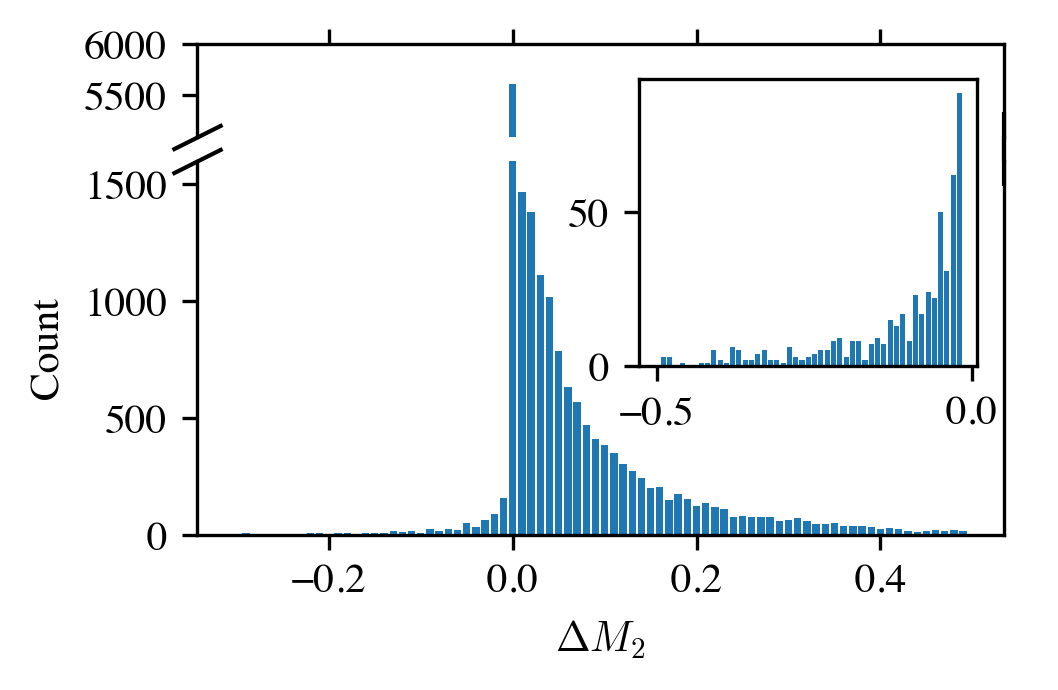}}
    \hfill
    \subfloat[\label{subfig:cfdiffs}]{
    \includegraphics[width=0.49\textwidth]{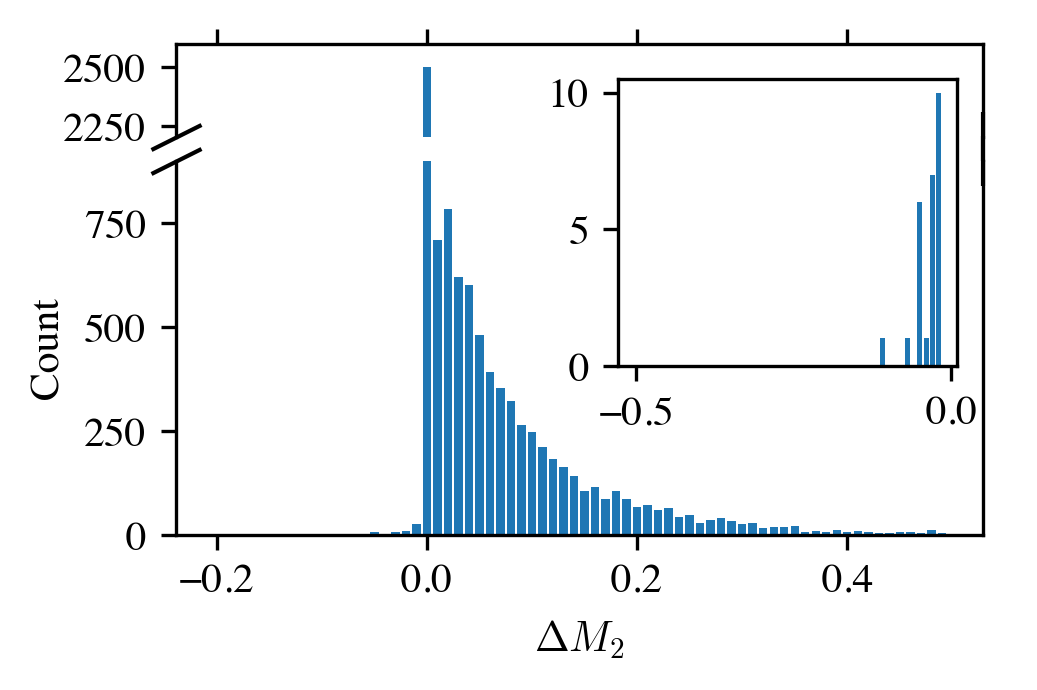}
    }
    \caption{(a) Distribution of the change in ground-state magic $\Delta M_2$ caused by adding a single qubit to the contextual subspace across all simulations. Consistent with the monotonicity argument outlined in Theorem \ref{th:gs_monotonicity}, adding a qubit to the contextual subspace (and thus increasing the maximum stabilizer nullity $\nu^{\vert\mathcal{W}\vert}$) increases or does not significantly change the magic in the overwhelming majority of cases. 
    (b) Distribution of the change in ground-state magic $\Delta M_2$ caused by adding a single qubit to the contextual subspace across simulations below the Coulson-Fischer bond length. We find that restricting to simulations below the Coulson-Fischer bond length results in a stronger monotonicity condition.}
    \label{fig:monotonicity}
    
\end{figure*}
In this section, we present numerical evidence for the linear relationship between SRE and overlap with the Hartree-Fock ground state established in Theorem \ref{th:overlaptheorem}.

In Figure \ref{fig:overlap_proof_eq}, we analyze the HF-FCI distance $\epsilon^\mathrm{FCI}=1-\left\vert\braket{\mathrm{HF}}{\psi_\mathrm{FCI}}\right\vert^2$ and FCI ground-state magic $M_2(\ket{\psi_\mathrm{FCI}})$ for 155 species from our dataset simulated at equilibrium bond length. In Figure \ref{subfig:fci_equ_high_overlap}, we consider the 144 species where $\epsilon^\mathrm{FCI}\leq0.3$ (corresponding to weakly- and moderately-correlated behavior). We find strong agreement with the $M_2(\ket{\psi_\mathrm{FCI}})=\frac{4}{\ln(2)}\epsilon^\mathrm{FCI}$ relationship predicted by Theorem \ref{th:overlaptheorem} (black dashed line, $R^2$=0.997). In Figure \ref{subfig:fci_equ_all}, we consider all 155 species simulated. We find that for 11 species with $\epsilon^\mathrm{FCI}>0.3$, the FCI ground-state magic $M_2(\ket{\psi_\mathrm{FCI}})$ is significantly lower than predicted by Theorem \ref{th:overlaptheorem}. This deviation is not particularly surprising; for $\epsilon\gtrsim0.3$ we expect the small-$\epsilon$ assumption to break down. We also note that all of these outlier species are above their Coulson-Fischer point at equilibrium, and thus we know the Hartree-Fock ground state $\ket{\mathrm{HF}}$ no longer accurately captures key features of the FCI ground state $\ket{\psi_\mathrm{FCI}}$. It is also notable that all 8 of these species exhibit $M_2\left(\ket{\psi_\mathrm{FCI}}\right)<\frac{4}{\ln(2)}\epsilon^\mathrm{FCI}$, which suggests that there may be another stabilizer state closer to the FCI ground state.

In Figure \ref{fig:overlap_proof_noneq}, we examine the behavior when molecules are simulated away from their equilibrium bond lengths according to the methodology described in Section \ref{subsec:database_overview}. For species which are monatomic or simulated below their Coulson-Fischer point (shown in Figure \ref{subfig:fci_belowcf}), we find continued strong agreement with the linear relationship predicted by Theorem \ref{th:overlaptheorem} with $R^2=$0.996. When species are simulated above their Coulson-Fischer point (shown in Figure \ref{subfig:fci_abovecf}), we again find several instances where the FCI ground-state magic $M_2(\ket{\psi_\mathrm{FCI}})$ is lower than would be expected from Theorem \ref{th:overlaptheorem}. As discussed above, this is in accordance with the Hartree-Fock ground state $\ket{\mathrm{HF}}$ no longer being a reliable approximation of $\ket{\psi_\mathrm{FCI}}$ beyond the Coulson-Fischer point.

We also extend our results to ground-states of the contextual subspace Hamiltonian $H_{cs}^\mathcal{W}=Q^\dagger_\mathcal{W}U^\dagger_\mathcal{W}H U_\mathcal{W}Q_\mathcal{W}$. In Figure \ref{fig:overlap_cs}, we consider the magic of CS ground-states $M_2(\ket{\mathrm{cs}^\mathcal{W}})$ and their distance from the Hartree-Fock ground-state $\epsilon^\mathcal{W}=1-\left\vert \braket{\mathrm{cs}^\mathcal{W}}{\mathrm{HF}^\mathcal{W}}\right\vert^2$. Here, $\ket{\mathrm{HF}^\mathcal{W}}=U_\mathcal{W}Q_\mathcal{W}\ket{\mathrm{HF}}$ is the HF ground state projected into the contextual subspace stabilized by $\mathcal{W}$. When species are simulated below their Coulson-Fischer point (shown in Figure \ref{subfig:cs_overlap_belowcf}), we continue to find strong agreement with Theorem \ref{th:overlaptheorem}, with $R^2=0.995.$ We thus conclude that below the Coulson-Fischer point, $2$-SRE and $\epsilon^\mathcal{W}$ are linear following $M_2(\ket{\mathrm{cs}^\mathcal{W}})=\frac{4}{\ln(2)}\epsilon^\mathcal{W}$. Above the Coulson-Fischer point (shown in Figure \ref{subfig:cs_overlap_abovecf}) we observe many cases where the CS ground-states have less magic than would be predicted by Theorem \ref{th:overlaptheorem}, as in Figure \ref{subfig:fci_abovecf}. Similarly to our analysis above, we interpret this as consistent with the Hartree-Fock ground state no longer being the best stabilizer reference beyond the Coulson-Fischer point.

We find strong simulation agreement with Theorem \ref{th:overlaptheorem}, with exceptions occurring when the selected stabilizer reference state is a poor reference for the simulated system. Therefore we conclude that when HF is a valid reference state for the ground state, there is a linear relationship between the HF-ground state distance and the amount of magic in the ground state for both FCI ground states and CS ground states.

\subsection{Magic monotonicity of the CS method}\label{subsec:monotone_evidence}
In this section, we present numerical results validating the contextual subspace method as a mechanism for manipulating the magic in a system. 

In Theorem \ref{th:gs_monotonicity}, we presented an argument that the magic of the ground-states of the contextual subspace Hamiltonian $H_{cs}^\mathcal{W}$ should increase monotonically as generators $\{G_k\}$ are removed from $\mathcal{W}$. We thus expect that if $\mathcal{W}=\mathcal{W}'/\{G_k\}$ then
\begin{equation}
    \Delta M_2 =M_2\left(\ket{\mathrm{cs}^\mathcal{W}}\right)-M_2\left(\ket{\mathrm{cs}^\mathcal{W'}}\right)\geq0.
\end{equation}
The action of removing a generator $G_k$ from $\mathcal{W}$ can also be understood as adding a qubit to the contextual subspace, or equivalently increasing the maximum stabilizer nullity $\nu^{\vert\mathcal{W}\vert}$. As discussed in Section \ref{subsec:methodology}, the contextual subspaces constructed by Symmer are not necessarily nested in the way generally assumed by CS literature. However, we show that we still expect the CS ground states to have monotonically increasing magic provided the heuristics Symmer uses to construct the subspaces are working as intended.

In Figure \ref{subfig:alldiffs}, we show the change in magic $\Delta M_2$ caused by adding a single qubit to the contextual subspace for the 1,301 parent Hamiltonians simulated in Figure \ref{fig:overlap_cs}. Of 18,786 individual instances of adding a qubit to the contextual subspace, we find that $\Delta M_2\geq0$ in 15,211 instances (81\%). Of the 3,575 instances where adding a single qubit to the contextual subspace resulted in $\Delta M_2<0$, in 2,439 of those instances (68.2\% of instances where $\Delta M_2<0$ and 13.0\% of all instances) the magnitude of the decrease in magic  $\vert\Delta M_2\vert$ is less than $10^{-3}$. In only 1,136 instances (31.8\% of instances where $\Delta M_2<0$ and 6.0\% of all instances) do we find $\Delta M_2\leq-10^{-3}$. In Figure \ref{subfig:cfdiffs}, we consider specifically the 673 Hamiltonians representing a species which is either monatomic or simulated below its Coulson-Fischer point. Of 9,380 individual instances of adding a qubit to the contextual subspace, we now find that $\Delta M_2\geq0$ in 8,059 instances (85.9\%) and that $\Delta M_2\leq -10^{-3}$ in only 136 instances (1.4\%). We thus conclude that Theorem \ref{th:gs_monotonicity} overwhelmingly holds, particularly when the CS method is provided with a high-quality reference state which enables the heuristics to be effective, and thus the CS method can be used to manipulate the magic of ground states.

\begin{figure*}[t!]
    \centering
\subfloat[\label{subfig:h2oplot}]{    \includegraphics[width=0.48\textwidth]{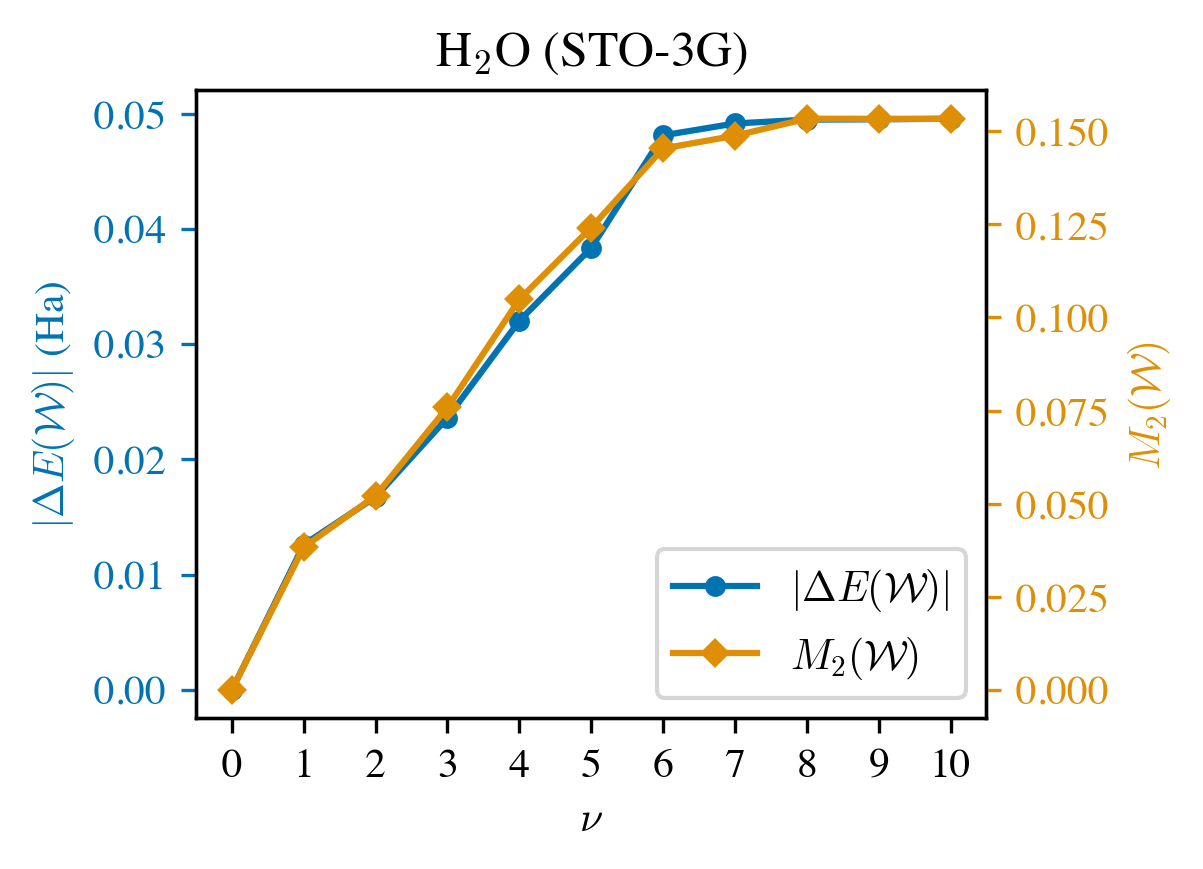}
}\hfill
\subfloat[\label{subfig:ch4plot}]{    \includegraphics[width=0.48\textwidth]{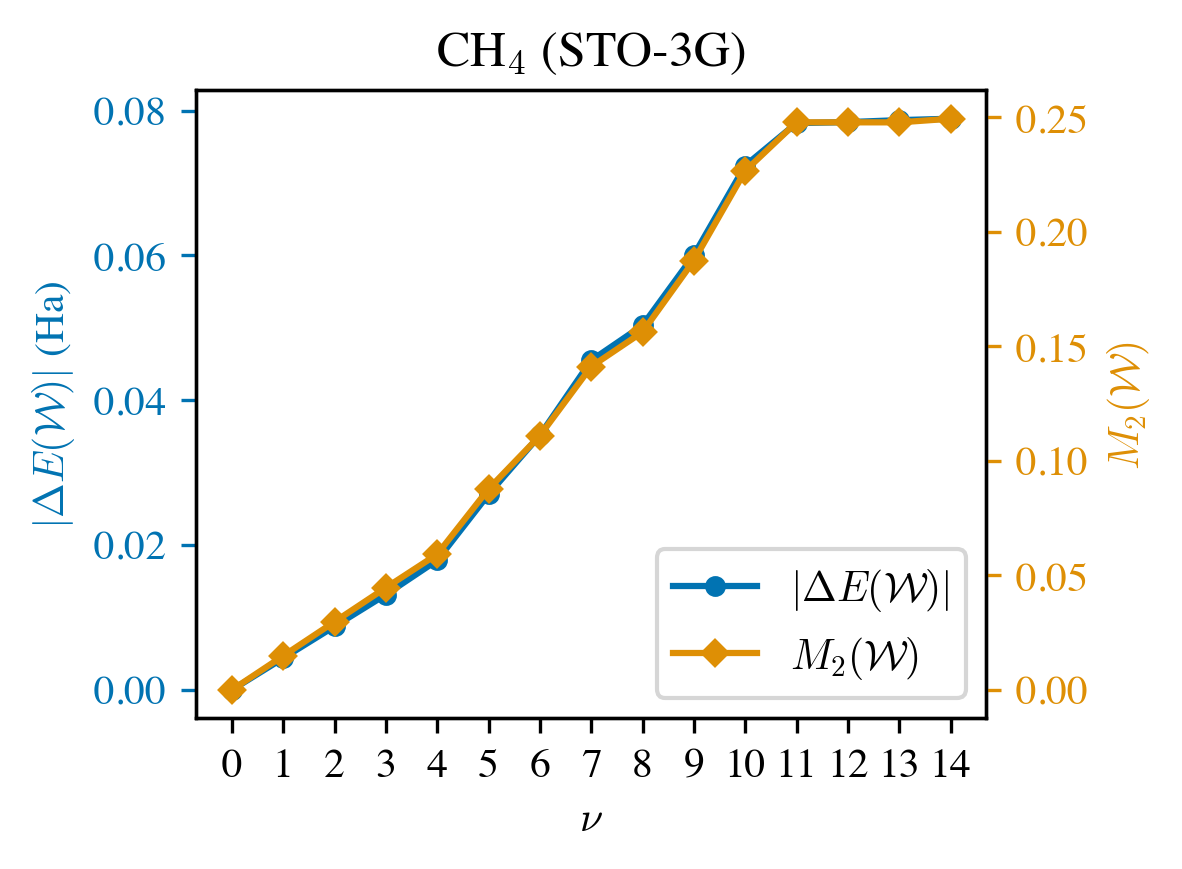}
}
    \caption{(a) and (b): Magnitude of correlation energy $\vert \Delta E^\mathcal{W}\vert$ (blue dots) and magic $M_2^\mathcal{W}$ (orange squares) of the contextual subspace Hamiltonian ground-states of (a) H$_2$O (STO-3G) and (b) CH$_4$ (STO-3G) with a varying number of qubits in the contextual subspace $\nu=n-|\mathcal{W}|$ at equilibrium bond length. In both cases, we see the magnitude of correlation energy recovered $\vert \Delta E^\mathcal{W}\vert$ and the magic $M_2^\mathcal{W}$ monotonically increase together with expanding contextual subspace size. As described in Section \ref{subsec:csp}, the monotonic increase in correlation energy recovered is an inherent feature of the CS method resulting from each contextual subspace including all states in smaller contextual subspaces.}\label{fig:H2OCH4}
    \hfill
\subfloat[\label{subfig:h2ofracfrac}]{\includegraphics[width=0.45\textwidth]{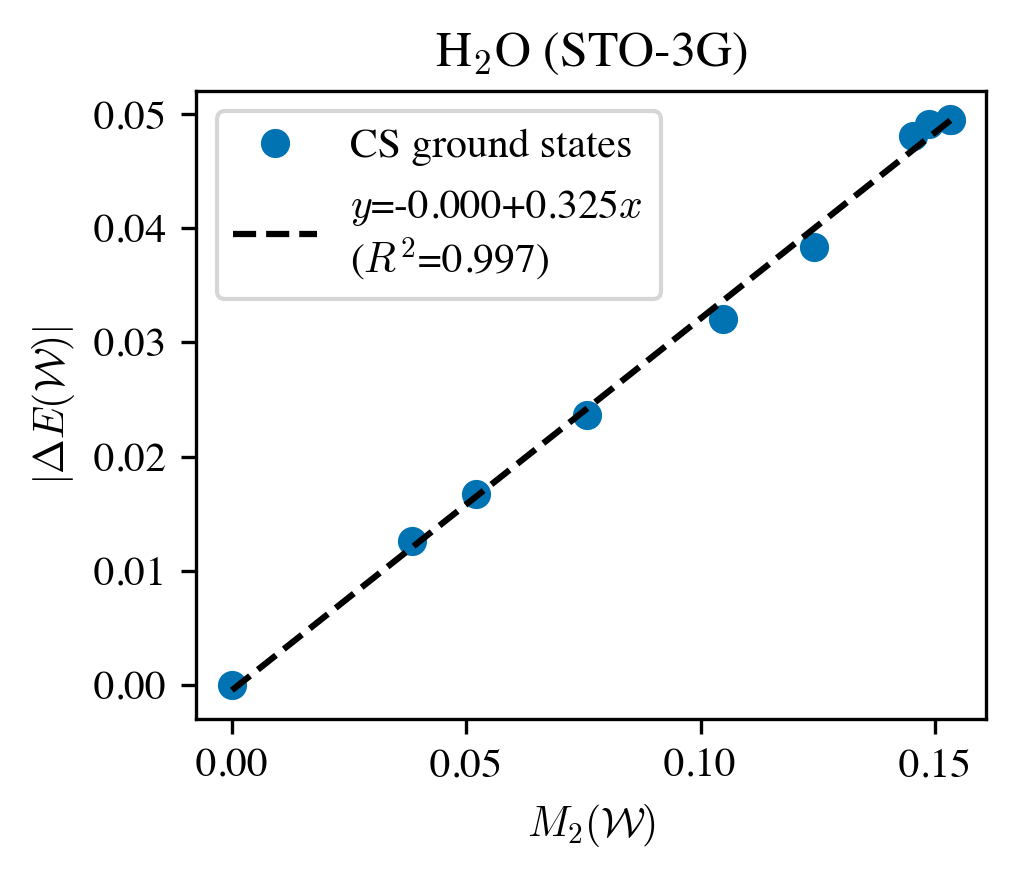}
}\hfill
\subfloat[\label{subfig:ch4fracfrac}]{    \includegraphics[width=0.45\textwidth]{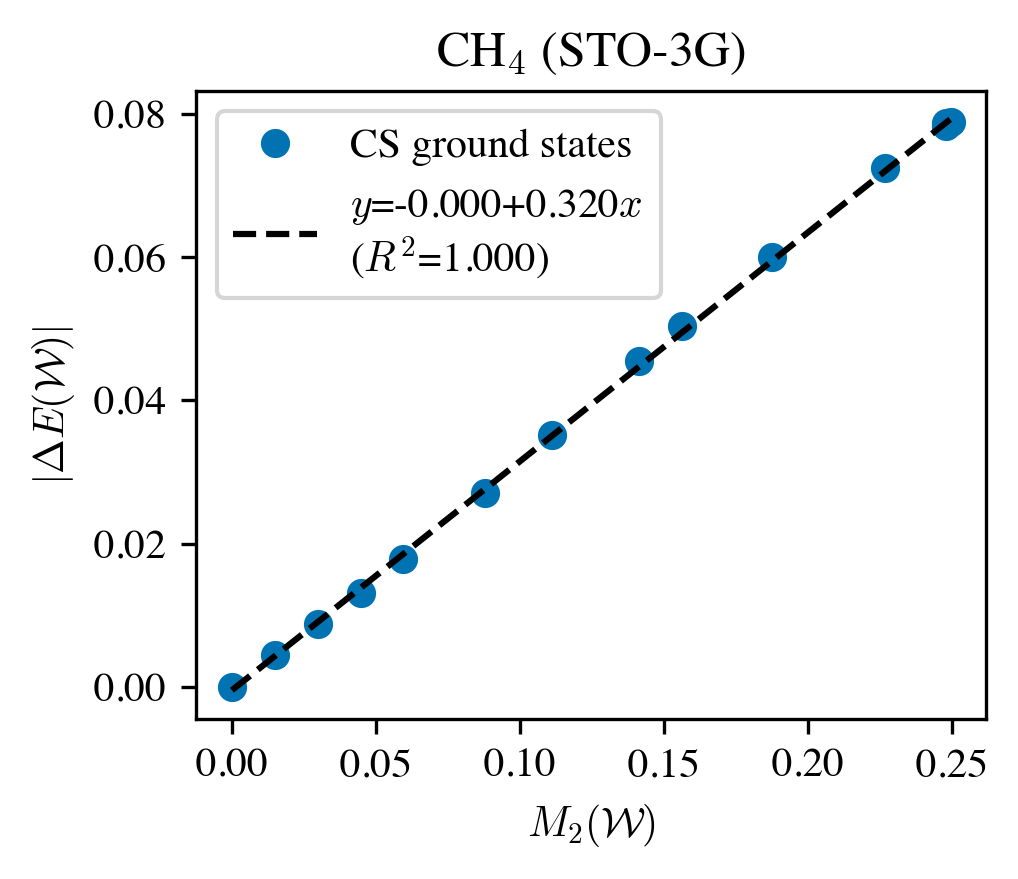}
}
\caption{Magnitude of correlation energy $\vert \Delta E^\mathcal{W}\vert$ as a function of magic $M_2^\mathcal{W}$ for (a) H$_2$O (STO-3G) and (b) CH$_4$ (STO-3G) simulated at equilibrium bond length. In both cases, we observe a roughly linear relationship between the magic of a CS ground-state and the correlation energy it recovers.}\label{fig:minifracfrac}
\end{figure*} 

\subsection{Linear relationship between magic and correlation energy}\label{subsec:linear_energy}
In this section, we provide numerical evidence for a linear relationship between the magic of post-Hartree-Fock ground states and the correlation energy they recover. 

In Theorem \ref{th:linearity}, we established that the correlation energy of a post-Hartree-Fock ground state is proportional to the magic of that state to leading order. In Section \ref{subsec:cs_magic}, we apply this result to the CS method described in Section \ref{subsec:csp}. The CS method has the advantage that it allows us to construct an entire family of approximate ground states as generators $G_k$ are sequentially removed from the stabilizer group $\mathcal{W}$. For an $n$-qubit Hamiltonian, removing an operator $G_k$ from $\mathcal{W}$ is equivalent to increasing the size of the contextual subspace $\nu=n-|\mathcal{W}|$ by one qubit. When $\nu=0$ (and thus $|\mathcal{W}|=n$), the CS ground state is equivalent to the non-contextual reference state (which we choose to be the Hartree-Fock state $\ket{\mathrm{HF}}$), while when $\nu=n$ (and thus $|\mathcal{W}|=0$) the CS ground state is equivalent to the FCI ground state $\ket{\psi_\mathrm{FCI}}$. For $0<\nu<n$, the CS method allows us to generate approximations to $\ket{\psi_\mathrm{FCI}}$ with reduced magic. 

In Figure \ref{fig:H2OCH4}, we plot the magnitude of correlation energy $\vert \Delta E^\mathcal{W}\vert$ and magic $M_2^\mathcal{W}$ of the contextual subspace Hamiltonian ground-states for two example molecules from our dataset, H$_2$O STO-3G (Figure \ref{subfig:h2oplot}) and CH$_4$ STO-3G (Figure \ref{subfig:ch4plot}). In both cases, we vary the number of qubits in the contextual subspace $\nu$ while holding bond length at equilibrium. For both cases, we observe the magnitude of correlation energy recovered $\vert \Delta E\vert$ and the magic $M_2$ monotonically increase with expanding contextual subspace size. As described in Section \ref{subsec:csp}, the monotonic increase in correlation energy recovered is an inherent feature of the CS method resulting from each contextual subspace including all states in smaller contextual subspaces. This monotonic increase in magic is predicted by Theorem \ref{th:gs_monotonicity}.

We also observe from Figure \ref{fig:H2OCH4} that the recovered correlation energy $\vert \Delta E^\mathcal{W}\vert$ and magic $M_2^\mathcal{W}$ appear to change together. To demonstrate this connection, in Figure \ref{fig:minifracfrac} we plot the magnitude of correlation energy $\Delta E(\mathcal{W})$ against the magic $M_2(\mathcal{W})$ for both molecules and in both cases find a highly linear relationship, though with slightly different slopes. While Theorem \ref{th:linearity} predicts that these quantities should be proportional, the linearity of the relationship implies that the constant of proportionality remains the same as operators are removed from $\mathcal{W}.$

While Figure \ref{fig:minifracfrac} suggests that the constant of proportionality between $M_2(\mathcal{W})$ and $\vert \Delta E(\mathcal{W})\vert$ is fixed for an individual species, the form of Equation \ref{eq:corr_vs_magic} makes clear that the constant of proportionality depends on an energy scale which we do not expect to be the same across species. Thus, to examine the relationship between the magic and correlation energy recovered by the ground states of the CS Hamiltonians $H_{cs}^{\mathcal{W}}$ across multiple species, for each CS ground state we consider the fractional correlation energy
    \begin{equation}
       \%\Delta E(\mathcal{W}):= \frac{E(\mathcal{W}) -E_{HF}}{E_{FCI}-E_{HF}}
    \end{equation}
and fractional magic
\begin{equation}
    \%M_2(\mathcal{W}):=\frac{M_2\left(\ket{\mathrm{cs}^\mathcal{W}}\right)}{M_2\left(\ket{\psi_\mathrm{FCI}}\right)}.
\end{equation}
By construction $\%\Delta E(|\mathcal{W}|=n)=\%M_2(|\mathcal{W}|=n)=0$ for the HF ground state, while $\%\Delta E(|\mathcal{W}|=0)=\%M_2(|\mathcal{W}|=0)=1$ for the FCI ground state, thus fixing the slope to be $1$ if the constant of proportionality between $M_2(\mathcal{W})$ and $\Delta E(\mathcal{W})$ remains constant.

\begin{figure*}[htp]
    \centering
    \subfloat[\label{subfig:frac_spaghett_plot_belowcf}]{
    \includegraphics[width=0.6\linewidth]{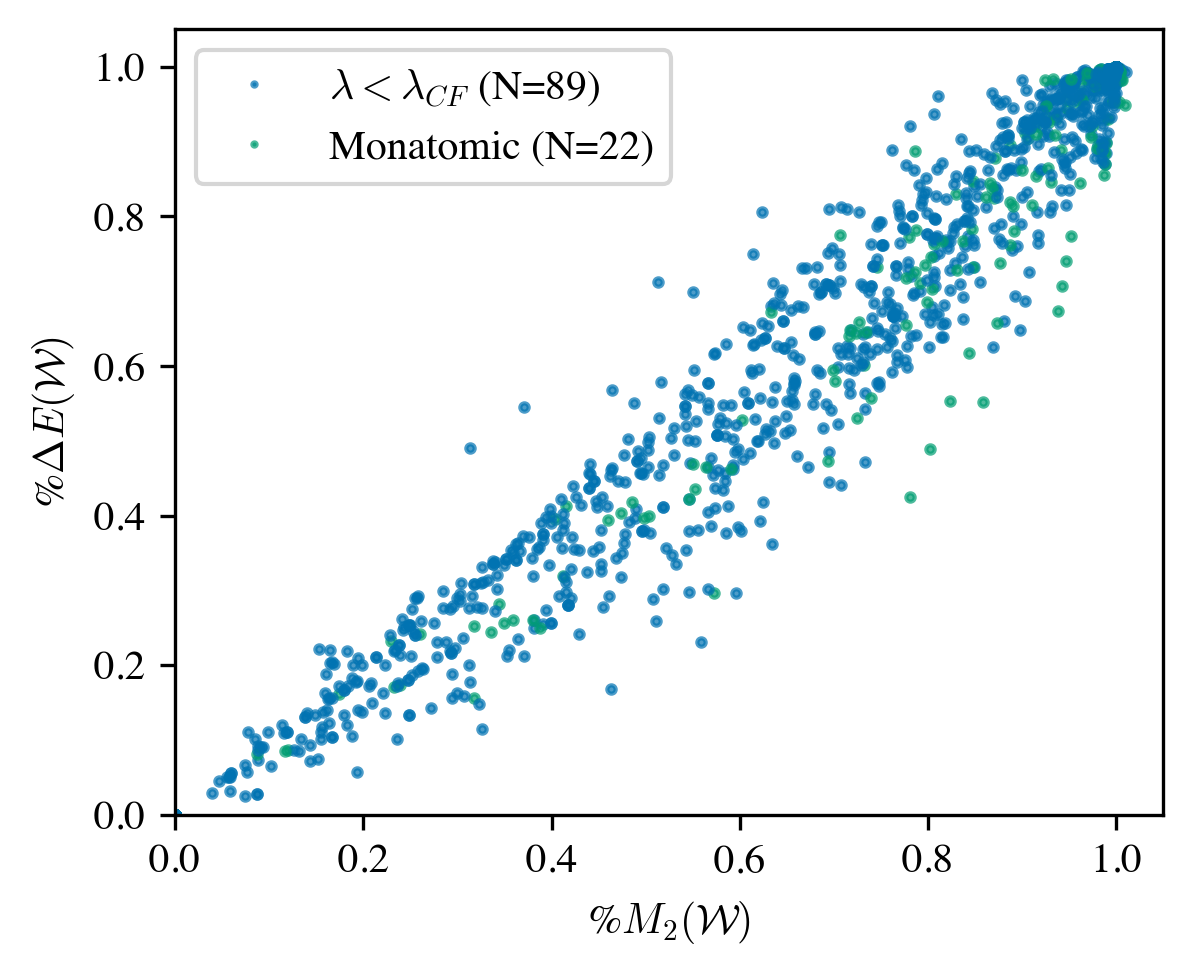}
    }
    \hfill
    \subfloat[\label{subfig:below_fithists}]{
    \includegraphics[width=0.35\linewidth]{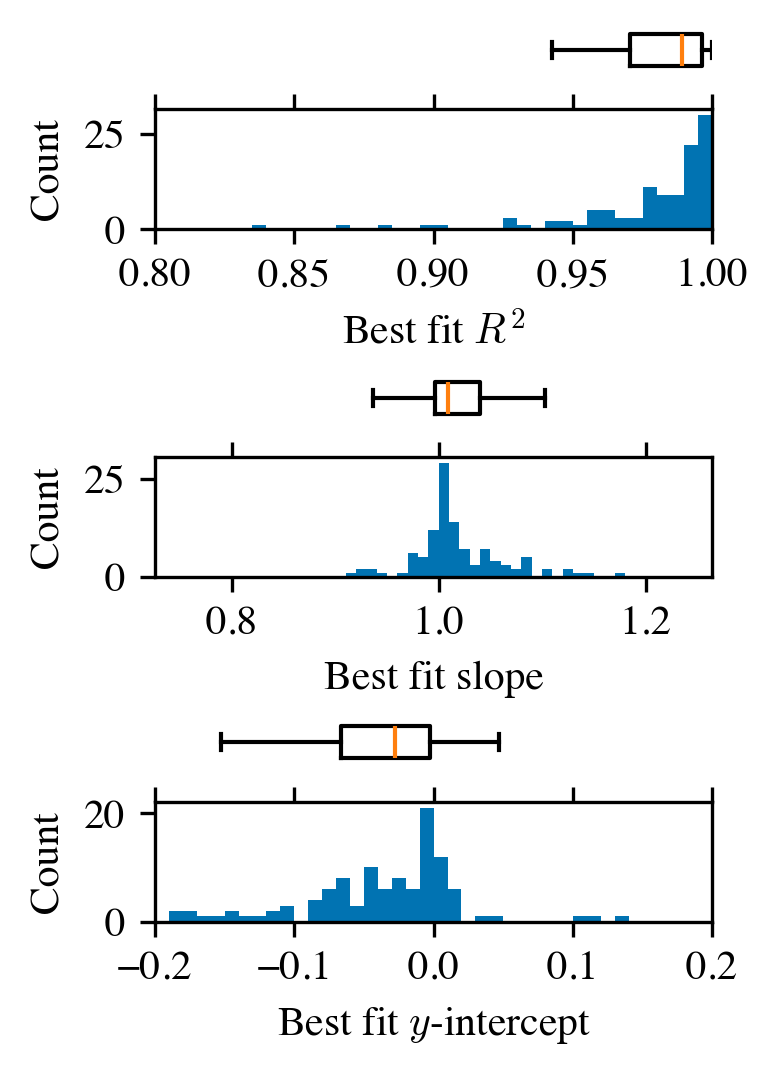}
    }
    \caption{(a) Fractional ground-state correlation energy $\%\Delta E(\mathcal{W})$ and fractional magic $\%M_2(\mathcal{W})$ for 89 multi-atomic species below their Coulson-Fischer point simulated at equilibrium bond length and 22 monatomic species. For each species, the number of qubits in the contextual subspace is varied from $0$ to $n$. (b) Histograms for the best fit $R^2$ values, slope, and y-intercept for the 111 species simulated in (a).}\label{fig:frac_spaghett_belowcf}
    \subfloat[\label{subfig:frac_spaghett_plot_abovecf}]{
    \includegraphics[width=0.6\linewidth]{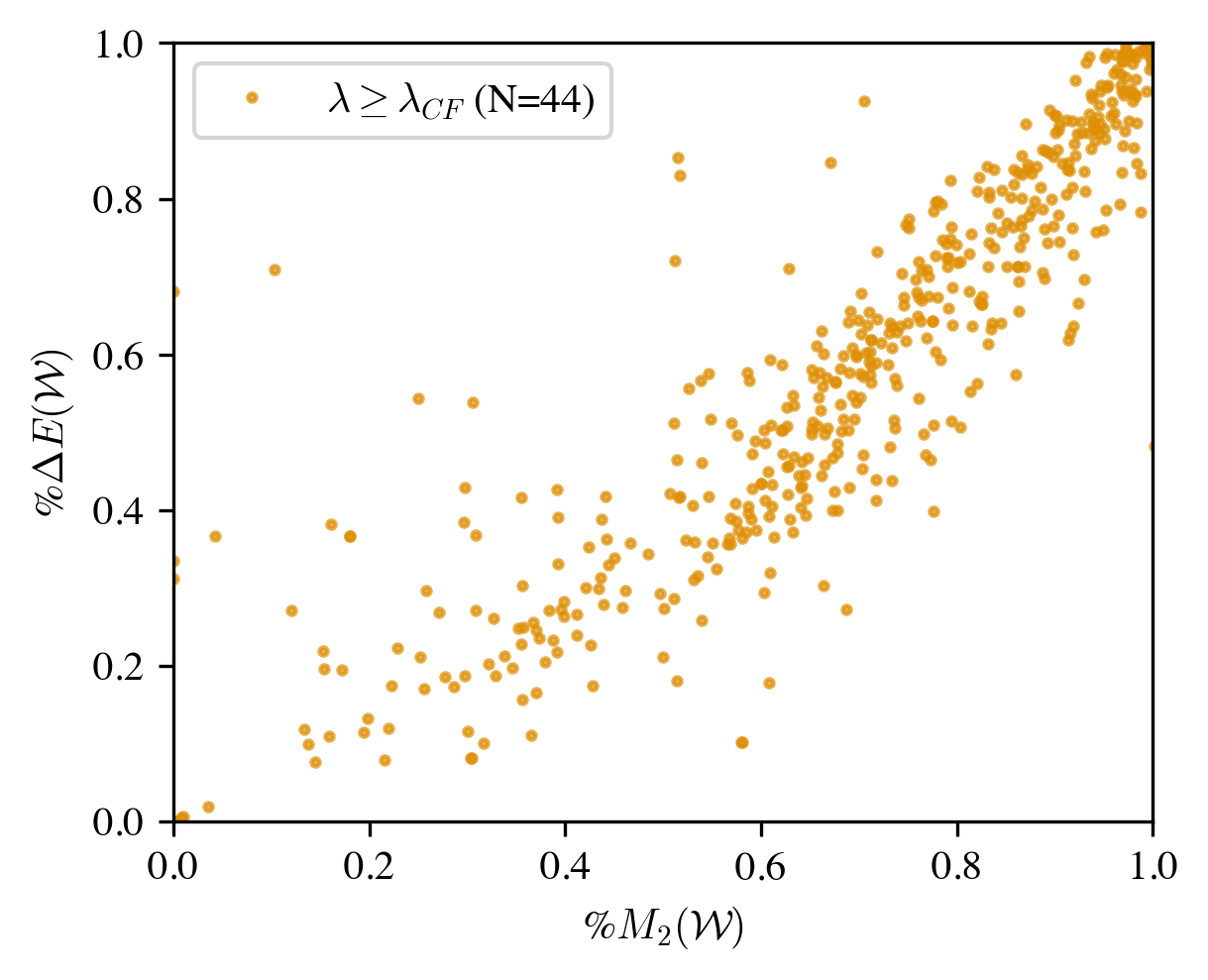}
    }
    \hfill
    \subfloat[\label{subfig:above_fithists}]{
    \includegraphics[width=0.35\linewidth]{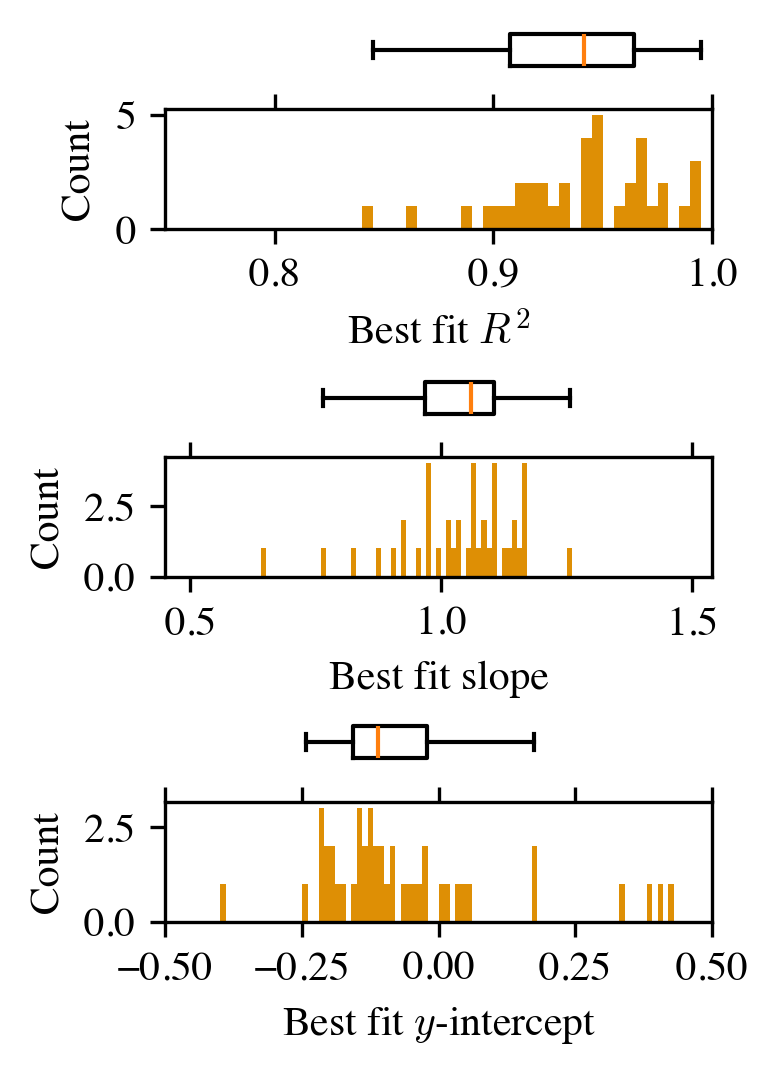}
    }
        \caption{(a) Fractional ground-state correlation energy $\%\Delta E(\mathcal{W})$ and fractional magic $\%M_2(\mathcal{W})$ for 44 multi-atomic species above their Coulson-Fischer point simulated at equilibrium bond length. For each species, the number of qubits in the contextual subspace is varied from $0$ to $n$. (b). Histograms for the best fit $R^2$ values, slope, and y-intercept for the 44 species simulated in (a).}
    \label{fig:frac_spaghett_abovecf}
\end{figure*}

\begin{figure*}[tp]
\subfloat[\label{subfig:bondlengthplots_belowcf}]{\includegraphics[width=\linewidth]{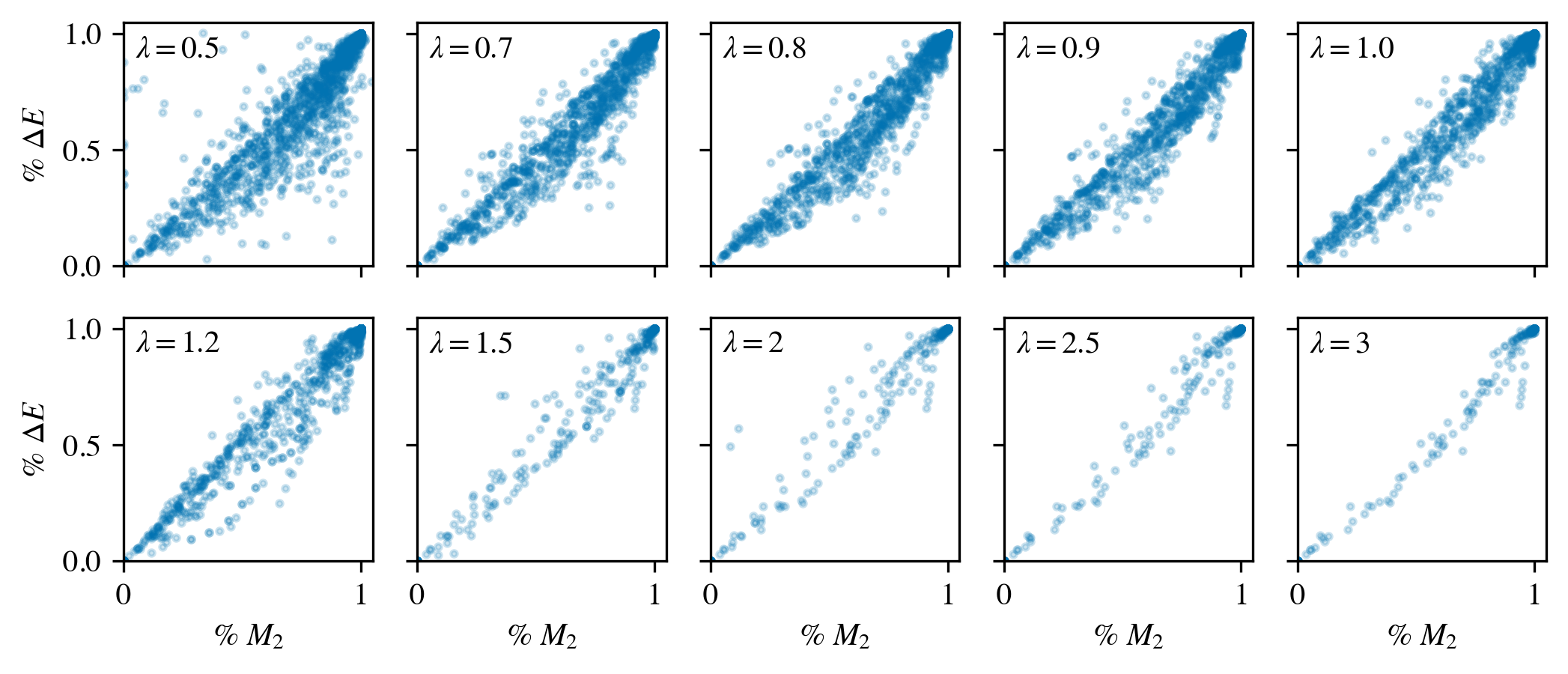}}
\\
\hfill\\
\subfloat[\label{subfig:bondlengthplots_abovecf}]{\includegraphics[width=\linewidth]{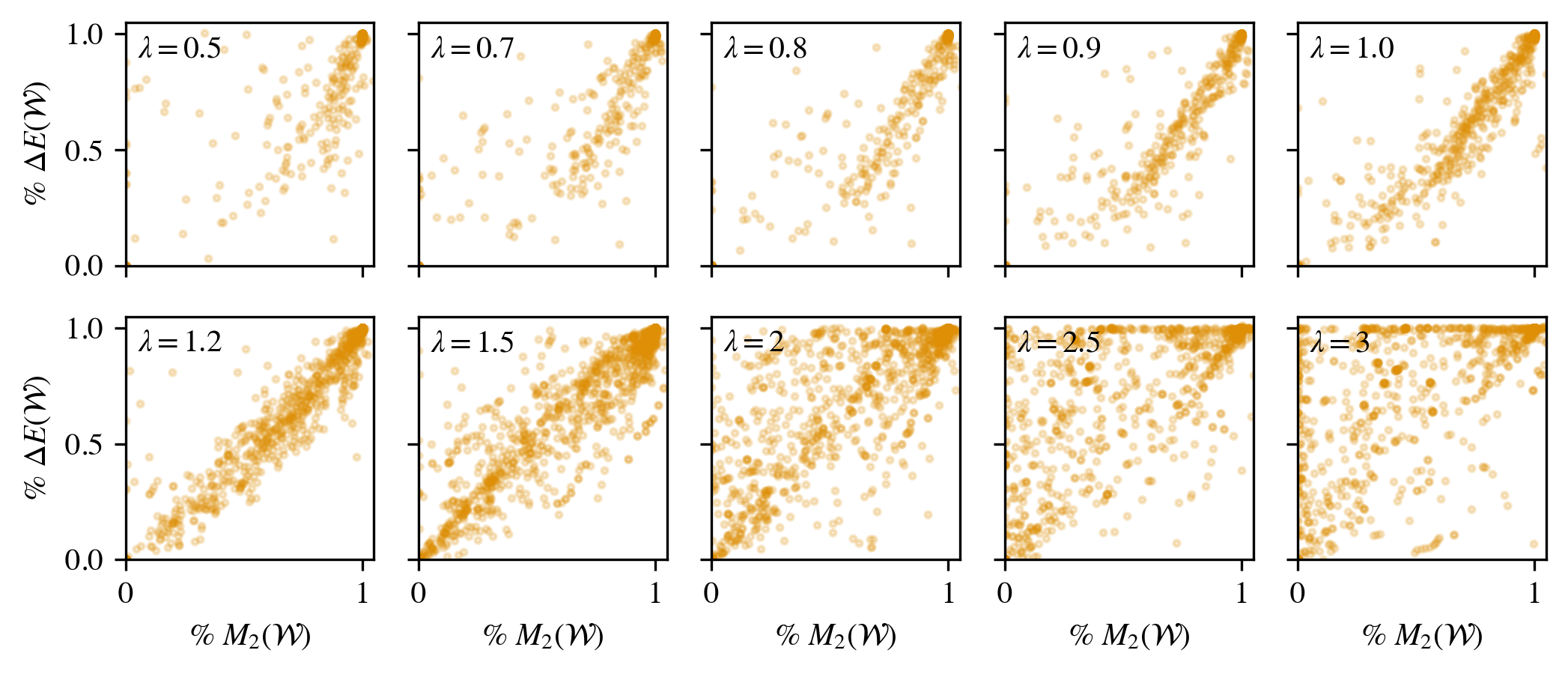}}
    \caption{Fractional correlation energy $\%\Delta E(\mathcal{W})$ and fractional magic $\%M_2(\mathcal{W})$ of the contextual subspace Hamiltonian ground-states for species simulated across a range of bond lengths. All bonds in each molecule are stretched by a factor $\lambda$, with $\lambda=1$ meaning all bonds are at equilibrium bond length. We range from $0.5 \leq\lambda  \leq 3$. For species simulated below their Coulson-Fischer point (shown in (a)), we find a strong linear relationship between the two quantities of $\%\Delta E(\mathcal{W})$ and $\% \Delta M_2(\mathcal{W})$, though the relationship is somewhat weaker at $\lambda=0.5$. For species simulated above their Coulson-Fischer point (shown in (b)), we find a linear relationship close to equilibrium but find that the two quantities of $\%\Delta E(\mathcal{W})$ and $\% \Delta M_2(\mathcal{W})$ are increasingly uncorrelated at large or short bond lengths far from equilibrium.}
    \label{fig:bondlengthplots}
\end{figure*}

\begin{figure*}[tp!]
\centering
    \subfloat[\label{subfig:fo-_rogue}]{\includegraphics[width=0.32\linewidth]{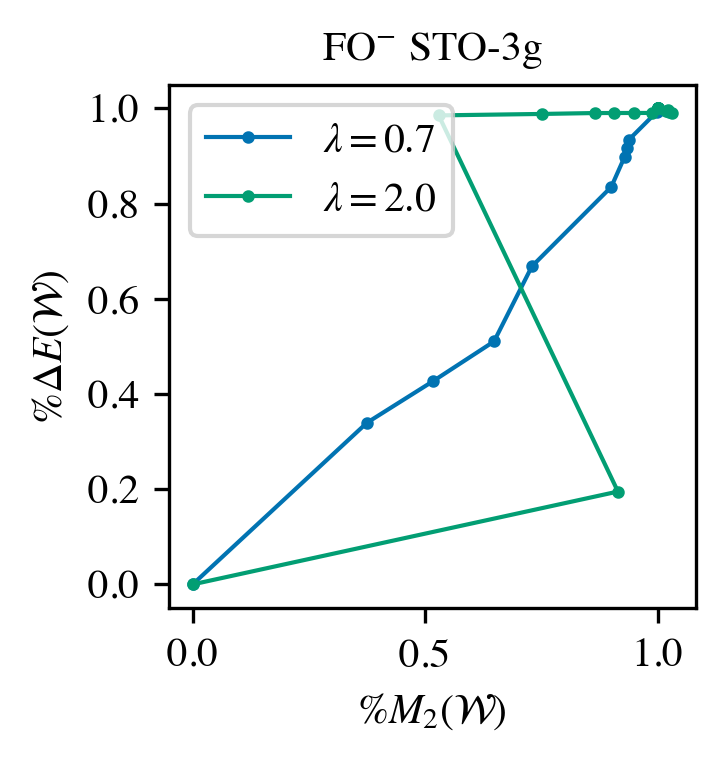}
    }\hfill
    \subfloat[\label{subfig:h4_rogue}]{\includegraphics[width=0.32\linewidth]{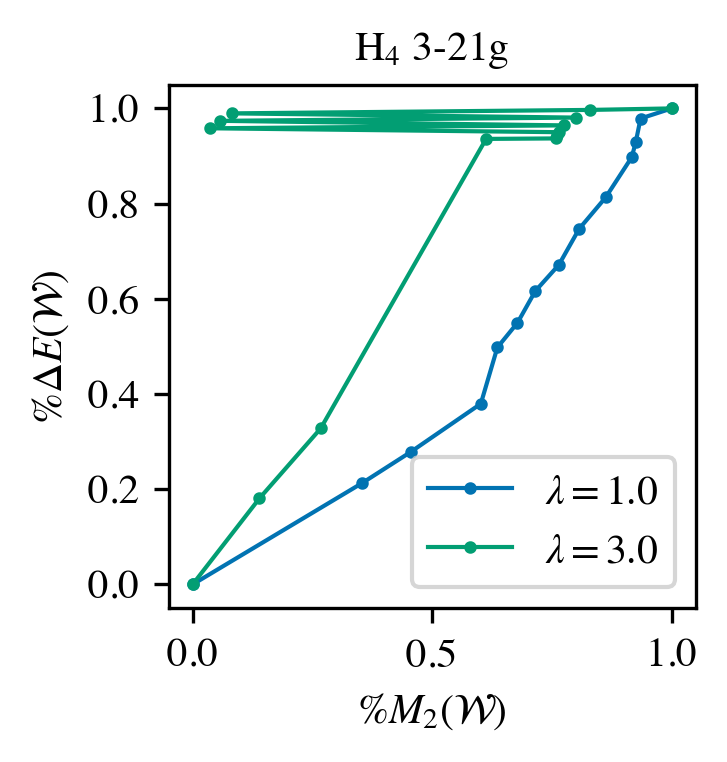}
    }
    \hfill
    \subfloat[\label{subfig:bh3_rogue}]{\includegraphics[width=0.32\linewidth]{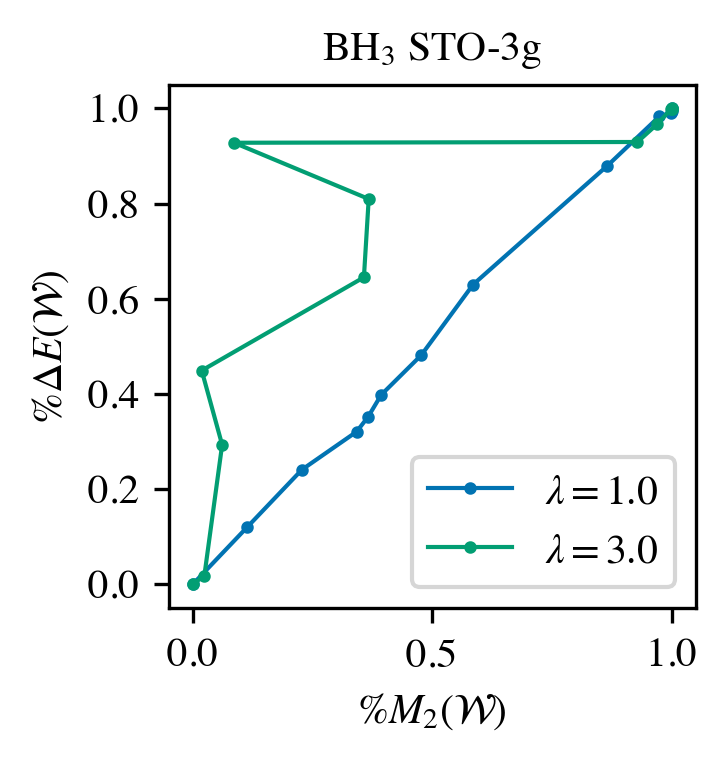}
    }\hfill
    \subfloat[\label{subfig:oh-_rogue}]{\includegraphics[width=0.32\linewidth]{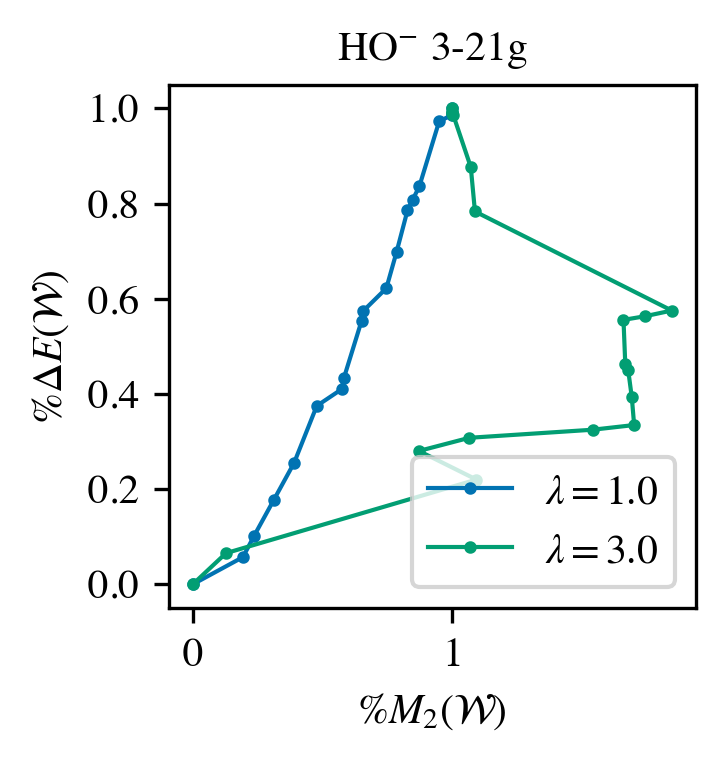}
    }\hfill
    \subfloat[\label{subfig:alh_rogue}]{\includegraphics[width=0.32\linewidth]{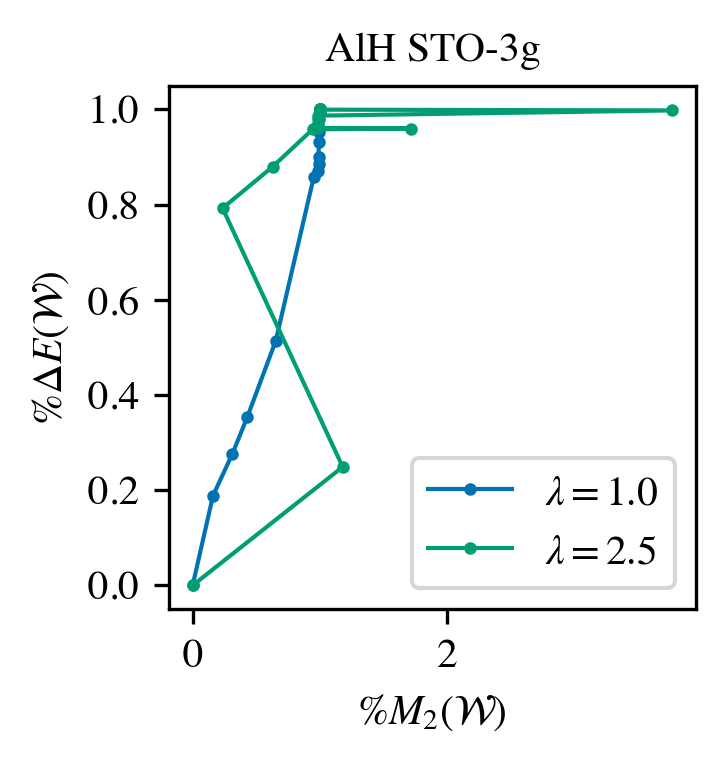}
    }\hfill
    \subfloat[\label{subfig:ch4_rogue}]{\includegraphics[width=0.32\linewidth]{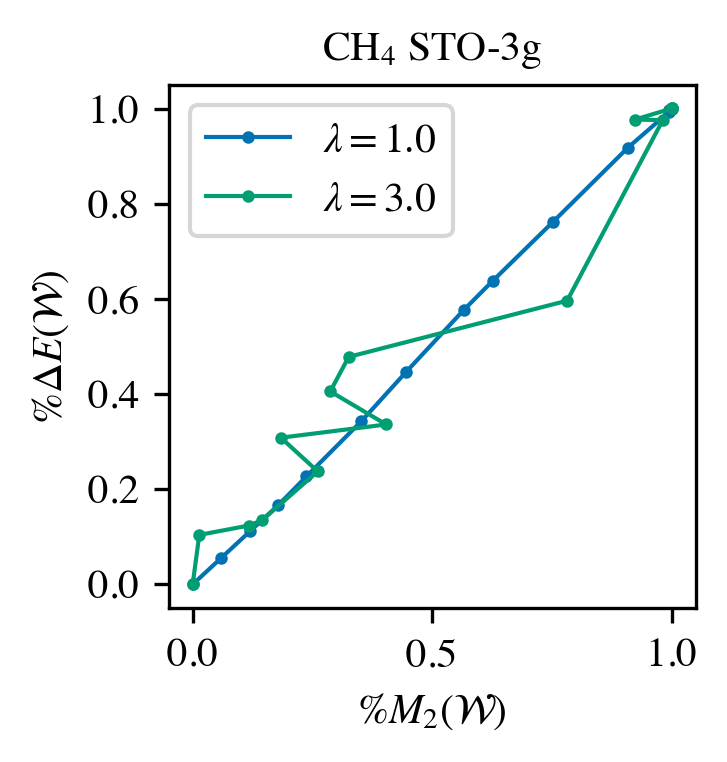}
    }\hfill
    \subfloat[\label{subfig:chn_rogue}]{\includegraphics[width=0.32\linewidth]{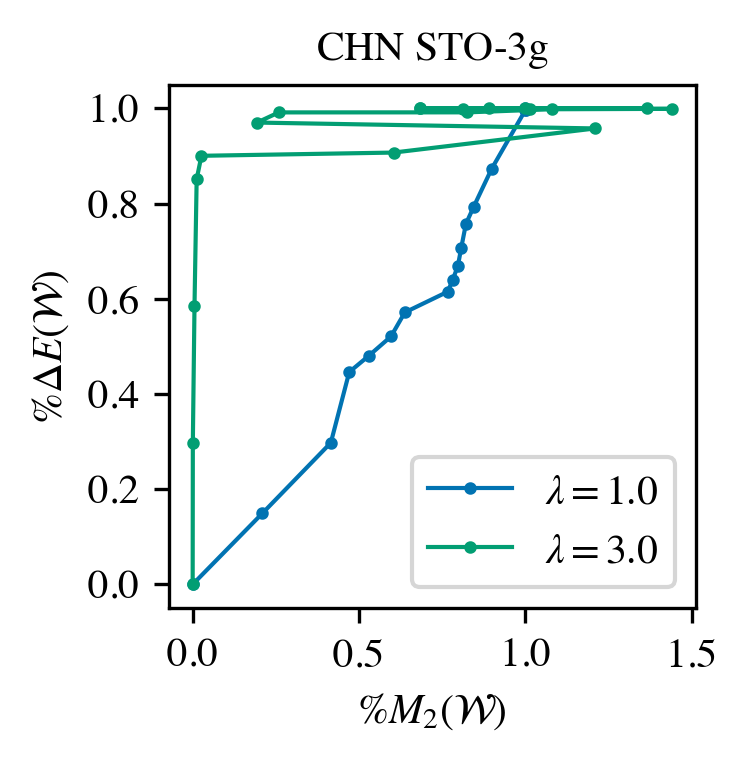}
    }\hfill
    \subfloat[\label{subfig:cn-_rogue}]{\includegraphics[width=0.32\linewidth]{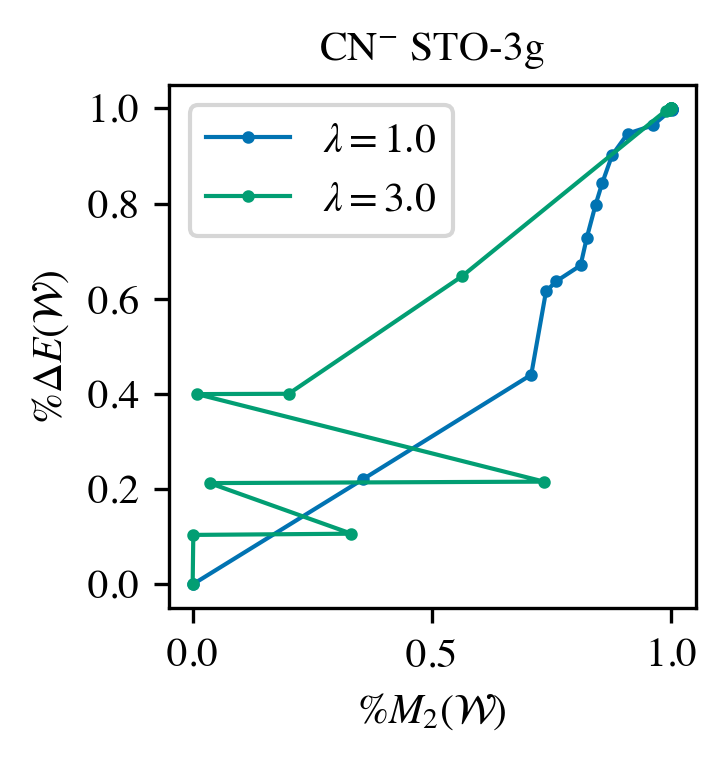}
    }\hfill
    \subfloat[\label{subfig:co_rogue}]{\includegraphics[width=0.32\linewidth]{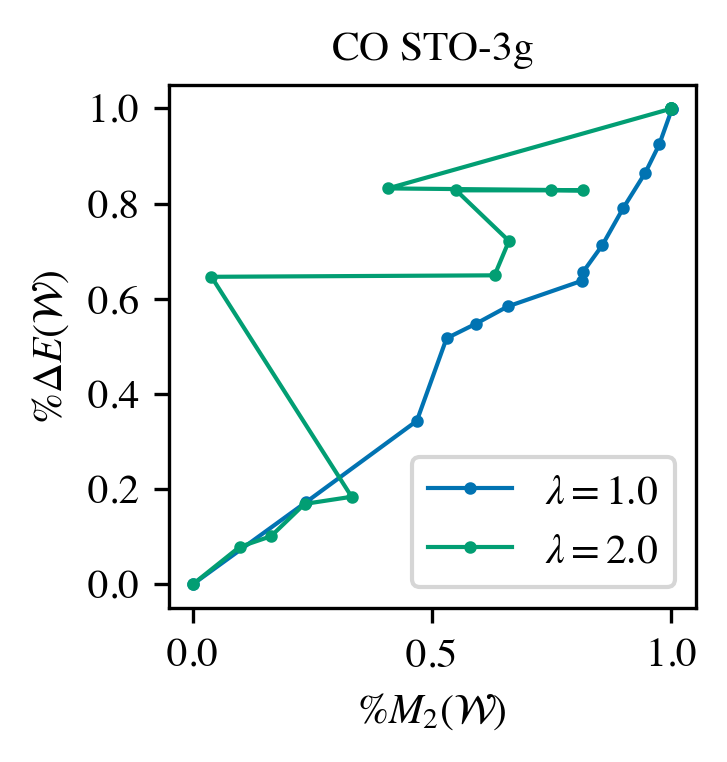}
    }
    
    \caption{(a)-(i) Simulations of selected molecules at a bond length below the Coulson-Fischer point (blue points) and at a bond length above the Coulson-Fischer point (green points). $\lambda$ is the ratio of simulated bond lengths to equilibrium. We find that for $\lambda<\lambda_{CF}$, the relationship between fractional correlation energy $\%\Delta E(\mathcal{W})$ and fractional magic $\%M_2(\mathcal{W})$ is monotonically increasing and roughly linear, while for $\lambda>\lambda_{CF}$ the relationship between fractional correlation energy $\%\Delta E(\mathcal{W})$ and fractional magic $\%M_2(\mathcal{W})$ is no longer monotonic and is often highly non-linear.}
    \label{fig:roguesgallery}
\end{figure*}

In Figures \ref{fig:frac_spaghett_belowcf} and \ref{fig:frac_spaghett_abovecf} we show the results of simulations on 133 multi-atomic species in our dataset simulated at equilibrium bond length and 22 monatomic species (which by definition do not have a bond length). The purpose of this simulation is to determine the linearity of $\%\Delta E(\mathcal{W})$ and $\% M_2(\mathcal{W})$, and to determine the relationship between choices in $\mathcal{W}$ and the proportionality constant presented in Theorem \ref{th:linearity}. 

In Figure \ref{fig:frac_spaghett_belowcf}, we examine the 111 species which have a Coulson-Fischer point above the equilibrium bond length or are monatomic. For these species, we observe a strong linear relationship between the fractional magic $\%M_2(\mathcal{W})$ and the fractional correlation energy $\%\Delta E(\mathcal{W})$. For each species, we find a linear best fit, with demographic information shown in Figure \ref{subfig:below_fithists}. We observe that for the overwhelming majority of molecules in our dataset the linear best fit is relatively robust, with a median $R^2$ value of 0.989 (interquartile range [0.971,0.996]). We also observe that the median slope is often very close to 1 (median 1.009, interquartile range [0.997,1.039]), while the $y$-intercept of the linear best fits is generally slightly negative and biased to the left (median -0.028, interquartile range [-0.067,-0.003]). This implies that in general the fractional magic $\%M_2(\mathcal{W})$ is a slight over-estimate of the fractional correlation energy $\%\Delta E(\mathcal{W})$, indicating the correlation energy could be understood as a lower bound for the quantum resources required for simulation. In general, we find that for species below their Coulson-Fischer point, the constant of proportionality in Theorem \ref{th:linearity} does not change significantly with the size of $\mathcal{W}$.

In Figure \ref{fig:frac_spaghett_abovecf}, we examine the 44 species which have a Coulson-Fischer point at or below their equilibrium bond length. We find a much weaker linear relationship between fractional magic $\%M_2(\mathcal{W})$ and fractional correlation energy $\%\Delta E(\mathcal{W})$ compared to the species below their Coulson-Fischer point shown in Figure \ref{fig:frac_spaghett_belowcf}, particularly for lower magic CS states. We again calculate a linear best fit for each species, with demographic data shown in Figure \ref{subfig:above_fithists}. We find the fits are less robust than for species below their Coulson-Fischer point, with a median $R^2$ of 0.942 (interquartile range [0.908,0.964]). We also find a significantly wider range of best fit slopes (median 1.059, interquartile range [0.967,1.106]) and $y$-intercepts (median -0.112, interquartile range [-0.156,-0.021]). This indicates that for species simulated above their Coulson-Fischer point, the constant of proportionality in Theorem \ref{th:linearity} changes with the size of $\mathcal{W}$.

As previous work has found that both the correlation energy \cite{helgaker2013molecular,ModernQuantumChemistry,kohanoff2006electronic,lewars2003computational} and ground-state magic \cite{sarkis2025are-602,gu2024zero}
change significantly with bond length, in Figure \ref{fig:bondlengthplots}, we examine the relationship between the fractional magic $\%M_2(\mathcal{W})$ and fractional correlation energy $\%\Delta E(\mathcal{W})$ for 137 multi-atomic species in our dataset as all bonds in the molecule are scaled by a factor $\lambda$. In Figure \ref{subfig:bondlengthplots_belowcf}, we plot the fractional magic $\%M_2(\mathcal{W})$ and fractional correlation energy $\%\Delta E(\mathcal{W})$ for those species which are below their Coulson-Fischer point at a given value of $\lambda$. We find a consistently linear relationship, though the relationship is slightly less linear for $\lambda=0.5$. This may indicate that at particularly short bond lengths, restricted Hartree-Fock does not appropriately reflect the dynamics of the molecule. In Figure \ref{subfig:bondlengthplots_abovecf}, we plot the fractional magic $\%M_2(\mathcal{W})$ and fractional correlation energy $\%\Delta E(\mathcal{W})$ for those species which are above their Coulson-Fischer point at a given value of $\lambda$. We find a somewhat robust linear relationship for bond lengths near equilibrium, but observe that for shorter and longer bond lengths, the fractional magic and fractional correlation energy become increasingly uncorrelated.

A potential mechanism for the breakdown of linearity above the Coulson-Fischer point is that, as described in Section \ref{subsec:correlation}, beyond the Coulson-Fischer point restricted Hartree-Fock is no longer the lowest energy single Slater determinant. If the HF ground state used as the non-contextual reference for the CS method does not capture key features of the FCI ground state the CS method may struggle to construct reasonable corrections. To further support the notion that a significant change in the performance of the CS method occurs around the Coulson-Fischer point, in Figure \ref{fig:roguesgallery} we plot the fractional magic $\%M_2(\mathcal{W})$ and fractional correlation energy $\%\Delta E(\mathcal{W})$ for 9 selected species in our dataset, each simulated at one bond length below the Coulson-Fischer point and one bond length above the Coulson-Fischer point. In all instances, we find significant changes between the CS ground states found below the Coulson-Fischer point (in blue) and the CS ground states found above the Coulson-Fischer point (in green), with many of the CS ground states above the Coulson-Fischer point violating the monotonicity prediction of Theorem \ref{th:gs_monotonicity}. As the heuristics Symmer uses to select generators to construct a contextual subspace are premised on structural similarity between the HF state and the FCI ground state, we would expect these heuristics to break down above the Coulson-Fischer point and result in a loss of monotonicity.

\section{Discussion}\label{sec:Discussion}

The identification of magic with correlation in electronic structure Hamiltonians provides an explanation for the historical success of classical methods for calculating the correlation energy in weakly-correlated systems \cite{helgaker2013molecular, kohanoff2006electronic}. Additionally, the linear relationship between SRE and the Hartree-Fock reference weight $\vert c_0\vert^2$ established in Theorem \ref{th:overlaptheorem} enables the efficient estimation of quantum resource costs for a wide range of weakly- and moderately-correlated molecules. The Hartree-Fock weight is frequently calculated by quantum chemistry simulation libraries, and thus can be used to reliably estimate the SRE of molecules with small correlation without the need for direct calculation for systems where the Hartree-Fock energy can be trusted. 

We note that because contextuality and magic are measures of non-classicality \cite{Howard2014,Mermin1993}, these results suggest that correlation energy can be viewed as a marker of non-classical behavior in chemical systems, an idea previously proposed through the lens of entanglement \cite{esquivel2015correlation,wang2007quantum-8bf}. When the electronic structure Hamiltonian is restricted to a non-contextual Hamiltonian, the associated Hartree-Fock ground-state is a stabilizer state and recovers zero correlation energy. When the full Hamiltonian is considered, the associated FCI ground state is generally a non-stabilizer state and recovers the full correlation energy by definition. When we utilize the CS method to construct Hamiltonians in between those two extremes, we find an increase in the non-classicality of the Hamiltonian measured by the ground-state magic is associated with a corresponding increase in the correlation energy. This suggests a relationship between correlation energy and non-classicality that goes beyond the weakly to moderately correlated regime where we can establish perturbative relationships.

These results provide validation for the contextual subspace method as a mechanism for manipulating the quantum resources in a given problem. Our numerical results indicate that, when provided with a reasonably accurate non-contextual approximation of the true ground-state, the CS method produces approximate ground states which trade some amount of energy error for a reduction in magic. While the CS method has thus far been primarily investigated in the context of quantum chemistry \cite{kirby2021contextual,weaving2025contextual,yao2025efficient,PhysRevResearch.5.043054}, this suggests that the CS method might have wider utility as a tool for probing the role of magic in various fields of physics. Prior investigations of magic include examples in condensed matter \cite{turkeshi2023pauli-14f,Oliviero2022,Robin2024,Viscardi2025}, nuclear and high energy physics \cite{Sam2025,White2024,munizzi2026magic,Robin2024,Chernyshev2025}, and quantum gravity \cite{Gargalionis2025,Cao2024}. 

We also believe that the connection between contextuality and correlation energy could be used to improve the CS method. Current approaches to the CS method in literature generally use a ``greedy search'' approach to selecting stabilizers to fix based on prioritizing stabilizers which are most heavily weighted in the Hamiltonian, but this could be improved by instead using the change in ground state SRE as a metric. We also believe that SRE could be used as a heuristic during the classical VQE optimization for CS-VQE. These also provide a motivation for future work seeking to reduce the computational cost of determining SRE. 

These results provide further evidence for SRE's utility as a measure of non-classical behavior in physical systems, and in particular quantum chemistry. As quantum simulation becomes a viable approach for simulating chemical systems, it will be vital to identify areas where there is a high likelihood of achieving an advantage over classical approaches, and SRE can be an important tool in identifying fruitful applications. Recent work \cite{sarkis2025are-602,gu2024zero} has examined ground state SRE in the context of molecular bonding and found evidence in small molecules for an increase in SRE where the extrinsic curvature of the binding energy curve is maximized and classical approaches have the most difficulty capturing electronic behavior.  Further investigation of how the SRE changes with bond length, as well as the impact of basis set choice, could prove useful for identifying opportunities for quantum advantage in quantum chemistry.

Finally, our results have implications for the classical simulation of chemistry. We note that in general the deviations from the linear relationship predicted by Theorem \ref{th:overlaptheorem} are negative, which implies the existence of stabilizer references which are closer to the FCI ground state than Hartree-Fock in the regime where Hartree-Fock is known to be unreliable. These stabilizer references could be identified with methods such as CAFQA \cite{cafqa,keecafqa} or GAS-SCF \cite{qascf} and provide an improvement over Hartree-Fock without increasing resource costs significantly. 

\section*{Data Availability Statement}
The data that support the findings of this article are openly available at \cite{magic-and-correlation-data}.

\begin{acknowledgments}

The authors wish to thank Alexis Ralli, William Simon, Hagar Abualazm, TJ Colvin, Tyler Thurtell, Ben Corbett, Cole Kelson-Packer, Leroy Fagan, Andrew Projansky, and Jason Necaise for useful advice and discussion. 

P.J.L. acknowledges support by the NSF STAQ project (PHY-1818914/232580) and by the NSF NQVL:QSTD:Pilot: Quantum Advantage-Class Trapped Ion system (QACTI) project NSF award number 2410675. A.M. acknowledges support by the National Science Foundation STAQ Project (Grants No. PHY-2325080). Support is acknowledged from the U.S. Department of Energy, Office of Science National Quantum Information Science Research Center, Quantum Systems Accelerator (Award No. DE-SCL0000121).

The authors also wish to acknowledge the 2nd International Workshop on Many-Body Quantum Magic held at the InQubator for Quantum Simulation (IQuS) at the University of Washington in Seattle, WA and the  2025 Southwest Quantum Information and Technology (SQuInT) Conference held at the University of New Mexico in Albuquerque, NM which both significantly influenced the conception and development of this work. IQuS is supported by U.S. Department of Energy, Office of Science, Office of Nuclear Physics, under Award Number DOE (NP) Award DE-SC0020970 via the program on Quantum Horizons: QIS Research and Innovation for Nuclear Science, and by the Department of Physics, and the College of Arts and Sciences at the University of Washington.

The authors acknowledge the \href{https://it.tufts.edu/high-performance-computing}{Tufts University High Performance Compute Cluster} which was utilized for the research reported in this paper.

Q.W. utilized Anthropic's Claude Opus 4.7 to assist in scraping and parsing molecular data from the NIST CCCBDB database. All AI code output was reviewed and modified as needed by the authors to ensure accuracy. All scientific claims, conclusions, and explanations are the original work of the authors.

\end{acknowledgments}

\appendix
\section{Non-Contextual Pauli Hamiltonians and the CS method}\label{app:CSoverview}
Previous work on the CS method was framed as the contextual subspace variational quantum eigensolver (CS-VQE) \cite{Ralli_UP_CS-VQE,weaving2025contextual, Kirby2020, weaving2023stabilizer}, focusing on reducing resources for quantum computation by leveraging the non-contextuality of electronic structure Hamiltonians. In this Appendix we will describe non-contextual Pauli Hamiltonians as is consistent with the existing literature (specifically \cite{ralli2025noncontextual}), as well as the unitary partitioning operator and the magic it induces on a stabilizer state.

\subsection{Non-contextual Pauli structure}\label{app:ncPauliHam}
The following definitions and structures are all defined in \cite{ralli2025noncontextual} and \cite{weaving2023stabilizer}, and are presented here as a reference for the reader. 

For a set of Pauli operators $\mathcal{J}$ to be non-contextual, commutation must be transitive on $\mathcal{J}$. Commutation being an equivalence relation means that for all Pauli operators $A,B,C \in \mathcal{J}$, if $[A,B]=0$, and $[B,C]=0$, then $[A,C]=0$. The Jordan product of Pauli operators $A,B \in \mathcal{J}$ is defined as $A\circ B = \frac{\{A,B\}}{2}$, and is guaranteed to generate Pauli operators which retain commutation as a transitive property. Then, any Pauli Hamiltonian is non-contextual if and only if it is a linear combination of Pauli operators generated by a set of non-contextual Pauli operators under the Jordan product. 

Let us say that $H_{nc}$ is a Pauli Hamiltonian with a non-contextual set of operators $\mathcal{J}$, then we can write $H$ as
\begin{equation}
    H = \sum_{P_j\in\mathcal{J}}a_jP_j.
\end{equation}
The generating set of Pauli operators for $\mathcal{J}$ is denoted as $\mathcal{S}$ and can be written as
\begin{equation}
    \mathcal{S}= \mathcal{G}\cup\{\mathcal{C}_i \vert i = 0,1,\dots M-1\}
\end{equation}
where $\mathcal{G}$ is the independent set of Pauli operators which commute with the entirety of $\mathcal{J}$, and $\{\mathcal{C}_i\}$, which we will denote as $\mathcal{A}$, is a set of mutually anti-commuting Pauli operators. For each $\mathcal{C}_i$, which we will call a \textit{clique}, all Pauli operators commute internally and anti-commute with every other clique. In other words, for $C_j,C_k \in \mathcal{C}_i$, $[C_j,C_k]=0$, but for all $C_j \in \mathcal{C}_j$ and $C_k \in \mathcal{C}_k$, $\{C_j,C_k\}=0$ for $j\neq k$. We can think of the set $\mathcal{G}$ as the set of symmetry generators, as they commute with the whole of $H_{nc}$. We select a representative Pauli operator $C_i$ from each clique $\mathcal{C}_i$, and construct the observable $C(\textbf{r})= \sum_{i=0}^{M-1}r_iC_i$. The vector $\textbf{r}$ is restricted to having $\vert \textbf{r} \vert = 1$, and the components $r_i$ are the expectation values of the clique representatives $r_i = \langle C_i \rangle $. Our set $\mathcal{S}$ can then be represented by $\mathcal{G}\cup C(\textbf{r})$.

The set $\mathcal{S}$ will generate the set $\mathcal{J}$ under the Jordan product as well as every Pauli term that is a product of commuting Pauli operators in $\mathcal{J}$ (known as the closure of $\mathcal{J}$, $\bar{\mathcal{J}}$ is the set such that for all commuting $A,B \in \mathcal{J}$, $AB \in \bar{\mathcal{J}}$). Another way of considering the set $\mathcal{S}$ is as a generating set for a stabilizer subspace. 

However $C(\textbf{r)}$ is not a traditional stabilizer in that it is a linear combination of anti-commuting operators. Therefore we use \textit{unitary partitioning} to rotate the observable $C(\textbf{r})$ by $R$ onto a single Pauli operator denoted $G_A = R^\dagger C(\textbf{r})R$ \cite{ZhaoMeasRedux, Ralli_UP_CS-VQE, weaving2023stabilizer}. The unitary partitioning operator $R$ takes on the form of either a sequence of rotations, or a linear combination of unitaries with complex coefficients. This operation is not Clifford, as it takes $M$ operators to a single Pauli operator, and can induce additional terms in $H_{nc}$. More details on the scaling of terms for a generic Hamiltonian can be found in \cite{Ralli_UP_CS-VQE}. Once the rotation $R$ has been applied to the entire generating set $\mathcal{S} = \mathcal{G}\cup C(\textbf{r}) \mapsto\mathcal{G}\cup G_A$, we obtain a true generating set for a stabilizer subspace on $n$-qubits associated with the $n$ generators. We denote the generator set as $\mathcal{S} = \{G_1, G_2, \dots G_n\}$ in the main text.

We use this stabilizer subspace formalism to find the ground state of our non-contextual Hamiltonian $H_{nc}$. To minimize the energy of the non-contextual Hamiltonian we need only minimize the value assignment of our generators in $\mathcal{S}$, and since we have a non-contextual structure we can have simultaneous value assignment \textit{without} contradiction. Therefore we optimize over all expectation value assignments for our symmetry generators as $G_i\in \mathcal{G}$ as $\langle G_i \rangle = \pm 1$. Notably, when $G_k$ is $G_A$, we have to treat the value assignment more carefully. The expectation value of $G_A$ can be set to always be $\langle G_A\rangle = +1$ without loss of generality, and we instead minimize the energy with respect to $\langle C_i \rangle =r_i\in \textbf{r}$ as $r_i \in [-1,1]$ such that $\vert \textbf{r}\vert = 1$. These optimized value assignments for the clique representatives $C_i$ will change the rotation $R$ in unitary partitioning, so when we apply the rotation to the entire Hamiltonian, we are rotating our problem space to the sector consistent with value assignment. 

\subsection{Unitary Partitioning Operator and Induced Magic}\label{app:taperingCS}

The use of the unitary partitioning technique makes the rotation operator $U_\mathcal{W}$ non-Clifford and increases the number of terms in the Hamiltonian by a factor of $\mathcal{O}(\vert H_\text{full}\vert (n^2-1))$ \cite{Ralli_UP_CS-VQE}. Additionally, the rotation from unitary partitioning (let us call the rotation operator for unitary partitioning $R$) induces magic on a state which scales $\mathcal{O}(\log(n))$ for an $n$ qubit system. To see this, let us examine a non-Clifford rotation of the form
\begin{equation}\label{eq:UPop}
    R = \cos(\theta)\mathbb{I} + i\sin (\theta)\sum_k a_kP_k, \quad \text{for}\ P_k \in \mathcal{A}, a_k \in \mathbb{R}.
\end{equation}
The Pauli set $\mathcal{A}$ is a linear combination of anti-commuting Pauli terms, which in the CS method represent the different cliques in the non-contextual Hamiltonian. 

From Equation \ref{eq:UPop} we can determine that the action of $R$ on a stabilizer state $\ket{\varphi}$ is
\begin{equation}\label{eq:UPstate}
    R\ket{\varphi} = \cos(\theta)\ket{\varphi} + i\sin(\theta) \sum_k a_kP_k\ket{\varphi}
\end{equation}
where there is a single term of the form $\cos(\theta)\ket{\varphi}$, and $\vert\mathcal{A}\vert$ terms which take on the form $i\sin(\theta)\left(a_kP_k\ket{\varphi}\right)$. This results in a stabilizer rank (the minimum terms required to express a state as a linear combination of stabilizer states) for the state $R\ket{\varphi}$ as
\begin{equation}
    \chi(R(\ket{\varphi}))\leq 1+ \vert\mathcal{A}\vert
\end{equation}
with the bound being satisfied if all $P_k\ket{\varphi}$ are stabilizer states. 

The size of the anti-commuting set $\mathcal{A}$ is upper bounded as $\vert\mathcal{A}\vert\leq2(n-\vert\mathcal{G}\vert)+1$ as established in \cite{ralli2025noncontextual}, where $\mathcal{G}$ is the generating set of commuting symmetries in the non-contextual Hamiltonian.
Therefore the stabilizer rank becomes
\begin{equation}
\begin{split}
    \chi(R\ket{\varphi}) &\leq 1+2(n-\vert\mathcal{G}\vert)+1
    \\&=2(n-\vert\mathcal{G}\vert+1).
\end{split}
\end{equation}
A stabilizer state has support on $2^n$ Pauli terms, therefore $R\ket{\varphi}$ has a maximum of $2^n\chi$ Paulis with non-zero expectation values. In other words, the support of Pauli operators on $R\ket{\varphi}$ is $2^n\chi$.

We know that the $\alpha-$SRE for $\alpha \geq 0$ is upper bounded by the Hartley Entropy ($\alpha\rightarrow0$)
\begin{equation}
    M_\alpha(\ket{\psi})\leq H_0(\ket{\psi})=\log(\text{card}\ket{\psi}/2^n), 
\end{equation}
for $\text{card}\ket{\psi}$ the number of non-zero Pauli terms in $\mathcal{P}^n$, for $R\ket{\varphi}$ card$R\ket{\varphi}=2^n\chi$. 

Finally we can upper bound SRE as
\begin{equation}
\begin{split}
    M_\alpha(R\ket{\varphi}) &\leq \log(2^n\chi/2^n) \\
    &= \log[2(n-\vert\mathcal{G}\vert + 1)] \\
    &= \mathcal{O}(\log(n)).
\end{split}
\end{equation}
Meaning that the effect of the unitary partitioning rotation $R$ on magic is an increase which scales at worst $\mathcal{O}(\log(n))$. For more details on unitary partitioning and the clique construction for the CS method please see Reference \cite{Ralli_UP_CS-VQE}.

\section{Efficient calculation of SRE for sparse pure states in the symplectic formalism}\label{app:SREcalc}
In this Appendix, we present an efficient algorithm for calculating the stabilizer R\'enyi entropy of sparse states utilizing the symplectic representation.

Any $n$-qubit Pauli string can be represented as
\begin{equation}
        P=\bigotimes_{j=1}^n i^{x_jz_j}X^{x_j}Z^{z_j},
\end{equation}
where
\begin{equation}
        x_j=\left\{\begin{array}{ll}
            1,& P_j=X \text{ or } Y\\
            0,& P_j=Z \text{ or } I
        \end{array}\right.,\quad z_j=\left\{\begin{array}{ll}
            1,& P_j=Z \text{ or } Y\\
            0,& P_j=X \text{ or } I
        \end{array}\right.
\end{equation}
We can then construct the symplectic vectors $\vec{X}$, $\vec{Z}\in\mathbb{F}^n_2$ as
\begin{equation}
    \vec{X}=[x_1,x_2,\ldots,x_n],\quad  \vec{Z}=[z_1,z_2,\ldots,z_n].
\end{equation}
The action of a Pauli string $P$ on a computational basis state $\ket{j}$ can then be written as
\begin{equation}
    P\ket{j}=i^{\vec{X}\cdot\vec{Z}}(-1)^{j\cdot\vec{Z}}\ket{j\oplus\vec{X}},
\end{equation}
where $\oplus$ represents bitwise XOR and j without a ket represents the bitstring of the computational basis state. Because of the orthonormality of computational basis states, we can then write the inner product of $P$ with any two computational basis states $\ket{j}$ and $\ket{j'}$ as 
\begin{equation}\label{eq:paulimel}
    \mel{j'}{P}{j}=i^{\vec{X}\cdot\vec{Z}}(-1)^{j\cdot\vec{Z}}\delta_{j',j\oplus\vec{X}}.
\end{equation}
Then for an arbitrary state $\ket{\psi}$ expressed as a superposition over computational basis states as $\ket{\psi}=\sum_j c_j \ket{j}$,
the expectation value $\mel{\psi}{P}{\psi}$ can be written as
\begin{align}
    \mel{\psi}{P}{\psi}&=\sum_{j,j'}\braket{\psi}{j'}\mel{j'}{P}{j}\braket{j}{\psi}\nonumber\\
    &=i^{\vec{X}\cdot\vec{Z}}\sum_{j} (-1)^{j\cdot \vec{Z}} c^*_{j}c_{j\oplus \vec{X}}.\label{eq:symppauliexp} 
\end{align}

Suppose now that we wish to calculate the SRE of a $k$-sparse $n$-qubit state which we write as a superposition of $k$ computational basis states $J=\{j_1,j_2,\cdots j_k\}$. In principle, there would be $4^N$ Pauli strings to measure expectation values of. But from \eqref{eq:symppauliexp}, we can see that the only Pauli strings which contribute will be those for which $\vec{X}\in J\oplus J$, i.e. $\vec{X}=j_1\oplus j_2$ for some $j_1,j_2\in J$, meaning at most $2^{n+k}$ Pauli strings are non-zero. Thus, we can write the stabilizer purity $\zeta_\alpha$ as 
\begin{equation}
    \zeta_\alpha =\frac{1}{2^n}\sum_{\vec{X}\in J\oplus J}\sum_{\vec{Z}\in\mathbb{F}^n_2}(-1)^{\alpha\vec{X}\cdot\vec{Z}}\left( \sum_{j}(-1)^{j\cdot\vec{Z}} c_jc^*_{j\oplus\vec{X}}  \right)^{2\alpha}
\end{equation}
As noted by \cite{xiao2026exponentially-96a,huang2026fast-9ec}, $\sum_{j}(-1)^{j\cdot\vec{Z}} c_jc^*_{j\oplus\vec{X}}$ can be calculated in $O(n\log n)$ time using the fast Walsh-Hadamard transform (FWHT). Thus, by combining sparsity screening and the FWHT, the SRE can be calculated in $O(2^k n\log n)$ time.

\section{Stabilizer R\'enyi entropy}\label{app:srefornerds}
It is evident, and proven in \cite{LeoneSREPhysRevLett.128.050402}, that the $\alpha$-SRE is able to satisfy all the requirements for a good measure of magic, namely
\begin{enumerate}
    \item Faithfulness: $M_{\alpha}(\ket{\psi}) = 0 $ iff $\ket{\psi} \in \text{ STAB}$, otherwise $M_{\alpha}(\ket{\psi}) > 0 $. 

    \item Stability under free operations (invariance under Clifford transformations): $C \in \mathcal{C}(\mathcal{H}): M_{\alpha}(C\ket{\psi}) = M_{\alpha}(\ket{\psi})$.

    \item Additivity: $M_{\alpha}(\ket{\psi}\otimes \ket{\phi}) = M_{\alpha}(\ket{\psi}) + M_{\alpha}(\ket{\phi})$.
\end{enumerate}

Furthermore, the $\alpha$-SRE has been shown to be a monotone of non-stabilizerness for $\alpha\geq2$ \cite{LeoneSREPhysRevLett.128.050402,HaugSRE10.22331/q-2023-08-28-1092,LeoneMagicMonotonePhysRevA.110.L040403}. A magic monotone is a real-valued function $M$ for all $n \in \mathbb{N}$ qubit systems such that 
\begin{enumerate}
    \item $M(\psi) = 0$ iff $\psi$ is a pure stabilizer state,
    \item $\forall$ pair $(\psi, \varepsilon)$, where $\varepsilon \in \mathcal{F}$ is a free stabilizer protocol $\varepsilon(\ket{\psi}\bra{\psi}) = \ket{\phi}\bra{\phi}$ it holds that 
\begin{equation}
    \mathcal{M}(\ket{\phi}) \leq \mathcal{M}(\ket{\psi}).
\end{equation}
\end{enumerate}

These requirements imply that stabilizer states possess 0 magic and that, under any Clifford operation or stabilizer protocol, the magic of the resulting state may only remain the same or decrease. Consequently, the non-contextuality measure $M(\psi)$ can increase only if the operation is not free. It has been shown that 2-SRE is a magic monotone \cite{LeoneMagicMonotonePhysRevA.110.L040403} providing us with the confidence to use 2-SRE in our work. Unlike other measures of magic, such as contextual separation, SRE can be directly calculated without requiring optimization procedures \cite{Kirby2019Contextuality}. For these reasons and due to the measurability of SRE, all simulations of magic presented in this work are based on the 2-SRE. 
\subsection{Expanding SRE to mixed states}\label{app:sreadditional}

Extending SRE to mixed states, we expand on 2-SRE in equation \eqref{eq:SREdef}. We look at the projector form of the entropy as well as an extension of SRE to the mixed state case. We have that the projector form of 2-SRE is given by
\begin{align} \label{eq:SREdef_proj}
    M_2(\ket{\psi}) &= -\text{log}\left(2^n\text{ tr}(Q\ket{\psi}\bra{\psi}^{\bigotimes4})\right) \\ Q&:= 2^{-2n}\sum_{P\in \mathcal{P}^{(n)}} P^{\bigotimes 4},
\end{align}
for $\mathcal{P}^{(n)}$ defined as the group of all n-qubit Pauli strings, including phases of $\pm 1, \pm i$, without the identity. This form is mostly useful in the standard randomized measurement schemes for 2-SRE \cite{LeoneSREPhysRevLett.128.050402}.

Extending 2-SRE to a mixed state representation is relatively straightforward by utilizing a renormalization of the pure state 2-SRE. Let us define our Clifford resource states (also referred to as free resources) as $\chi = 2^{-n} (\mathbb{I}_{n} + \sum_{P \in G} \phi_P P)$, with $G \subset \mathcal{P}^{(n)}$ and $0 \leq |G| \leq 2^n-1$. The 2-SRE is then given by

\begin{align}\label{eq:SREdef_mixedstates}
    \Tilde{M}_2(\rho) &:= M_2(\rho) - S_2(\rho) 
\end{align}

for $M_2(\rho) = -\text{log}\left(2^n\text{ tr}(Q\rho^{\bigotimes4})\right)$ the 2-SRE and $S_2(\rho) := -\text{log Pur}(\rho)$ the 2-R{\'e}nyi entanglement entropy \cite{LeoneSREPhysRevLett.128.050402}. In the same form as in equation \eqref{eq:SREdef} we find that 
\begin{equation}\label{eq:SREdef_mixed_tr}
    \Tilde{M}_2(\rho) = - \frac{\sum_{P\in\mathcal{P}^{(n)}} \text{tr}^4(P\rho)}{\sum_{P\in\mathcal{P}^{(n)}} \text{tr}^2(P\rho)}.
\end{equation}
The same properties of pure state SRE are found in this mixed state form.

\section{Hamiltonian dataset}\label{app:hamiltonian_database}

The molecular Hamiltonians used in this work are drawn from the Symmer-Hamiltonian dataset~\cite{symmer_hamiltonian_database}, a collection of Jordan-Wigner-encoded qubit operators compiled for quantum computing benchmarks. Table~\ref{tab:dataset_summary} summarizes the dataset.
The generated Hamiltonians were cross-checked against the Symmer reference operators and their Hartree-Fock energies against NIST CCCBDB reference energies, and found consistent with both.

\begin{table}[tp]
\centering
\caption{Summary of the Symmer-Hamiltonian dataset.}
\label{tab:dataset_summary}
\begin{tabular}{ll}
\hline\hline
Property & Value \\
\hline
Species (without basis)       & 135 \\
Species (with basis)        & 190 \\
Hamiltonians            & 1{,}594  \\
Target spin sector      & Singlet ($2S{+}1 = 1$) \\
Qubit range (untapered) & 2--24 \\
Qubit range (tapered)   & 1--20 \\
Encoding                & Jordan-Wigner  \\
Basis sets              & STO-3G, 3-21G, 6-31G**, \\
                        & 6-311G**, cc-pVDZ \\
\hline\hline
\end{tabular}
\end{table}

For diatomic and polyatomic species, the equilibrium geometry is uniformly scaled by a factor $\alpha \in \{0.5, 0.7, 0.8, 0.9, 1.0, 1.2, 1.5, 2.0, 2.5, 3.0\}$, where $\alpha$ multiplies all atomic Cartesian coordinates relative to the center of mass at equilibrium. For diatomics this is equivalent to bond-length scaling. For polyatomic species it uniformly scales all interatomic distances. Single-atom species are computed at $\alpha = 1.0$ only.

Each Hamiltonian is produced by the following pipeline, implemented using PySCF~2.11.0~\cite{sun2020recent} and symmerpyscf~0.1.0~\cite{symmerpyscf}:

\begin{enumerate}
    \item \emph{Geometry sourcing.} Equilibrium nuclear coordinates are retrieved from NIST CCCBDB~\cite{CCCBDB}, PennyLane molecular datasets, or the Symmer reference collection, and the first available source is used. Note that equilibrium geometries are inherently dependent on the specific theory and basis set, and may not precisely match each other or the experimental geometry.
    \item \emph{Geometry scaling.} Each diatomic and polyatomic geometry is scaled by the ten $\alpha$ values listed above.
    \item \emph{Self-consistent field.} A restricted Hartree-Fock calculation is performed. Geometries are processed in two sweeps from equilibrium: a stretching sweep ($\alpha = 1.0 \to 1.2 \to 1.5 \to 2.0 \to 2.5 \to 3.0$) followed by a compression sweep ($\alpha = 1.0 \to 0.9 \to 0.8 \to 0.7 \to 0.5$), and the converged density matrix from each geometry seeds the initial guess for the next, ensuring continuity of the electronic state at distorted geometries.
    \item \emph{Jordan-Wigner encoding.} The second-quantized fermionic Hamiltonian is mapped to a qubit operator via the Jordan-Wigner transformation, implemented by Symmer~\cite{Ralli_UP_CS-VQE}.
    \item \emph{$\mathbb{Z}_2$ tapering metadata.} Symmetry generators are identified and the achievable tapered qubit count is recorded. The stored Hamiltonians remain in the full, untapered form. In our dataset, only species whose tapered qubit count satisfies $n_{\mathrm{tap}} \leq 20$ are included. 
\end{enumerate}

\bibliography{references}

\begin{thebibliography}{109}%
\makeatletter
\providecommand \@ifxundefined [1]{%
 \@ifx{#1\undefined}
}%
\providecommand \@ifnum [1]{%
 \ifnum #1\expandafter \@firstoftwo
 \else \expandafter \@secondoftwo
 \fi
}%
\providecommand \@ifx [1]{%
 \ifx #1\expandafter \@firstoftwo
 \else \expandafter \@secondoftwo
 \fi
}%
\providecommand \natexlab [1]{#1}%
\providecommand \enquote  [1]{``#1''}%
\providecommand \bibnamefont  [1]{#1}%
\providecommand \bibfnamefont [1]{#1}%
\providecommand \citenamefont [1]{#1}%
\providecommand \href@noop [0]{\@secondoftwo}%
\providecommand \href [0]{\begingroup \@sanitize@url \@href}%
\providecommand \@href[1]{\@@startlink{#1}\@@href}%
\providecommand \@@href[1]{\endgroup#1\@@endlink}%
\providecommand \@sanitize@url [0]{\catcode `\\12\catcode `\$12\catcode `\&12\catcode `\#12\catcode `\^12\catcode `\_12\catcode `\%12\relax}%
\providecommand \@@startlink[1]{}%
\providecommand \@@endlink[0]{}%
\providecommand \url  [0]{\begingroup\@sanitize@url \@url }%
\providecommand \@url [1]{\endgroup\@href {#1}{\urlprefix }}%
\providecommand \urlprefix  [0]{URL }%
\providecommand \Eprint [0]{\href }%
\providecommand \doibase [0]{https://doi.org/}%
\providecommand \selectlanguage [0]{\@gobble}%
\providecommand \bibinfo  [0]{\@secondoftwo}%
\providecommand \bibfield  [0]{\@secondoftwo}%
\providecommand \translation [1]{[#1]}%
\providecommand \BibitemOpen [0]{}%
\providecommand \bibitemStop [0]{}%
\providecommand \bibitemNoStop [0]{.\EOS\space}%
\providecommand \EOS [0]{\spacefactor3000\relax}%
\providecommand \BibitemShut  [1]{\csname bibitem#1\endcsname}%
\let\auto@bib@innerbib\@empty
\bibitem [{\citenamefont {Lewars}(2003)}]{lewars2003computational}%
  \BibitemOpen
  \bibfield  {author} {\bibinfo {author} {\bibfnamefont {E.}~\bibnamefont {Lewars}},\ }\href@noop {} {\emph {\bibinfo {title} {Computational Chemistry: Introduction to the Theory and Applications of Molecular and Quantum Mechanics}}}\ (\bibinfo  {publisher} {Springer},\ \bibinfo {year} {2003})\BibitemShut {NoStop}%
\bibitem [{\citenamefont {Szabo}\ and\ \citenamefont {Ostlund}(2012)}]{ModernQuantumChemistry}%
  \BibitemOpen
  \bibfield  {author} {\bibinfo {author} {\bibfnamefont {A.}~\bibnamefont {Szabo}}\ and\ \bibinfo {author} {\bibfnamefont {N.~S.}\ \bibnamefont {Ostlund}},\ }\href@noop {} {\emph {\bibinfo {title} {Modern Quantum Chemistry: Introduction to Advanced Electronic Structure Theory}}}\ (\bibinfo  {publisher} {Macmillan},\ \bibinfo {year} {2012})\BibitemShut {NoStop}%
\bibitem [{\citenamefont {Helgaker}\ \emph {et~al.}(2013)\citenamefont {Helgaker}, \citenamefont {Jorgensen},\ and\ \citenamefont {Olsen}}]{helgaker2013molecular}%
  \BibitemOpen
  \bibfield  {author} {\bibinfo {author} {\bibfnamefont {T.}~\bibnamefont {Helgaker}}, \bibinfo {author} {\bibfnamefont {P.}~\bibnamefont {Jorgensen}},\ and\ \bibinfo {author} {\bibfnamefont {J.}~\bibnamefont {Olsen}},\ }\href@noop {} {\emph {\bibinfo {title} {Molecular Electronic-Structure Theory}}}\ (\bibinfo  {publisher} {John Wiley \& Sons},\ \bibinfo {year} {2013})\BibitemShut {NoStop}%
\bibitem [{\citenamefont {L\"owdin}(1954)}]{lwdin1954quantum-ebe}%
  \BibitemOpen
  \bibfield  {author} {\bibinfo {author} {\bibfnamefont {P.-O.}\ \bibnamefont {L\"owdin}},\ }\bibfield  {title} {\bibinfo {title} {Quantum theory of many-particle systems. {III}. {E}xtension of the {H}artree-{F}ock scheme to include degenerate systems and correlation effects},\ }\href {https://doi.org/10.1103/physrev.97.1509} {\bibfield  {journal} {\bibinfo  {journal} {Phys. Rev.}\ }\textbf {\bibinfo {volume} {97}},\ \bibinfo {pages} {1509} (\bibinfo {year} {1954})}\BibitemShut {NoStop}%
\bibitem [{\citenamefont {L\"{o}wdin}(1995)}]{lowdincorrelation}%
  \BibitemOpen
  \bibfield  {author} {\bibinfo {author} {\bibfnamefont {P.-O.}\ \bibnamefont {L\"{o}wdin}},\ }\bibfield  {title} {\bibinfo {title} {The historical development of the electron correlation problem},\ }\href {https://doi.org/10.1002/qua.560550203} {\bibfield  {journal} {\bibinfo  {journal} {Int. J. Quantum Chem.}\ }\textbf {\bibinfo {volume} {55}},\ \bibinfo {pages} {77} (\bibinfo {year} {1995})}\BibitemShut {NoStop}%
\bibitem [{\citenamefont {Aspuru-Guzik}\ \emph {et~al.}(2005)\citenamefont {Aspuru-Guzik}, \citenamefont {Dutoi}, \citenamefont {Love},\ and\ \citenamefont {Head-Gordon}}]{doi:10.1126/science.1113479}%
  \BibitemOpen
  \bibfield  {author} {\bibinfo {author} {\bibfnamefont {A.}~\bibnamefont {Aspuru-Guzik}}, \bibinfo {author} {\bibfnamefont {A.~D.}\ \bibnamefont {Dutoi}}, \bibinfo {author} {\bibfnamefont {P.~J.}\ \bibnamefont {Love}},\ and\ \bibinfo {author} {\bibfnamefont {M.}~\bibnamefont {Head-Gordon}},\ }\bibfield  {title} {\bibinfo {title} {Simulated quantum computation of molecular energies},\ }\href {https://doi.org/10.1126/science.1113479} {\bibfield  {journal} {\bibinfo  {journal} {Science}\ }\textbf {\bibinfo {volume} {309}},\ \bibinfo {pages} {1704} (\bibinfo {year} {2005})}\BibitemShut {NoStop}%
\bibitem [{\citenamefont {Lanyon}\ \emph {et~al.}(2010)\citenamefont {Lanyon}, \citenamefont {Whitfield}, \citenamefont {Gillett}, \citenamefont {Goggin}, \citenamefont {Almeida}, \citenamefont {Kassal}, \citenamefont {Biamonte}, \citenamefont {Mohseni}, \citenamefont {Powell}, \citenamefont {Barbieri}, \citenamefont {Aspuru-Guzik},\ and\ \citenamefont {White}}]{lanyon2010towards}%
  \BibitemOpen
  \bibfield  {author} {\bibinfo {author} {\bibfnamefont {B.~P.}\ \bibnamefont {Lanyon}}, \bibinfo {author} {\bibfnamefont {J.~D.}\ \bibnamefont {Whitfield}}, \bibinfo {author} {\bibfnamefont {G.~G.}\ \bibnamefont {Gillett}}, \bibinfo {author} {\bibfnamefont {M.~E.}\ \bibnamefont {Goggin}}, \bibinfo {author} {\bibfnamefont {M.~P.}\ \bibnamefont {Almeida}}, \bibinfo {author} {\bibfnamefont {I.}~\bibnamefont {Kassal}}, \bibinfo {author} {\bibfnamefont {J.~D.}\ \bibnamefont {Biamonte}}, \bibinfo {author} {\bibfnamefont {M.}~\bibnamefont {Mohseni}}, \bibinfo {author} {\bibfnamefont {B.~J.}\ \bibnamefont {Powell}}, \bibinfo {author} {\bibfnamefont {M.}~\bibnamefont {Barbieri}}, \bibinfo {author} {\bibfnamefont {A.}~\bibnamefont {Aspuru-Guzik}},\ and\ \bibinfo {author} {\bibfnamefont {A.~G.}\ \bibnamefont {White}},\ }\bibfield  {title} {\bibinfo {title} {Towards quantum chemistry on a quantum computer},\ }\href {https://doi.org/10.1038/nchem.483} {\bibfield  {journal} {\bibinfo  {journal} {Nat. Chem.}\ }\textbf
  {\bibinfo {volume} {2}},\ \bibinfo {pages} {106} (\bibinfo {year} {2010})}\BibitemShut {NoStop}%
\bibitem [{\citenamefont {Whitfield}\ \emph {et~al.}(2011)\citenamefont {Whitfield}, \citenamefont {Biamonte},\ and\ \citenamefont {Aspuru-Guzik}}]{whitfield2011simulation}%
  \BibitemOpen
  \bibfield  {author} {\bibinfo {author} {\bibfnamefont {J.~D.}\ \bibnamefont {Whitfield}}, \bibinfo {author} {\bibfnamefont {J.}~\bibnamefont {Biamonte}},\ and\ \bibinfo {author} {\bibfnamefont {A.}~\bibnamefont {Aspuru-Guzik}},\ }\bibfield  {title} {\bibinfo {title} {Simulation of electronic structure {H}amiltonians using quantum computers},\ }\href {https://doi.org/10.1080/00268976.2011.552441} {\bibfield  {journal} {\bibinfo  {journal} {Mol. Phys.}\ }\textbf {\bibinfo {volume} {109}},\ \bibinfo {pages} {735} (\bibinfo {year} {2011})}\BibitemShut {NoStop}%
\bibitem [{\citenamefont {Kassal}\ \emph {et~al.}(2011)\citenamefont {Kassal}, \citenamefont {Whitfield}, \citenamefont {Perdomo-Ortiz}, \citenamefont {Yung},\ and\ \citenamefont {Aspuru-Guzik}}]{kassal2011simulating}%
  \BibitemOpen
  \bibfield  {author} {\bibinfo {author} {\bibfnamefont {I.}~\bibnamefont {Kassal}}, \bibinfo {author} {\bibfnamefont {J.~D.}\ \bibnamefont {Whitfield}}, \bibinfo {author} {\bibfnamefont {A.}~\bibnamefont {Perdomo-Ortiz}}, \bibinfo {author} {\bibfnamefont {M.-H.}\ \bibnamefont {Yung}},\ and\ \bibinfo {author} {\bibfnamefont {A.}~\bibnamefont {Aspuru-Guzik}},\ }\bibfield  {title} {\bibinfo {title} {Simulating chemistry using quantum computers},\ }\href {https://doi.org/10.1146/annurev-physchem-032210-103512} {\bibfield  {journal} {\bibinfo  {journal} {Annu. Rev. Phys. Chem.}\ }\textbf {\bibinfo {volume} {62}},\ \bibinfo {pages} {185} (\bibinfo {year} {2011})}\BibitemShut {NoStop}%
\bibitem [{\citenamefont {{McArdle}}\ \emph {et~al.}(2020)\citenamefont {{McArdle}}, \citenamefont {Endo}, \citenamefont {Aspuru-Guzik}, \citenamefont {Benjamin},\ and\ \citenamefont {Yuan}}]{McArdle2020}%
  \BibitemOpen
  \bibfield  {author} {\bibinfo {author} {\bibfnamefont {S.}~\bibnamefont {{McArdle}}}, \bibinfo {author} {\bibfnamefont {S.}~\bibnamefont {Endo}}, \bibinfo {author} {\bibfnamefont {A.}~\bibnamefont {Aspuru-Guzik}}, \bibinfo {author} {\bibfnamefont {S.~C.}\ \bibnamefont {Benjamin}},\ and\ \bibinfo {author} {\bibfnamefont {X.}~\bibnamefont {Yuan}},\ }\bibfield  {title} {\bibinfo {title} {Quantum computational chemistry},\ }\href {https://doi.org/10.1103/revmodphys.92.015003} {\bibfield  {journal} {\bibinfo  {journal} {Rev. Mod. Phys.}\ }\textbf {\bibinfo {volume} {92}},\ \bibinfo {pages} {015003} (\bibinfo {year} {2020})}\BibitemShut {NoStop}%
\bibitem [{\citenamefont {Schleich}\ and\ \citenamefont {Aspuru-Guzik}(2025)}]{schleich2025cracking}%
  \BibitemOpen
  \bibfield  {author} {\bibinfo {author} {\bibfnamefont {P.}~\bibnamefont {Schleich}}\ and\ \bibinfo {author} {\bibfnamefont {A.}~\bibnamefont {Aspuru-Guzik}},\ }\bibfield  {title} {\bibinfo {title} {Cracking chemistry with quantum simulations},\ }\href {https://doi.org/10.1126/science.ado6686} {\bibfield  {journal} {\bibinfo  {journal} {Science}\ }\textbf {\bibinfo {volume} {390}},\ \bibinfo {pages} {1002} (\bibinfo {year} {2025})}\BibitemShut {NoStop}%
\bibitem [{\citenamefont {Lee}\ \emph {et~al.}(2023)\citenamefont {Lee}, \citenamefont {Lee}, \citenamefont {Zhai}, \citenamefont {Tong}, \citenamefont {Dalzell}, \citenamefont {Kumar}, \citenamefont {Helms}, \citenamefont {Gray}, \citenamefont {Cui}, \citenamefont {Liu} \emph {et~al.}}]{lee2023evaluating}%
  \BibitemOpen
  \bibfield  {author} {\bibinfo {author} {\bibfnamefont {S.}~\bibnamefont {Lee}}, \bibinfo {author} {\bibfnamefont {J.}~\bibnamefont {Lee}}, \bibinfo {author} {\bibfnamefont {H.}~\bibnamefont {Zhai}}, \bibinfo {author} {\bibfnamefont {Y.}~\bibnamefont {Tong}}, \bibinfo {author} {\bibfnamefont {A.~M.}\ \bibnamefont {Dalzell}}, \bibinfo {author} {\bibfnamefont {A.}~\bibnamefont {Kumar}}, \bibinfo {author} {\bibfnamefont {P.}~\bibnamefont {Helms}}, \bibinfo {author} {\bibfnamefont {J.}~\bibnamefont {Gray}}, \bibinfo {author} {\bibfnamefont {Z.-H.}\ \bibnamefont {Cui}}, \bibinfo {author} {\bibfnamefont {W.}~\bibnamefont {Liu}}, \emph {et~al.},\ }\bibfield  {title} {\bibinfo {title} {Evaluating the evidence for exponential quantum advantage in ground-state quantum chemistry},\ }\href {https://doi.org/10.1038/s41467-023-37587-6} {\bibfield  {journal} {\bibinfo  {journal} {Nat. Commun.}\ }\textbf {\bibinfo {volume} {14}},\ \bibinfo {pages} {1952} (\bibinfo {year} {2023})}\BibitemShut {NoStop}%
\bibitem [{\citenamefont {Bauer}\ \emph {et~al.}(2020)\citenamefont {Bauer}, \citenamefont {Bravyi}, \citenamefont {Motta},\ and\ \citenamefont {Chan}}]{bauer2020quantum}%
  \BibitemOpen
  \bibfield  {author} {\bibinfo {author} {\bibfnamefont {B.}~\bibnamefont {Bauer}}, \bibinfo {author} {\bibfnamefont {S.}~\bibnamefont {Bravyi}}, \bibinfo {author} {\bibfnamefont {M.}~\bibnamefont {Motta}},\ and\ \bibinfo {author} {\bibfnamefont {G.~K.-L.}\ \bibnamefont {Chan}},\ }\bibfield  {title} {\bibinfo {title} {Quantum algorithms for quantum chemistry and quantum materials science},\ }\href {https://doi.org/10.1021/acs.chemrev.9b00829} {\bibfield  {journal} {\bibinfo  {journal} {Chem. Rev.}\ }\textbf {\bibinfo {volume} {120}},\ \bibinfo {pages} {12685} (\bibinfo {year} {2020})}\BibitemShut {NoStop}%
\bibitem [{\citenamefont {Chen}\ and\ \citenamefont {Chan}()}]{chen2025framework}%
  \BibitemOpen
  \bibfield  {author} {\bibinfo {author} {\bibfnamefont {J.}~\bibnamefont {Chen}}\ and\ \bibinfo {author} {\bibfnamefont {G.~K.-L.}\ \bibnamefont {Chan}},\ }\bibfield  {title} {\bibinfo {title} {A framework for robust quantum speedups in practical correlated electronic structure and dynamics},\ }\Eprint {https://arxiv.org/abs/2508.15765} {arXiv:2508.15765} \BibitemShut {NoStop}%
\bibitem [{\citenamefont {Schr{\"o}dinger}(1935)}]{schrodinger1935discussion}%
  \BibitemOpen
  \bibfield  {author} {\bibinfo {author} {\bibfnamefont {E.}~\bibnamefont {Schr{\"o}dinger}},\ }\bibfield  {title} {\bibinfo {title} {Discussion of probability relations between separated systems},\ }in\ \href@noop {} {\emph {\bibinfo {booktitle} {Mathematical Proceedings of the Cambridge Philosophical Society}}},\ Vol.~\bibinfo {volume} {31}\ (\bibinfo {organization} {Cambridge University Press},\ \bibinfo {year} {1935})\ pp.\ \bibinfo {pages} {555--563}\BibitemShut {NoStop}%
\bibitem [{\citenamefont {Nielsen}\ and\ \citenamefont {Chuang}(2010)}]{MikeNIke}%
  \BibitemOpen
  \bibfield  {author} {\bibinfo {author} {\bibfnamefont {M.~A.}\ \bibnamefont {Nielsen}}\ and\ \bibinfo {author} {\bibfnamefont {I.~L.}\ \bibnamefont {Chuang}},\ }\href@noop {} {\emph {\bibinfo {title} {Quantum Computation and Quantum Information}}}\ (\bibinfo  {publisher} {Cambridge University Press},\ \bibinfo {year} {2010})\BibitemShut {NoStop}%
\bibitem [{\citenamefont {Bravyi}\ and\ \citenamefont {Gosset}(2016)}]{bravyi2016improved}%
  \BibitemOpen
  \bibfield  {author} {\bibinfo {author} {\bibfnamefont {S.}~\bibnamefont {Bravyi}}\ and\ \bibinfo {author} {\bibfnamefont {D.}~\bibnamefont {Gosset}},\ }\bibfield  {title} {\bibinfo {title} {Improved classical simulation of quantum circuits dominated by {C}lifford gates},\ }\href {https://doi.org/10.1103/PhysRevLett.116.250501} {\bibfield  {journal} {\bibinfo  {journal} {Phys. Rev. Lett.}\ }\textbf {\bibinfo {volume} {116}},\ \bibinfo {pages} {250501} (\bibinfo {year} {2016})}\BibitemShut {NoStop}%
\bibitem [{\citenamefont {Bravyi}\ \emph {et~al.}(2019)\citenamefont {Bravyi}, \citenamefont {Browne}, \citenamefont {Calpin}, \citenamefont {Campbell}, \citenamefont {Gosset},\ and\ \citenamefont {Howard}}]{bravyi2019simulation}%
  \BibitemOpen
  \bibfield  {author} {\bibinfo {author} {\bibfnamefont {S.}~\bibnamefont {Bravyi}}, \bibinfo {author} {\bibfnamefont {D.}~\bibnamefont {Browne}}, \bibinfo {author} {\bibfnamefont {P.}~\bibnamefont {Calpin}}, \bibinfo {author} {\bibfnamefont {E.}~\bibnamefont {Campbell}}, \bibinfo {author} {\bibfnamefont {D.}~\bibnamefont {Gosset}},\ and\ \bibinfo {author} {\bibfnamefont {M.}~\bibnamefont {Howard}},\ }\bibfield  {title} {\bibinfo {title} {Simulation of quantum circuits by low-rank stabilizer decompositions},\ }\href {https://doi.org/10.22331/q-2019-09-02-181} {\bibfield  {journal} {\bibinfo  {journal} {Quantum}\ }\textbf {\bibinfo {volume} {3}},\ \bibinfo {pages} {181} (\bibinfo {year} {2019})}\BibitemShut {NoStop}%
\bibitem [{\citenamefont {Howard}\ \emph {et~al.}(2014)\citenamefont {Howard}, \citenamefont {Wallman}, \citenamefont {Veitch},\ and\ \citenamefont {Emerson}}]{Howard2014}%
  \BibitemOpen
  \bibfield  {author} {\bibinfo {author} {\bibfnamefont {M.}~\bibnamefont {Howard}}, \bibinfo {author} {\bibfnamefont {J.}~\bibnamefont {Wallman}}, \bibinfo {author} {\bibfnamefont {V.}~\bibnamefont {Veitch}},\ and\ \bibinfo {author} {\bibfnamefont {J.}~\bibnamefont {Emerson}},\ }\bibfield  {title} {\bibinfo {title} {{Contextuality supplies the `magic' for quantum computation}},\ }\href {https://doi.org/10.1038/nature13460} {\bibfield  {journal} {\bibinfo  {journal} {Nature}\ }\textbf {\bibinfo {volume} {510}},\ \bibinfo {pages} {351} (\bibinfo {year} {2014})}\BibitemShut {NoStop}%
\bibitem [{\citenamefont {Bermejo-Vega}\ \emph {et~al.}(2017)\citenamefont {Bermejo-Vega}, \citenamefont {Delfosse}, \citenamefont {Browne}, \citenamefont {Okay},\ and\ \citenamefont {Raussendorf}}]{BermejoVega2017}%
  \BibitemOpen
  \bibfield  {author} {\bibinfo {author} {\bibfnamefont {J.}~\bibnamefont {Bermejo-Vega}}, \bibinfo {author} {\bibfnamefont {N.}~\bibnamefont {Delfosse}}, \bibinfo {author} {\bibfnamefont {D.~E.}\ \bibnamefont {Browne}}, \bibinfo {author} {\bibfnamefont {C.}~\bibnamefont {Okay}},\ and\ \bibinfo {author} {\bibfnamefont {R.}~\bibnamefont {Raussendorf}},\ }\bibfield  {title} {\bibinfo {title} {{Contextuality as a resource for models of quantum computation with qubits}},\ }\href {https://doi.org/10.1103/physrevlett.119.120505} {\bibfield  {journal} {\bibinfo  {journal} {Phys. Rev. Lett.}\ }\textbf {\bibinfo {volume} {119}},\ \bibinfo {pages} {120505} (\bibinfo {year} {2017})}\BibitemShut {NoStop}%
\bibitem [{\citenamefont {Spekkens}(2008)}]{spekkens2008negativity}%
  \BibitemOpen
  \bibfield  {author} {\bibinfo {author} {\bibfnamefont {R.~W.}\ \bibnamefont {Spekkens}},\ }\bibfield  {title} {\bibinfo {title} {Negativity and contextuality are equivalent notions of nonclassicality},\ }\href {https://doi.org/10.1103/physrevlett.101.020401} {\bibfield  {journal} {\bibinfo  {journal} {Phys. Rev. Lett.}\ }\textbf {\bibinfo {volume} {101}},\ \bibinfo {pages} {020401} (\bibinfo {year} {2008})}\BibitemShut {NoStop}%
\bibitem [{\citenamefont {Kocia}\ and\ \citenamefont {Love}(2017)}]{kocia2017discrete}%
  \BibitemOpen
  \bibfield  {author} {\bibinfo {author} {\bibfnamefont {L.}~\bibnamefont {Kocia}}\ and\ \bibinfo {author} {\bibfnamefont {P.}~\bibnamefont {Love}},\ }\bibfield  {title} {\bibinfo {title} {Discrete {W}igner formalism for qubits and noncontextuality of {C}lifford gates on qubit stabilizer states},\ }\href {https://doi.org/10.1103/PhysRevA.96.062134} {\bibfield  {journal} {\bibinfo  {journal} {Phys. Rev. A}\ }\textbf {\bibinfo {volume} {96}},\ \bibinfo {pages} {062134} (\bibinfo {year} {2017})}\BibitemShut {NoStop}%
\bibitem [{\citenamefont {Veitch}\ \emph {et~al.}(2014)\citenamefont {Veitch}, \citenamefont {Mousavian}, \citenamefont {Gottesman},\ and\ \citenamefont {Emerson}}]{Veitch2014}%
  \BibitemOpen
  \bibfield  {author} {\bibinfo {author} {\bibfnamefont {V.}~\bibnamefont {Veitch}}, \bibinfo {author} {\bibfnamefont {S.~A.~H.}\ \bibnamefont {Mousavian}}, \bibinfo {author} {\bibfnamefont {D.}~\bibnamefont {Gottesman}},\ and\ \bibinfo {author} {\bibfnamefont {J.}~\bibnamefont {Emerson}},\ }\bibfield  {title} {\bibinfo {title} {{The resource theory of stabilizer quantum computation}},\ }\href {https://doi.org/10.1088/1367-2630/16/1/013009} {\bibfield  {journal} {\bibinfo  {journal} {New J. Phys.}\ }\textbf {\bibinfo {volume} {16}},\ \bibinfo {pages} {013009} (\bibinfo {year} {2014})}\BibitemShut {NoStop}%
\bibitem [{\citenamefont {Leone}\ \emph {et~al.}(2022)\citenamefont {Leone}, \citenamefont {Oliviero},\ and\ \citenamefont {Hamma}}]{LeoneSREPhysRevLett.128.050402}%
  \BibitemOpen
  \bibfield  {author} {\bibinfo {author} {\bibfnamefont {L.}~\bibnamefont {Leone}}, \bibinfo {author} {\bibfnamefont {S.~F.~E.}\ \bibnamefont {Oliviero}},\ and\ \bibinfo {author} {\bibfnamefont {A.}~\bibnamefont {Hamma}},\ }\bibfield  {title} {\bibinfo {title} {Stabilizer {R}\'enyi entropy},\ }\href {https://doi.org/10.1103/PhysRevLett.128.050402} {\bibfield  {journal} {\bibinfo  {journal} {Phys. Rev. Lett.}\ }\textbf {\bibinfo {volume} {128}},\ \bibinfo {pages} {050402} (\bibinfo {year} {2022})}\BibitemShut {NoStop}%
\bibitem [{\citenamefont {Leone}\ and\ \citenamefont {Bittel}(2024)}]{LeoneMagicMonotonePhysRevA.110.L040403}%
  \BibitemOpen
  \bibfield  {author} {\bibinfo {author} {\bibfnamefont {L.}~\bibnamefont {Leone}}\ and\ \bibinfo {author} {\bibfnamefont {L.}~\bibnamefont {Bittel}},\ }\bibfield  {title} {\bibinfo {title} {Stabilizer entropies are monotones for magic-state resource theory},\ }\href {https://doi.org/10.1103/PhysRevA.110.L040403} {\bibfield  {journal} {\bibinfo  {journal} {Phys. Rev. A}\ }\textbf {\bibinfo {volume} {110}},\ \bibinfo {pages} {L040403} (\bibinfo {year} {2024})}\BibitemShut {NoStop}%
\bibitem [{\citenamefont {Bittel}\ and\ \citenamefont {Leone}(2026)}]{bittel2025operational}%
  \BibitemOpen
  \bibfield  {author} {\bibinfo {author} {\bibfnamefont {L.}~\bibnamefont {Bittel}}\ and\ \bibinfo {author} {\bibfnamefont {L.}~\bibnamefont {Leone}},\ }\bibfield  {title} {\bibinfo {title} {Operational interpretation of the stabilizer entropy},\ }\href {https://doi.org/10.22331/q-2026-04-15-2069} {\bibfield  {journal} {\bibinfo  {journal} {Quantum}\ }\textbf {\bibinfo {volume} {10}},\ \bibinfo {pages} {2069} (\bibinfo {year} {2026})}\BibitemShut {NoStop}%
\bibitem [{\citenamefont {Haug}\ \emph {et~al.}(2025)\citenamefont {Haug}, \citenamefont {Aolita},\ and\ \citenamefont {Kim}}]{haug2025probing-d19}%
  \BibitemOpen
  \bibfield  {author} {\bibinfo {author} {\bibfnamefont {T.}~\bibnamefont {Haug}}, \bibinfo {author} {\bibfnamefont {L.}~\bibnamefont {Aolita}},\ and\ \bibinfo {author} {\bibfnamefont {M.}~\bibnamefont {Kim}},\ }\bibfield  {title} {\bibinfo {title} {Probing quantum complexity via universal saturation of stabilizer entropies},\ }\href {https://doi.org/10.22331/q-2025-07-21-1801} {\bibfield  {journal} {\bibinfo  {journal} {Quantum}\ }\textbf {\bibinfo {volume} {9}},\ \bibinfo {pages} {1801} (\bibinfo {year} {2025})}\BibitemShut {NoStop}%
\bibitem [{\citenamefont {Oliviero}\ \emph {et~al.}(2022{\natexlab{a}})\citenamefont {Oliviero}, \citenamefont {Leone}, \citenamefont {Hamma},\ and\ \citenamefont {Lloyd}}]{Oliviero2022measuringmagic}%
  \BibitemOpen
  \bibfield  {author} {\bibinfo {author} {\bibfnamefont {S.~F.~E.}\ \bibnamefont {Oliviero}}, \bibinfo {author} {\bibfnamefont {L.}~\bibnamefont {Leone}}, \bibinfo {author} {\bibfnamefont {A.}~\bibnamefont {Hamma}},\ and\ \bibinfo {author} {\bibfnamefont {S.}~\bibnamefont {Lloyd}},\ }\bibfield  {title} {\bibinfo {title} {Measuring magic on a quantum processor},\ }\href {https://doi.org/10.1038/s41534-022-00666-5} {\bibfield  {journal} {\bibinfo  {journal} {npj Quantum Inf.}\ }\textbf {\bibinfo {volume} {8}},\ \bibinfo {pages} {148} (\bibinfo {year} {2022}{\natexlab{a}})}\BibitemShut {NoStop}%
\bibitem [{\citenamefont {Beverland}\ \emph {et~al.}(2020)\citenamefont {Beverland}, \citenamefont {Campbell}, \citenamefont {Howard},\ and\ \citenamefont {Kliuchnikov}}]{beverland2020lower}%
  \BibitemOpen
  \bibfield  {author} {\bibinfo {author} {\bibfnamefont {M.}~\bibnamefont {Beverland}}, \bibinfo {author} {\bibfnamefont {E.}~\bibnamefont {Campbell}}, \bibinfo {author} {\bibfnamefont {M.}~\bibnamefont {Howard}},\ and\ \bibinfo {author} {\bibfnamefont {V.}~\bibnamefont {Kliuchnikov}},\ }\bibfield  {title} {\bibinfo {title} {Lower bounds on the non-{C}lifford resources for quantum computations},\ }\href {https://doi.org/10.1088/2058-9565/ab8963} {\bibfield  {journal} {\bibinfo  {journal} {Quantum Sci. Technol.}\ }\textbf {\bibinfo {volume} {5}},\ \bibinfo {pages} {035009} (\bibinfo {year} {2020})}\BibitemShut {NoStop}%
\bibitem [{\citenamefont {White}\ \emph {et~al.}(2021)\citenamefont {White}, \citenamefont {Cao},\ and\ \citenamefont {Swingle}}]{PhysRevB.103.075145}%
  \BibitemOpen
  \bibfield  {author} {\bibinfo {author} {\bibfnamefont {C.~D.}\ \bibnamefont {White}}, \bibinfo {author} {\bibfnamefont {C.}~\bibnamefont {Cao}},\ and\ \bibinfo {author} {\bibfnamefont {B.}~\bibnamefont {Swingle}},\ }\bibfield  {title} {\bibinfo {title} {Conformal field theories are magical},\ }\href {https://doi.org/10.1103/PhysRevB.103.075145} {\bibfield  {journal} {\bibinfo  {journal} {Phys. Rev. B}\ }\textbf {\bibinfo {volume} {103}},\ \bibinfo {pages} {075145} (\bibinfo {year} {2021})}\BibitemShut {NoStop}%
\bibitem [{\citenamefont {Robin}\ and\ \citenamefont {Savage}(2025)}]{Robin2024}%
  \BibitemOpen
  \bibfield  {author} {\bibinfo {author} {\bibfnamefont {C.~E.~P.}\ \bibnamefont {Robin}}\ and\ \bibinfo {author} {\bibfnamefont {M.~J.}\ \bibnamefont {Savage}},\ }\bibfield  {title} {\bibinfo {title} {Quantum complexity fluctuations from nuclear and hypernuclear forces},\ }\href {https://doi.org/10.1103/r8rq-y9tb} {\bibfield  {journal} {\bibinfo  {journal} {Phys. Rev. C}\ }\textbf {\bibinfo {volume} {112}},\ \bibinfo {pages} {044004} (\bibinfo {year} {2025})}\BibitemShut {NoStop}%
\bibitem [{\citenamefont {Chernyshev}\ \emph {et~al.}(2025)\citenamefont {Chernyshev}, \citenamefont {Robin},\ and\ \citenamefont {Savage}}]{Chernyshev2025}%
  \BibitemOpen
  \bibfield  {author} {\bibinfo {author} {\bibfnamefont {I.}~\bibnamefont {Chernyshev}}, \bibinfo {author} {\bibfnamefont {C.~E.~P.}\ \bibnamefont {Robin}},\ and\ \bibinfo {author} {\bibfnamefont {M.~J.}\ \bibnamefont {Savage}},\ }\bibfield  {title} {\bibinfo {title} {Quantum magic and computational complexity in the neutrino sector},\ }\href {https://doi.org/10.1103/PhysRevResearch.7.023228} {\bibfield  {journal} {\bibinfo  {journal} {Phys. Rev. Research}\ }\textbf {\bibinfo {volume} {7}},\ \bibinfo {pages} {023228} (\bibinfo {year} {2025})}\BibitemShut {NoStop}%
\bibitem [{\citenamefont {Falc\~ao}\ \emph {et~al.}(2025)\citenamefont {Falc\~ao}, \citenamefont {Tarabunga}, \citenamefont {Frau}, \citenamefont {Tirrito}, \citenamefont {Zakrzewski},\ and\ \citenamefont {Dalmonte}}]{PhysRevB.111.L081102}%
  \BibitemOpen
  \bibfield  {author} {\bibinfo {author} {\bibfnamefont {P.~R.~N.}\ \bibnamefont {Falc\~ao}}, \bibinfo {author} {\bibfnamefont {P.~S.}\ \bibnamefont {Tarabunga}}, \bibinfo {author} {\bibfnamefont {M.}~\bibnamefont {Frau}}, \bibinfo {author} {\bibfnamefont {E.}~\bibnamefont {Tirrito}}, \bibinfo {author} {\bibfnamefont {J.}~\bibnamefont {Zakrzewski}},\ and\ \bibinfo {author} {\bibfnamefont {M.}~\bibnamefont {Dalmonte}},\ }\bibfield  {title} {\bibinfo {title} {Nonstabilizerness in {U(1)} lattice gauge theory},\ }\href {https://doi.org/10.1103/PhysRevB.111.L081102} {\bibfield  {journal} {\bibinfo  {journal} {Phys. Rev. B}\ }\textbf {\bibinfo {volume} {111}},\ \bibinfo {pages} {L081102} (\bibinfo {year} {2025})}\BibitemShut {NoStop}%
\bibitem [{\citenamefont {Cao}\ \emph {et~al.}(2025)\citenamefont {Cao}, \citenamefont {Cheng}, \citenamefont {Hamma}, \citenamefont {Leone}, \citenamefont {Munizzi},\ and\ \citenamefont {Oliviero}}]{Cao2024}%
  \BibitemOpen
  \bibfield  {author} {\bibinfo {author} {\bibfnamefont {C.}~\bibnamefont {Cao}}, \bibinfo {author} {\bibfnamefont {G.}~\bibnamefont {Cheng}}, \bibinfo {author} {\bibfnamefont {A.}~\bibnamefont {Hamma}}, \bibinfo {author} {\bibfnamefont {L.}~\bibnamefont {Leone}}, \bibinfo {author} {\bibfnamefont {W.}~\bibnamefont {Munizzi}},\ and\ \bibinfo {author} {\bibfnamefont {S.~F.}\ \bibnamefont {Oliviero}},\ }\bibfield  {title} {\bibinfo {title} {Gravitational backreaction is magical},\ }\href {https://doi.org/10.1103/z3vr-w5c5} {\bibfield  {journal} {\bibinfo  {journal} {PRX Quantum}\ }\textbf {\bibinfo {volume} {6}},\ \bibinfo {pages} {040375} (\bibinfo {year} {2025})}\BibitemShut {NoStop}%
\bibitem [{\citenamefont {White}\ and\ \citenamefont {White}(2024)}]{White2024}%
  \BibitemOpen
  \bibfield  {author} {\bibinfo {author} {\bibfnamefont {C.~D.}\ \bibnamefont {White}}\ and\ \bibinfo {author} {\bibfnamefont {M.~J.}\ \bibnamefont {White}},\ }\bibfield  {title} {\bibinfo {title} {Magic states of top quarks},\ }\href {https://doi.org/10.1103/PhysRevD.110.116016} {\bibfield  {journal} {\bibinfo  {journal} {Phys. Rev. D}\ }\textbf {\bibinfo {volume} {110}},\ \bibinfo {pages} {116016} (\bibinfo {year} {2024})}\BibitemShut {NoStop}%
\bibitem [{\citenamefont {Gargalionis}\ \emph {et~al.}()\citenamefont {Gargalionis}, \citenamefont {Moynihan}, \citenamefont {Trifinopoulos}, \citenamefont {Wallace}, \citenamefont {White},\ and\ \citenamefont {White}}]{Gargalionis2025}%
  \BibitemOpen
  \bibfield  {author} {\bibinfo {author} {\bibfnamefont {J.}~\bibnamefont {Gargalionis}}, \bibinfo {author} {\bibfnamefont {N.}~\bibnamefont {Moynihan}}, \bibinfo {author} {\bibfnamefont {S.}~\bibnamefont {Trifinopoulos}}, \bibinfo {author} {\bibfnamefont {E.~N.~V.}\ \bibnamefont {Wallace}}, \bibinfo {author} {\bibfnamefont {C.~D.}\ \bibnamefont {White}},\ and\ \bibinfo {author} {\bibfnamefont {M.~J.}\ \bibnamefont {White}},\ }\bibfield  {title} {\bibinfo {title} {{Spin versus Magic: Lessons from Gluon and Graviton Scattering}},\ }\Eprint {https://arxiv.org/abs/2508.14967} {arXiv:2508.14967} \BibitemShut {NoStop}%
\bibitem [{\citenamefont {Aoude}\ \emph {et~al.}()\citenamefont {Aoude}, \citenamefont {Banks}, \citenamefont {White},\ and\ \citenamefont {White}}]{Aoude2025}%
  \BibitemOpen
  \bibfield  {author} {\bibinfo {author} {\bibfnamefont {R.}~\bibnamefont {Aoude}}, \bibinfo {author} {\bibfnamefont {H.}~\bibnamefont {Banks}}, \bibinfo {author} {\bibfnamefont {C.~D.}\ \bibnamefont {White}},\ and\ \bibinfo {author} {\bibfnamefont {M.~J.}\ \bibnamefont {White}},\ }\bibfield  {title} {\bibinfo {title} {{Probing new physics in the top sector using quantum information}},\ }\Eprint {https://arxiv.org/abs/2505.12522} {arXiv:2505.12522} \BibitemShut {NoStop}%
\bibitem [{\citenamefont {Tarabunga}\ \emph {et~al.}(2023)\citenamefont {Tarabunga}, \citenamefont {Tirrito}, \citenamefont {Chanda},\ and\ \citenamefont {Dalmonte}}]{tarabunga2023many}%
  \BibitemOpen
  \bibfield  {author} {\bibinfo {author} {\bibfnamefont {P.~S.}\ \bibnamefont {Tarabunga}}, \bibinfo {author} {\bibfnamefont {E.}~\bibnamefont {Tirrito}}, \bibinfo {author} {\bibfnamefont {T.}~\bibnamefont {Chanda}},\ and\ \bibinfo {author} {\bibfnamefont {M.}~\bibnamefont {Dalmonte}},\ }\bibfield  {title} {\bibinfo {title} {Many-body magic via {P}auli-{M}arkov chains--from criticality to gauge theories},\ }\href {https://doi.org/10.1103/PRXQuantum.4.040317} {\bibfield  {journal} {\bibinfo  {journal} {PRX Quantum}\ }\textbf {\bibinfo {volume} {4}},\ \bibinfo {pages} {040317} (\bibinfo {year} {2023})}\BibitemShut {NoStop}%
\bibitem [{\citenamefont {Oliviero}\ \emph {et~al.}(2022{\natexlab{b}})\citenamefont {Oliviero}, \citenamefont {Leone},\ and\ \citenamefont {Hamma}}]{Oliviero2022}%
  \BibitemOpen
  \bibfield  {author} {\bibinfo {author} {\bibfnamefont {S.~F.}\ \bibnamefont {Oliviero}}, \bibinfo {author} {\bibfnamefont {L.}~\bibnamefont {Leone}},\ and\ \bibinfo {author} {\bibfnamefont {A.}~\bibnamefont {Hamma}},\ }\bibfield  {title} {\bibinfo {title} {Magic-state resource theory for the ground state of the transverse-field {I}sing model},\ }\href {https://doi.org/10.1103/PhysRevA.106.042426} {\bibfield  {journal} {\bibinfo  {journal} {Phys. Rev. A}\ }\textbf {\bibinfo {volume} {106}},\ \bibinfo {pages} {042426} (\bibinfo {year} {2022}{\natexlab{b}})}\BibitemShut {NoStop}%
\bibitem [{\citenamefont {Turkeshi}\ \emph {et~al.}(2025)\citenamefont {Turkeshi}, \citenamefont {Dymarsky},\ and\ \citenamefont {Sierant}}]{turkeshi2023pauli-14f}%
  \BibitemOpen
  \bibfield  {author} {\bibinfo {author} {\bibfnamefont {X.}~\bibnamefont {Turkeshi}}, \bibinfo {author} {\bibfnamefont {A.}~\bibnamefont {Dymarsky}},\ and\ \bibinfo {author} {\bibfnamefont {P.}~\bibnamefont {Sierant}},\ }\bibfield  {title} {\bibinfo {title} {Pauli spectrum and nonstabilizerness of typical quantum many-body states},\ }\href {https://doi.org/10.1103/PhysRevB.111.054301} {\bibfield  {journal} {\bibinfo  {journal} {Phys. Rev. B}\ }\textbf {\bibinfo {volume} {111}},\ \bibinfo {pages} {054301} (\bibinfo {year} {2025})}\BibitemShut {NoStop}%
\bibitem [{\citenamefont {{CMS Collaboration}}(2025)}]{CERNMagic}%
  \BibitemOpen
  \bibfield  {author} {\bibinfo {author} {\bibnamefont {{CMS Collaboration}}},\ }\href {https://cds.cern.ch/record/2926751} {\emph {\bibinfo {title} {{Observation of magic states of top quark pairs produced in proton-proton collisions at $\sqrt{s}=13~\mathrm{TeV}$}}}},\ \bibinfo {type} {CMS Physics Analysis Summary}\ \bibinfo {number} {CMS-PAS-TOP-25-001}\ (\bibinfo  {institution} {CERN},\ \bibinfo {year} {2025})\BibitemShut {NoStop}%
\bibitem [{\citenamefont {Xu}\ \emph {et~al.}(2023)\citenamefont {Xu}, \citenamefont {Soriano-Agueda}, \citenamefont {L{\'o}pez}, \citenamefont {Ramos-Cordoba},\ and\ \citenamefont {Matito}}]{xu2023all}%
  \BibitemOpen
  \bibfield  {author} {\bibinfo {author} {\bibfnamefont {X.}~\bibnamefont {Xu}}, \bibinfo {author} {\bibfnamefont {L.}~\bibnamefont {Soriano-Agueda}}, \bibinfo {author} {\bibfnamefont {X.}~\bibnamefont {L{\'o}pez}}, \bibinfo {author} {\bibfnamefont {E.}~\bibnamefont {Ramos-Cordoba}},\ and\ \bibinfo {author} {\bibfnamefont {E.}~\bibnamefont {Matito}},\ }\bibfield  {title} {\bibinfo {title} {All-purpose measure of electron correlation for multireference diagnostics},\ }\href {https://doi.org/10.1021/acs.jctc.3c01073} {\bibfield  {journal} {\bibinfo  {journal} {J. Chem. Theory Comput.}\ }\textbf {\bibinfo {volume} {20}},\ \bibinfo {pages} {721} (\bibinfo {year} {2023})}\BibitemShut {NoStop}%
\bibitem [{\citenamefont {Lee}\ and\ \citenamefont {Taylor}(1989)}]{lee1989diagnostic-bca}%
  \BibitemOpen
  \bibfield  {author} {\bibinfo {author} {\bibfnamefont {T.~J.}\ \bibnamefont {Lee}}\ and\ \bibinfo {author} {\bibfnamefont {P.~R.}\ \bibnamefont {Taylor}},\ }\bibfield  {title} {\bibinfo {title} {A diagnostic for determining the quality of single-reference electron correlation methods},\ }\href {https://doi.org/10.1002/qua.560360824} {\bibfield  {journal} {\bibinfo  {journal} {Int. J. Quantum Chem.}\ }\textbf {\bibinfo {volume} {36}},\ \bibinfo {pages} {199} (\bibinfo {year} {1989})}\BibitemShut {NoStop}%
\bibitem [{\citenamefont {Izs{\'a}k}\ \emph {et~al.}(2023)\citenamefont {Izs{\'a}k}, \citenamefont {Ivanov}, \citenamefont {Blunt}, \citenamefont {Holzmann},\ and\ \citenamefont {Neese}}]{izsak2023measuring}%
  \BibitemOpen
  \bibfield  {author} {\bibinfo {author} {\bibfnamefont {R.}~\bibnamefont {Izs{\'a}k}}, \bibinfo {author} {\bibfnamefont {A.~V.}\ \bibnamefont {Ivanov}}, \bibinfo {author} {\bibfnamefont {N.~S.}\ \bibnamefont {Blunt}}, \bibinfo {author} {\bibfnamefont {N.}~\bibnamefont {Holzmann}},\ and\ \bibinfo {author} {\bibfnamefont {F.}~\bibnamefont {Neese}},\ }\bibfield  {title} {\bibinfo {title} {Measuring electron correlation: The impact of symmetry and orbital transformations},\ }\href {https://doi.org/10.1021/acs.jctc.3c00122} {\bibfield  {journal} {\bibinfo  {journal} {J. Chem. Theory Comput.}\ }\textbf {\bibinfo {volume} {19}},\ \bibinfo {pages} {2703} (\bibinfo {year} {2023})}\BibitemShut {NoStop}%
\bibitem [{\citenamefont {Kirby}\ \emph {et~al.}(2021)\citenamefont {Kirby}, \citenamefont {Tranter},\ and\ \citenamefont {Love}}]{kirby2021contextual}%
  \BibitemOpen
  \bibfield  {author} {\bibinfo {author} {\bibfnamefont {W.~M.}\ \bibnamefont {Kirby}}, \bibinfo {author} {\bibfnamefont {A.}~\bibnamefont {Tranter}},\ and\ \bibinfo {author} {\bibfnamefont {P.~J.}\ \bibnamefont {Love}},\ }\bibfield  {title} {\bibinfo {title} {Contextual subspace variational quantum eigensolver},\ }\href {https://doi.org/10.22331/q-2021-05-14-456} {\bibfield  {journal} {\bibinfo  {journal} {Quantum}\ }\textbf {\bibinfo {volume} {5}},\ \bibinfo {pages} {456} (\bibinfo {year} {2021})}\BibitemShut {NoStop}%
\bibitem [{\citenamefont {Ralli}\ \emph {et~al.}(2023)\citenamefont {Ralli}, \citenamefont {Weaving}, \citenamefont {Tranter}, \citenamefont {Kirby}, \citenamefont {Love},\ and\ \citenamefont {Coveney}}]{Ralli_UP_CS-VQE}%
  \BibitemOpen
  \bibfield  {author} {\bibinfo {author} {\bibfnamefont {A.}~\bibnamefont {Ralli}}, \bibinfo {author} {\bibfnamefont {T.}~\bibnamefont {Weaving}}, \bibinfo {author} {\bibfnamefont {A.}~\bibnamefont {Tranter}}, \bibinfo {author} {\bibfnamefont {W.~M.}\ \bibnamefont {Kirby}}, \bibinfo {author} {\bibfnamefont {P.~J.}\ \bibnamefont {Love}},\ and\ \bibinfo {author} {\bibfnamefont {P.~V.}\ \bibnamefont {Coveney}},\ }\bibfield  {title} {\bibinfo {title} {Unitary partitioning and the contextual subspace variational quantum eigensolver},\ }\href {https://doi.org/10.1103/PhysRevResearch.5.013095} {\bibfield  {journal} {\bibinfo  {journal} {Phys. Rev. Research}\ }\textbf {\bibinfo {volume} {5}},\ \bibinfo {pages} {013095} (\bibinfo {year} {2023})}\BibitemShut {NoStop}%
\bibitem [{\citenamefont {Weaving}\ \emph {et~al.}(2023{\natexlab{a}})\citenamefont {Weaving}, \citenamefont {Ralli}, \citenamefont {Kirby}, \citenamefont {Tranter}, \citenamefont {Love},\ and\ \citenamefont {Coveney}}]{weaving2023stabilizer}%
  \BibitemOpen
  \bibfield  {author} {\bibinfo {author} {\bibfnamefont {T.}~\bibnamefont {Weaving}}, \bibinfo {author} {\bibfnamefont {A.}~\bibnamefont {Ralli}}, \bibinfo {author} {\bibfnamefont {W.~M.}\ \bibnamefont {Kirby}}, \bibinfo {author} {\bibfnamefont {A.}~\bibnamefont {Tranter}}, \bibinfo {author} {\bibfnamefont {P.~J.}\ \bibnamefont {Love}},\ and\ \bibinfo {author} {\bibfnamefont {P.~V.}\ \bibnamefont {Coveney}},\ }\bibfield  {title} {\bibinfo {title} {A stabilizer framework for the contextual subspace variational quantum eigensolver and the noncontextual projection ansatz},\ }\href {https://doi.org/10.1021/acs.jctc.2c00910} {\bibfield  {journal} {\bibinfo  {journal} {J. Chem. Theory Comput.}\ }\textbf {\bibinfo {volume} {19}},\ \bibinfo {pages} {808} (\bibinfo {year} {2023}{\natexlab{a}})}\BibitemShut {NoStop}%
\bibitem [{\citenamefont {Coulson}\ and\ \citenamefont {Fischer}(1949)}]{coulson1949xxxiv}%
  \BibitemOpen
  \bibfield  {author} {\bibinfo {author} {\bibfnamefont {C.~A.}\ \bibnamefont {Coulson}}\ and\ \bibinfo {author} {\bibfnamefont {I.}~\bibnamefont {Fischer}},\ }\bibfield  {title} {\bibinfo {title} {{XXXIV}. {N}otes on the molecular orbital treatment of the hydrogen molecule},\ }\href {https://doi.org/10.1080/14786444908521726} {\bibfield  {journal} {\bibinfo  {journal} {Philos. Mag.}\ }\textbf {\bibinfo {volume} {40}},\ \bibinfo {pages} {386} (\bibinfo {year} {1949})}\BibitemShut {NoStop}%
\bibitem [{\citenamefont {Gao}\ \emph {et~al.}(2024)\citenamefont {Gao}, \citenamefont {Imamura}, \citenamefont {Kasagi},\ and\ \citenamefont {Yoshida}}]{gao2024distributed}%
  \BibitemOpen
  \bibfield  {author} {\bibinfo {author} {\bibfnamefont {H.}~\bibnamefont {Gao}}, \bibinfo {author} {\bibfnamefont {S.}~\bibnamefont {Imamura}}, \bibinfo {author} {\bibfnamefont {A.}~\bibnamefont {Kasagi}},\ and\ \bibinfo {author} {\bibfnamefont {E.}~\bibnamefont {Yoshida}},\ }\bibfield  {title} {\bibinfo {title} {Distributed implementation of full configuration interaction for one trillion determinants},\ }\href {https://doi.org/10.1021/acs.jctc.3c01190} {\bibfield  {journal} {\bibinfo  {journal} {J. Chem. Theory Comput.}\ }\textbf {\bibinfo {volume} {20}},\ \bibinfo {pages} {1185} (\bibinfo {year} {2024})}\BibitemShut {NoStop}%
\bibitem [{\citenamefont {Hartree}(1928)}]{hartree1928wave}%
  \BibitemOpen
  \bibfield  {author} {\bibinfo {author} {\bibfnamefont {D.~R.}\ \bibnamefont {Hartree}},\ }\bibfield  {title} {\bibinfo {title} {The wave mechanics of an atom with a non-{C}oulomb central field. {P}art {II}. {S}ome results and discussion},\ }in\ \href {https://doi.org/10.1017/S0305004100011920} {\emph {\bibinfo {booktitle} {Mathematical Proceedings of the Cambridge Philosophical Society}}},\ Vol.~\bibinfo {volume} {24}\ (\bibinfo {organization} {Cambridge University Press},\ \bibinfo {year} {1928})\ pp.\ \bibinfo {pages} {111--132}\BibitemShut {NoStop}%
\bibitem [{\citenamefont {Fock}(1930)}]{fock1930naherungsmethode}%
  \BibitemOpen
  \bibfield  {author} {\bibinfo {author} {\bibfnamefont {V.}~\bibnamefont {Fock}},\ }\bibfield  {title} {\bibinfo {title} {N{\"a}herungsmethode zur {L}{\"o}sung des quantenmechanischen {M}ehrk{\"o}rperproblems},\ }\href {https://doi.org/10.1007/BF01340294} {\bibfield  {journal} {\bibinfo  {journal} {Z. f{\"u}r Physik}\ }\textbf {\bibinfo {volume} {61}},\ \bibinfo {pages} {126} (\bibinfo {year} {1930})}\BibitemShut {NoStop}%
\bibitem [{\citenamefont {Roothaan}(1951)}]{roothaan1951new}%
  \BibitemOpen
  \bibfield  {author} {\bibinfo {author} {\bibfnamefont {C.~C.~J.}\ \bibnamefont {Roothaan}},\ }\bibfield  {title} {\bibinfo {title} {New developments in molecular orbital theory},\ }\href {https://doi.org/10.1103/RevModPhys.23.69} {\bibfield  {journal} {\bibinfo  {journal} {Rev. Mod. Phys.}\ }\textbf {\bibinfo {volume} {23}},\ \bibinfo {pages} {69} (\bibinfo {year} {1951})}\BibitemShut {NoStop}%
\bibitem [{\citenamefont {Hall}(1951)}]{hall1951molecular}%
  \BibitemOpen
  \bibfield  {author} {\bibinfo {author} {\bibfnamefont {G.~G.}\ \bibnamefont {Hall}},\ }\bibfield  {title} {\bibinfo {title} {The molecular orbital theory of chemical valency {VIII.} {A} method of calculating ionization potentials},\ }\href {https://doi.org/10.1098/rspa.1951.0048} {\bibfield  {journal} {\bibinfo  {journal} {Proc. R. Soc. Lond. A}\ }\textbf {\bibinfo {volume} {205}},\ \bibinfo {pages} {541} (\bibinfo {year} {1951})}\BibitemShut {NoStop}%
\bibitem [{\citenamefont {Limacher}\ \emph {et~al.}(2013)\citenamefont {Limacher}, \citenamefont {Ayers}, \citenamefont {Johnson}, \citenamefont {De~Baerdemacker}, \citenamefont {Van~Neck},\ and\ \citenamefont {Bultinck}}]{Limacher2013}%
  \BibitemOpen
  \bibfield  {author} {\bibinfo {author} {\bibfnamefont {P.~A.}\ \bibnamefont {Limacher}}, \bibinfo {author} {\bibfnamefont {P.~W.}\ \bibnamefont {Ayers}}, \bibinfo {author} {\bibfnamefont {P.~A.}\ \bibnamefont {Johnson}}, \bibinfo {author} {\bibfnamefont {S.}~\bibnamefont {De~Baerdemacker}}, \bibinfo {author} {\bibfnamefont {D.}~\bibnamefont {Van~Neck}},\ and\ \bibinfo {author} {\bibfnamefont {P.}~\bibnamefont {Bultinck}},\ }\bibfield  {title} {\bibinfo {title} {A new mean-field method suitable for strongly correlated electrons: Computationally facile antisymmetric products of nonorthogonal geminals},\ }\href {https://doi.org/10.1021/ct300902c} {\bibfield  {journal} {\bibinfo  {journal} {J. Chem. Theory Comput.}\ }\textbf {\bibinfo {volume} {9}},\ \bibinfo {pages} {1394} (\bibinfo {year} {2013})}\BibitemShut {NoStop}%
\bibitem [{\citenamefont {Pople}(1999)}]{pople2003quantum}%
  \BibitemOpen
  \bibfield  {author} {\bibinfo {author} {\bibfnamefont {J.~A.}\ \bibnamefont {Pople}},\ }\bibfield  {title} {\bibinfo {title} {Nobel {L}ecture: Quantum chemical models},\ }\href {https://doi.org/10.1103/RevModPhys.71.1267} {\bibfield  {journal} {\bibinfo  {journal} {Rev. Mod. Phys.}\ }\textbf {\bibinfo {volume} {71}},\ \bibinfo {pages} {1267} (\bibinfo {year} {1999})}\BibitemShut {NoStop}%
\bibitem [{\citenamefont {Kohanoff}(2006)}]{kohanoff2006electronic}%
  \BibitemOpen
  \bibfield  {author} {\bibinfo {author} {\bibfnamefont {J.}~\bibnamefont {Kohanoff}},\ }\href@noop {} {\emph {\bibinfo {title} {Electronic Structure Calculations for Solids and Molecules: Theory and Computational methods}}}\ (\bibinfo  {publisher} {Cambridge University Press},\ \bibinfo {year} {2006})\BibitemShut {NoStop}%
\bibitem [{\citenamefont {Esquivel}\ \emph {et~al.}(2015)\citenamefont {Esquivel}, \citenamefont {L{\'o}pez-Rosa},\ and\ \citenamefont {Dehesa}}]{esquivel2015correlation}%
  \BibitemOpen
  \bibfield  {author} {\bibinfo {author} {\bibfnamefont {R.}~\bibnamefont {Esquivel}}, \bibinfo {author} {\bibfnamefont {S.}~\bibnamefont {L{\'o}pez-Rosa}},\ and\ \bibinfo {author} {\bibfnamefont {J.}~\bibnamefont {Dehesa}},\ }\bibfield  {title} {\bibinfo {title} {Correlation energy as a measure of non-locality: {Q}uantum entanglement of helium-like systems},\ }\href {https://doi.org/10.1209/0295-5075/111/40009} {\bibfield  {journal} {\bibinfo  {journal} {Europhys. Lett.}\ }\textbf {\bibinfo {volume} {111}},\ \bibinfo {pages} {40009} (\bibinfo {year} {2015})}\BibitemShut {NoStop}%
\bibitem [{\citenamefont {Wang}\ and\ \citenamefont {Kais}(2007)}]{wang2007quantum-8bf}%
  \BibitemOpen
  \bibfield  {author} {\bibinfo {author} {\bibfnamefont {H.}~\bibnamefont {Wang}}\ and\ \bibinfo {author} {\bibfnamefont {S.}~\bibnamefont {Kais}},\ }\bibfield  {title} {\bibinfo {title} {Quantum entanglement and electron correlation in molecular systems},\ }\href {https://doi.org/10.1560/ijc.47.1.59} {\bibfield  {journal} {\bibinfo  {journal} {Israel J. Chem.}\ }\textbf {\bibinfo {volume} {47}},\ \bibinfo {pages} {59} (\bibinfo {year} {2007})}\BibitemShut {NoStop}%
\bibitem [{\citenamefont {Bell}(1964)}]{Belltheorem1964}%
  \BibitemOpen
  \bibfield  {author} {\bibinfo {author} {\bibfnamefont {J.~S.}\ \bibnamefont {Bell}},\ }\bibfield  {title} {\bibinfo {title} {On the {E}instein {P}odolsky {R}osen paradox},\ }\href {https://doi.org/10.1103/PhysicsPhysiqueFizika.1.195} {\bibfield  {journal} {\bibinfo  {journal} {Physics Physique Fizika}\ }\textbf {\bibinfo {volume} {1}},\ \bibinfo {pages} {195} (\bibinfo {year} {1964})}\BibitemShut {NoStop}%
\bibitem [{\citenamefont {Einstein}\ \emph {et~al.}(1935)\citenamefont {Einstein}, \citenamefont {Podolsky},\ and\ \citenamefont {Rosen}}]{EPR}%
  \BibitemOpen
  \bibfield  {author} {\bibinfo {author} {\bibfnamefont {A.}~\bibnamefont {Einstein}}, \bibinfo {author} {\bibfnamefont {B.}~\bibnamefont {Podolsky}},\ and\ \bibinfo {author} {\bibfnamefont {N.}~\bibnamefont {Rosen}},\ }\bibfield  {title} {\bibinfo {title} {Can quantum-mechanical description of physical reality be considered complete?},\ }\href {https://doi.org/10.1103/physrev.47.777} {\bibfield  {journal} {\bibinfo  {journal} {Phys. Rev.}\ }\textbf {\bibinfo {volume} {47}},\ \bibinfo {pages} {777} (\bibinfo {year} {1935})}\BibitemShut {NoStop}%
\bibitem [{\citenamefont {Bohr}(1935)}]{BohrEPR1935}%
  \BibitemOpen
  \bibfield  {author} {\bibinfo {author} {\bibfnamefont {N.}~\bibnamefont {Bohr}},\ }\bibfield  {title} {\bibinfo {title} {Can quantum-mechanical description of physical reality be considered complete?},\ }\href {https://doi.org/10.1103/PhysRev.48.696} {\bibfield  {journal} {\bibinfo  {journal} {Phys. Rev.}\ }\textbf {\bibinfo {volume} {48}},\ \bibinfo {pages} {696} (\bibinfo {year} {1935})}\BibitemShut {NoStop}%
\bibitem [{\citenamefont {Gottesman}\ and\ \citenamefont {Chuang}(1999)}]{Gottesman1999}%
  \BibitemOpen
  \bibfield  {author} {\bibinfo {author} {\bibfnamefont {D.}~\bibnamefont {Gottesman}}\ and\ \bibinfo {author} {\bibfnamefont {I.~L.}\ \bibnamefont {Chuang}},\ }\bibfield  {title} {\bibinfo {title} {{Demonstrating the viability of universal quantum computation using teleportation and single-qubit operations}},\ }\href {https://doi.org/10.1038/46503} {\bibfield  {journal} {\bibinfo  {journal} {Nature}\ }\textbf {\bibinfo {volume} {402}},\ \bibinfo {pages} {390} (\bibinfo {year} {1999})}\BibitemShut {NoStop}%
\bibitem [{\citenamefont {Pusey}\ \emph {et~al.}(2012)\citenamefont {Pusey}, \citenamefont {Barrett},\ and\ \citenamefont {Rudolph}}]{pusey2012reality}%
  \BibitemOpen
  \bibfield  {author} {\bibinfo {author} {\bibfnamefont {M.~F.}\ \bibnamefont {Pusey}}, \bibinfo {author} {\bibfnamefont {J.}~\bibnamefont {Barrett}},\ and\ \bibinfo {author} {\bibfnamefont {T.}~\bibnamefont {Rudolph}},\ }\bibfield  {title} {\bibinfo {title} {On the reality of the quantum state},\ }\href {https://doi.org/10.1038/nphys2309} {\bibfield  {journal} {\bibinfo  {journal} {Nature Phys.}\ }\textbf {\bibinfo {volume} {8}},\ \bibinfo {pages} {475} (\bibinfo {year} {2012})}\BibitemShut {NoStop}%
\bibitem [{\citenamefont {Mermin}(1993)}]{Mermin1993}%
  \BibitemOpen
  \bibfield  {author} {\bibinfo {author} {\bibfnamefont {N.~D.}\ \bibnamefont {Mermin}},\ }\bibfield  {title} {\bibinfo {title} {Hidden variables and the two theorems of {J}ohn {B}ell},\ }\href {https://doi.org/10.1103/revmodphys.65.803} {\bibfield  {journal} {\bibinfo  {journal} {Rev. Mod. Phys.}\ }\textbf {\bibinfo {volume} {65}},\ \bibinfo {pages} {803} (\bibinfo {year} {1993})}\BibitemShut {NoStop}%
\bibitem [{\citenamefont {Kwon}\ \emph {et~al.}(2019)\citenamefont {Kwon}, \citenamefont {Tan}, \citenamefont {Volkoff},\ and\ \citenamefont {Jeong}}]{KwonNonclassPhysRevLett.122.040503}%
  \BibitemOpen
  \bibfield  {author} {\bibinfo {author} {\bibfnamefont {H.}~\bibnamefont {Kwon}}, \bibinfo {author} {\bibfnamefont {K.~C.}\ \bibnamefont {Tan}}, \bibinfo {author} {\bibfnamefont {T.}~\bibnamefont {Volkoff}},\ and\ \bibinfo {author} {\bibfnamefont {H.}~\bibnamefont {Jeong}},\ }\bibfield  {title} {\bibinfo {title} {Nonclassicality as a quantifiable resource for quantum metrology},\ }\href {https://doi.org/10.1103/PhysRevLett.122.040503} {\bibfield  {journal} {\bibinfo  {journal} {Phys. Rev. Lett.}\ }\textbf {\bibinfo {volume} {122}},\ \bibinfo {pages} {040503} (\bibinfo {year} {2019})}\BibitemShut {NoStop}%
\bibitem [{\citenamefont {Gokhale}\ \emph {et~al.}(2020)\citenamefont {Gokhale}, \citenamefont {Angiuli}, \citenamefont {Ding}, \citenamefont {Gui}, \citenamefont {Tomesh}, \citenamefont {Suchara}, \citenamefont {Martonosi},\ and\ \citenamefont {Chong}}]{GokhaleMeasCostVQEMolecHamArxiv}%
  \BibitemOpen
  \bibfield  {author} {\bibinfo {author} {\bibfnamefont {P.}~\bibnamefont {Gokhale}}, \bibinfo {author} {\bibfnamefont {O.}~\bibnamefont {Angiuli}}, \bibinfo {author} {\bibfnamefont {Y.}~\bibnamefont {Ding}}, \bibinfo {author} {\bibfnamefont {K.}~\bibnamefont {Gui}}, \bibinfo {author} {\bibfnamefont {T.}~\bibnamefont {Tomesh}}, \bibinfo {author} {\bibfnamefont {M.}~\bibnamefont {Suchara}}, \bibinfo {author} {\bibfnamefont {M.}~\bibnamefont {Martonosi}},\ and\ \bibinfo {author} {\bibfnamefont {F.~T.}\ \bibnamefont {Chong}},\ }\bibfield  {title} {\bibinfo {title} {{$O(N^3)$} measurement cost for variational quantum eigensolver on molecular {H}amiltonians},\ }\href {https://doi.org/10.1109/TQE.2020.3035814} {\bibfield  {journal} {\bibinfo  {journal} {IEEE Trans. Quantum Eng.}\ }\textbf {\bibinfo {volume} {1}},\ \bibinfo {pages} {1} (\bibinfo {year} {2020})}\BibitemShut {NoStop}%
\bibitem [{\citenamefont {Bennett}\ \emph {et~al.}(1993)\citenamefont {Bennett}, \citenamefont {Brassard}, \citenamefont {Cr\'epeau}, \citenamefont {Jozsa}, \citenamefont {Peres},\ and\ \citenamefont {Wootters}}]{BennettTeleportationPhysRevLett.70.1895}%
  \BibitemOpen
  \bibfield  {author} {\bibinfo {author} {\bibfnamefont {C.~H.}\ \bibnamefont {Bennett}}, \bibinfo {author} {\bibfnamefont {G.}~\bibnamefont {Brassard}}, \bibinfo {author} {\bibfnamefont {C.}~\bibnamefont {Cr\'epeau}}, \bibinfo {author} {\bibfnamefont {R.}~\bibnamefont {Jozsa}}, \bibinfo {author} {\bibfnamefont {A.}~\bibnamefont {Peres}},\ and\ \bibinfo {author} {\bibfnamefont {W.~K.}\ \bibnamefont {Wootters}},\ }\bibfield  {title} {\bibinfo {title} {Teleporting an unknown quantum state via dual classical and {Einstein-Podolsky-Rosen} channels},\ }\href {https://doi.org/10.1103/PhysRevLett.70.1895} {\bibfield  {journal} {\bibinfo  {journal} {Phys. Rev. Lett.}\ }\textbf {\bibinfo {volume} {70}},\ \bibinfo {pages} {1895} (\bibinfo {year} {1993})}\BibitemShut {NoStop}%
\bibitem [{\citenamefont {Kenfack}\ and\ \citenamefont {\.{Z}yczkowski}(2004)}]{kenfack2004negativity}%
  \BibitemOpen
  \bibfield  {author} {\bibinfo {author} {\bibfnamefont {A.}~\bibnamefont {Kenfack}}\ and\ \bibinfo {author} {\bibfnamefont {K.}~\bibnamefont {\.{Z}yczkowski}},\ }\bibfield  {title} {\bibinfo {title} {Negativity of the {W}igner function as an indicator of non-classicality},\ }\href {https://doi.org/10.1088/1464-4266/6/10/003} {\bibfield  {journal} {\bibinfo  {journal} {J. Opt. B: Quantum Semiclassical Opt.}\ }\textbf {\bibinfo {volume} {6}},\ \bibinfo {pages} {396} (\bibinfo {year} {2004})}\BibitemShut {NoStop}%
\bibitem [{\citenamefont {Kochen}\ and\ \citenamefont {Specker}(1967)}]{KS1967}%
  \BibitemOpen
  \bibfield  {author} {\bibinfo {author} {\bibfnamefont {S.}~\bibnamefont {Kochen}}\ and\ \bibinfo {author} {\bibfnamefont {E.}~\bibnamefont {Specker}},\ }\bibfield  {title} {\bibinfo {title} {{The problem of hidden variables in quantum mechanics}},\ }\href {https://doi.org/10.1512/iumj.1968.17.17004} {\bibfield  {journal} {\bibinfo  {journal} {J. Math. Mech.}\ }\textbf {\bibinfo {volume} {17}},\ \bibinfo {pages} {59} (\bibinfo {year} {1967})}\BibitemShut {NoStop}%
\bibitem [{\citenamefont {Mermin}(1990)}]{MerminBKS10.1103/physrevlett.65.3373}%
  \BibitemOpen
  \bibfield  {author} {\bibinfo {author} {\bibfnamefont {N.~D.}\ \bibnamefont {Mermin}},\ }\bibfield  {title} {\bibinfo {title} {{Simple unified form for the major no-hidden-variables theorems}},\ }\href {https://doi.org/10.1103/physrevlett.65.3373} {\bibfield  {journal} {\bibinfo  {journal} {Phys. Rev. Lett.}\ }\textbf {\bibinfo {volume} {65}},\ \bibinfo {pages} {3373} (\bibinfo {year} {1990})}\BibitemShut {NoStop}%
\bibitem [{\citenamefont {Haug}\ and\ \citenamefont {Piroli}(2023)}]{HaugSRE10.22331/q-2023-08-28-1092}%
  \BibitemOpen
  \bibfield  {author} {\bibinfo {author} {\bibfnamefont {T.}~\bibnamefont {Haug}}\ and\ \bibinfo {author} {\bibfnamefont {L.}~\bibnamefont {Piroli}},\ }\bibfield  {title} {\bibinfo {title} {{Stabilizer entropies and nonstabilizerness monotones}},\ }\href {https://doi.org/10.22331/q-2023-08-28-1092} {\bibfield  {journal} {\bibinfo  {journal} {Quantum}\ }\textbf {\bibinfo {volume} {7}},\ \bibinfo {pages} {1092} (\bibinfo {year} {2023})}\BibitemShut {NoStop}%
\bibitem [{\citenamefont {Rajabpour}()}]{rajabpour2025stabilizershannonrenyiequivalenceexact}%
  \BibitemOpen
  \bibfield  {author} {\bibinfo {author} {\bibfnamefont {M.~A.}\ \bibnamefont {Rajabpour}},\ }\bibfield  {title} {\bibinfo {title} {Stabilizer-{S}hannon {R}enyi equivalence: Exact results for quantum critical chains},\ }\Eprint {https://arxiv.org/abs/2509.10700} {arXiv:2509.10700} \BibitemShut {NoStop}%
\bibitem [{\citenamefont {Sternberg}(1994)}]{Sternberg1994groupsphysics}%
  \BibitemOpen
  \bibfield  {author} {\bibinfo {author} {\bibfnamefont {S.}~\bibnamefont {Sternberg}},\ }\href@noop {} {\emph {\bibinfo {title} {Group Theory and Physics}}}\ (\bibinfo  {publisher} {Cambridge University Press},\ \bibinfo {address} {Cambridge},\ \bibinfo {year} {1994})\BibitemShut {NoStop}%
\bibitem [{\citenamefont {\.{Z}yczkowski}\ and\ \citenamefont {Sommers}(2001)}]{ZyczkowskiSommers2001haarrandom}%
  \BibitemOpen
  \bibfield  {author} {\bibinfo {author} {\bibfnamefont {K.}~\bibnamefont {\.{Z}yczkowski}}\ and\ \bibinfo {author} {\bibfnamefont {H.-J.}\ \bibnamefont {Sommers}},\ }\bibfield  {title} {\bibinfo {title} {Induced measures in the space of mixed quantum states},\ }\href {https://doi.org/10.1088/0305-4470/34/35/335} {\bibfield  {journal} {\bibinfo  {journal} {J. Phys. A: Math. Gen.}\ }\textbf {\bibinfo {volume} {34}},\ \bibinfo {pages} {7111} (\bibinfo {year} {2001})}\BibitemShut {NoStop}%
\bibitem [{\citenamefont {Rattacaso}\ \emph {et~al.}(2023)\citenamefont {Rattacaso}, \citenamefont {Leone}, \citenamefont {Oliviero},\ and\ \citenamefont {Hamma}}]{rattacaso2023stabilizer}%
  \BibitemOpen
  \bibfield  {author} {\bibinfo {author} {\bibfnamefont {D.}~\bibnamefont {Rattacaso}}, \bibinfo {author} {\bibfnamefont {L.}~\bibnamefont {Leone}}, \bibinfo {author} {\bibfnamefont {S.~F.}\ \bibnamefont {Oliviero}},\ and\ \bibinfo {author} {\bibfnamefont {A.}~\bibnamefont {Hamma}},\ }\bibfield  {title} {\bibinfo {title} {Stabilizer entropy dynamics after a quantum quench},\ }\href {https://doi.org/10.1103/PhysRevA.108.042407} {\bibfield  {journal} {\bibinfo  {journal} {Phys. Rev. A}\ }\textbf {\bibinfo {volume} {108}},\ \bibinfo {pages} {042407} (\bibinfo {year} {2023})}\BibitemShut {NoStop}%
\bibitem [{\citenamefont {Alterman}\ and\ \citenamefont {Love}(2026)}]{Sam2025}%
  \BibitemOpen
  \bibfield  {author} {\bibinfo {author} {\bibfnamefont {S.}~\bibnamefont {Alterman}}\ and\ \bibinfo {author} {\bibfnamefont {P.~J.}\ \bibnamefont {Love}},\ }\bibfield  {title} {\bibinfo {title} {Entanglement and magic on the light-front},\ }\href {https://doi.org/10.1088/1367-2630/ae4fea} {\bibfield  {journal} {\bibinfo  {journal} {New J. Phys.}\ }\textbf {\bibinfo {volume} {28}},\ \bibinfo {pages} {034513} (\bibinfo {year} {2026})}\BibitemShut {NoStop}%
\bibitem [{\citenamefont {Br\"okemeier}\ \emph {et~al.}(2025)\citenamefont {Br\"okemeier}, \citenamefont {Hengstenberg}, \citenamefont {Keeble}, \citenamefont {Robin}, \citenamefont {Rocco},\ and\ \citenamefont {Savage}}]{Brokemeier2025}%
  \BibitemOpen
  \bibfield  {author} {\bibinfo {author} {\bibfnamefont {F.}~\bibnamefont {Br\"okemeier}}, \bibinfo {author} {\bibfnamefont {S.~M.}\ \bibnamefont {Hengstenberg}}, \bibinfo {author} {\bibfnamefont {J.~W.~T.}\ \bibnamefont {Keeble}}, \bibinfo {author} {\bibfnamefont {C.~E.~P.}\ \bibnamefont {Robin}}, \bibinfo {author} {\bibfnamefont {F.}~\bibnamefont {Rocco}},\ and\ \bibinfo {author} {\bibfnamefont {M.~J.}\ \bibnamefont {Savage}},\ }\bibfield  {title} {\bibinfo {title} {Quantum magic and multipartite entanglement in the structure of nuclei},\ }\href {https://doi.org/10.1103/PhysRevC.111.034317} {\bibfield  {journal} {\bibinfo  {journal} {Phys. Rev. C}\ }\textbf {\bibinfo {volume} {111}},\ \bibinfo {pages} {034317} (\bibinfo {year} {2025})}\BibitemShut {NoStop}%
\bibitem [{\citenamefont {Gu}\ \emph {et~al.}(2024)\citenamefont {Gu}, \citenamefont {Hu}, \citenamefont {Luo}, \citenamefont {Patti}, \citenamefont {Rubin},\ and\ \citenamefont {Yelin}}]{gu2024zero}%
  \BibitemOpen
  \bibfield  {author} {\bibinfo {author} {\bibfnamefont {A.}~\bibnamefont {Gu}}, \bibinfo {author} {\bibfnamefont {H.-Y.}\ \bibnamefont {Hu}}, \bibinfo {author} {\bibfnamefont {D.}~\bibnamefont {Luo}}, \bibinfo {author} {\bibfnamefont {T.~L.}\ \bibnamefont {Patti}}, \bibinfo {author} {\bibfnamefont {N.~C.}\ \bibnamefont {Rubin}},\ and\ \bibinfo {author} {\bibfnamefont {S.~F.}\ \bibnamefont {Yelin}},\ }\bibfield  {title} {\bibinfo {title} {Zero and finite temperature quantum simulations powered by quantum magic},\ }\href {https://doi.org/10.22331/q-2024-07-23-1422} {\bibfield  {journal} {\bibinfo  {journal} {Quantum}\ }\textbf {\bibinfo {volume} {8}},\ \bibinfo {pages} {1422} (\bibinfo {year} {2024})}\BibitemShut {NoStop}%
\bibitem [{\citenamefont {Sarkis}\ and\ \citenamefont {Tkatchenko}()}]{sarkis2025are-602}%
  \BibitemOpen
  \bibfield  {author} {\bibinfo {author} {\bibfnamefont {M.}~\bibnamefont {Sarkis}}\ and\ \bibinfo {author} {\bibfnamefont {A.}~\bibnamefont {Tkatchenko}},\ }\bibfield  {title} {\bibinfo {title} {Are molecules magical? {N}on-stabilizerness in molecular bonding},\ }\Eprint {https://arxiv.org/abs/2504.06673} {arXiv:2504.06673} \BibitemShut {NoStop}%
\bibitem [{\citenamefont {Jordan}\ and\ \citenamefont {Wigner}(1928)}]{jordan1928ber-654}%
  \BibitemOpen
  \bibfield  {author} {\bibinfo {author} {\bibfnamefont {P.}~\bibnamefont {Jordan}}\ and\ \bibinfo {author} {\bibfnamefont {E.}~\bibnamefont {Wigner}},\ }\bibfield  {title} {\bibinfo {title} {\"{U}ber das {P}aulische \"{A}quivalenzverbot},\ }\href {https://doi.org/10.1007/bf01331938} {\bibfield  {journal} {\bibinfo  {journal} {Z. f\"ur Physik}\ }\textbf {\bibinfo {volume} {47}},\ \bibinfo {pages} {631} (\bibinfo {year} {1928})}\BibitemShut {NoStop}%
\bibitem [{\citenamefont {Bravyi}\ and\ \citenamefont {Kitaev}(2002)}]{bravyi2002fermionic}%
  \BibitemOpen
  \bibfield  {author} {\bibinfo {author} {\bibfnamefont {S.~B.}\ \bibnamefont {Bravyi}}\ and\ \bibinfo {author} {\bibfnamefont {A.~Y.}\ \bibnamefont {Kitaev}},\ }\bibfield  {title} {\bibinfo {title} {Fermionic quantum computation},\ }\href {https://doi.org/10.1006/aphy.2002.6254} {\bibfield  {journal} {\bibinfo  {journal} {Ann. Phys.}\ }\textbf {\bibinfo {volume} {298}},\ \bibinfo {pages} {210} (\bibinfo {year} {2002})}\BibitemShut {NoStop}%
\bibitem [{\citenamefont {Seeley}\ \emph {et~al.}(2012)\citenamefont {Seeley}, \citenamefont {Richard},\ and\ \citenamefont {Love}}]{seeley2012bravyi}%
  \BibitemOpen
  \bibfield  {author} {\bibinfo {author} {\bibfnamefont {J.~T.}\ \bibnamefont {Seeley}}, \bibinfo {author} {\bibfnamefont {M.~J.}\ \bibnamefont {Richard}},\ and\ \bibinfo {author} {\bibfnamefont {P.~J.}\ \bibnamefont {Love}},\ }\bibfield  {title} {\bibinfo {title} {The {B}ravyi-{K}itaev transformation for quantum computation of electronic structure},\ }\bibfield  {journal} {\bibinfo  {journal} {J. Chem. Phys.}\ }\textbf {\bibinfo {volume} {137}},\ \href {https://doi.org/10.1063/1.4768229} {10.1063/1.4768229} (\bibinfo {year} {2012})\BibitemShut {NoStop}%
\bibitem [{\citenamefont {Bravyi}\ \emph {et~al.}()\citenamefont {Bravyi}, \citenamefont {Gambetta}, \citenamefont {Mezzacapo},\ and\ \citenamefont {Temme}}]{bravyi2017taperingqubitssimulatefermionic}%
  \BibitemOpen
  \bibfield  {author} {\bibinfo {author} {\bibfnamefont {S.}~\bibnamefont {Bravyi}}, \bibinfo {author} {\bibfnamefont {J.~M.}\ \bibnamefont {Gambetta}}, \bibinfo {author} {\bibfnamefont {A.}~\bibnamefont {Mezzacapo}},\ and\ \bibinfo {author} {\bibfnamefont {K.}~\bibnamefont {Temme}},\ }\bibfield  {title} {\bibinfo {title} {Tapering off qubits to simulate fermionic {H}amiltonians},\ }\Eprint {https://arxiv.org/abs/1701.08213} {arXiv:1701.08213} \BibitemShut {NoStop}%
\bibitem [{\citenamefont {Setia}\ \emph {et~al.}(2020)\citenamefont {Setia}, \citenamefont {Chen}, \citenamefont {Rice}, \citenamefont {Mezzacapo}, \citenamefont {Pistoia},\ and\ \citenamefont {Whitfield}}]{doi:10.1021/acs.jctc.0c00113}%
  \BibitemOpen
  \bibfield  {author} {\bibinfo {author} {\bibfnamefont {K.}~\bibnamefont {Setia}}, \bibinfo {author} {\bibfnamefont {R.}~\bibnamefont {Chen}}, \bibinfo {author} {\bibfnamefont {J.~E.}\ \bibnamefont {Rice}}, \bibinfo {author} {\bibfnamefont {A.}~\bibnamefont {Mezzacapo}}, \bibinfo {author} {\bibfnamefont {M.}~\bibnamefont {Pistoia}},\ and\ \bibinfo {author} {\bibfnamefont {J.~D.}\ \bibnamefont {Whitfield}},\ }\bibfield  {title} {\bibinfo {title} {Reducing qubit requirements for quantum simulations using molecular point group symmetries},\ }\href {https://doi.org/10.1021/acs.jctc.0c00113} {\bibfield  {journal} {\bibinfo  {journal} {J. Chem. Theory Comput.}\ }\textbf {\bibinfo {volume} {16}},\ \bibinfo {pages} {6091} (\bibinfo {year} {2020})}\BibitemShut {NoStop}%
\bibitem [{\citenamefont {Ralli}\ \emph {et~al.}(2025)\citenamefont {Ralli}, \citenamefont {Weaving},\ and\ \citenamefont {Love}}]{ralli2025noncontextual}%
  \BibitemOpen
  \bibfield  {author} {\bibinfo {author} {\bibfnamefont {A.}~\bibnamefont {Ralli}}, \bibinfo {author} {\bibfnamefont {T.}~\bibnamefont {Weaving}},\ and\ \bibinfo {author} {\bibfnamefont {P.~J.}\ \bibnamefont {Love}},\ }\bibfield  {title} {\bibinfo {title} {Noncontextual {P}auli {H}amiltonians},\ }\href {https://doi.org/10.1088/1751-8121/adf677} {\bibfield  {journal} {\bibinfo  {journal} {J. Phys. A: Math. Theor.}\ }\textbf {\bibinfo {volume} {58}},\ \bibinfo {pages} {335301} (\bibinfo {year} {2025})}\BibitemShut {NoStop}%
\bibitem [{\citenamefont {Kirby}\ and\ \citenamefont {Love}(2019)}]{Kirby2019Contextuality}%
  \BibitemOpen
  \bibfield  {author} {\bibinfo {author} {\bibfnamefont {W.~M.}\ \bibnamefont {Kirby}}\ and\ \bibinfo {author} {\bibfnamefont {P.~J.}\ \bibnamefont {Love}},\ }\bibfield  {title} {\bibinfo {title} {Contextuality test of the nonclassicality of variational quantum eigensolvers},\ }\href {https://doi.org/10.1103/PhysRevLett.123.200501} {\bibfield  {journal} {\bibinfo  {journal} {Phys. Rev. Lett.}\ }\textbf {\bibinfo {volume} {123}},\ \bibinfo {pages} {200501} (\bibinfo {year} {2019})}\BibitemShut {NoStop}%
\bibitem [{\citenamefont {Kirby}\ and\ \citenamefont {Love}(2020)}]{Kirby2020}%
  \BibitemOpen
  \bibfield  {author} {\bibinfo {author} {\bibfnamefont {W.~M.}\ \bibnamefont {Kirby}}\ and\ \bibinfo {author} {\bibfnamefont {P.~J.}\ \bibnamefont {Love}},\ }\bibfield  {title} {\bibinfo {title} {Classical simulation of noncontextual {P}auli {H}amiltonians},\ }\href {https://doi.org/10.1103/PhysRevA.102.032418} {\bibfield  {journal} {\bibinfo  {journal} {Phys. Rev. A}\ }\textbf {\bibinfo {volume} {102}},\ \bibinfo {pages} {032418} (\bibinfo {year} {2020})}\BibitemShut {NoStop}%
\bibitem [{\citenamefont {Zhao}\ \emph {et~al.}(2020)\citenamefont {Zhao}, \citenamefont {Tranter}, \citenamefont {Kirby}, \citenamefont {Ung}, \citenamefont {Miyake},\ and\ \citenamefont {Love}}]{ZhaoMeasRedux}%
  \BibitemOpen
  \bibfield  {author} {\bibinfo {author} {\bibfnamefont {A.}~\bibnamefont {Zhao}}, \bibinfo {author} {\bibfnamefont {A.}~\bibnamefont {Tranter}}, \bibinfo {author} {\bibfnamefont {W.~M.}\ \bibnamefont {Kirby}}, \bibinfo {author} {\bibfnamefont {S.~F.}\ \bibnamefont {Ung}}, \bibinfo {author} {\bibfnamefont {A.}~\bibnamefont {Miyake}},\ and\ \bibinfo {author} {\bibfnamefont {P.~J.}\ \bibnamefont {Love}},\ }\bibfield  {title} {\bibinfo {title} {Measurement reduction in variational quantum algorithms},\ }\href {https://doi.org/10.1103/PhysRevA.101.062322} {\bibfield  {journal} {\bibinfo  {journal} {Phys. Rev. A}\ }\textbf {\bibinfo {volume} {101}},\ \bibinfo {pages} {062322} (\bibinfo {year} {2020})}\BibitemShut {NoStop}%
\bibitem [{\citenamefont {{\relax QMatter Labs}}(2025)}]{symmer}%
  \BibitemOpen
  \bibfield  {author} {\bibinfo {author} {\bibnamefont {{\relax QMatter Labs}}},\ }\href@noop {} {\bibinfo {title} {Symmer version 0.0.9}} (\bibinfo {year} {2025}),\ \bibinfo {note} {\url{https://github.com/qmatter-labs/symmer/}}\BibitemShut {NoStop}%
\bibitem [{\citenamefont {Weaving}\ \emph {et~al.}(2025)\citenamefont {Weaving}, \citenamefont {Ralli}, \citenamefont {Love}, \citenamefont {Succi},\ and\ \citenamefont {Coveney}}]{weaving2025contextual}%
  \BibitemOpen
  \bibfield  {author} {\bibinfo {author} {\bibfnamefont {T.}~\bibnamefont {Weaving}}, \bibinfo {author} {\bibfnamefont {A.}~\bibnamefont {Ralli}}, \bibinfo {author} {\bibfnamefont {P.~J.}\ \bibnamefont {Love}}, \bibinfo {author} {\bibfnamefont {S.}~\bibnamefont {Succi}},\ and\ \bibinfo {author} {\bibfnamefont {P.~V.}\ \bibnamefont {Coveney}},\ }\bibfield  {title} {\bibinfo {title} {Contextual subspace variational quantum eigensolver calculation of the dissociation curve of molecular nitrogen on a superconducting quantum computer},\ }\href {https://doi.org/10.1038/s41534-024-00952-4} {\bibfield  {journal} {\bibinfo  {journal} {npj Quantum Information}\ }\textbf {\bibinfo {volume} {11}},\ \bibinfo {pages} {25} (\bibinfo {year} {2025})}\BibitemShut {NoStop}%
\bibitem [{\citenamefont {Liu}\ and\ \citenamefont {Winter}(2022)}]{liu2022many}%
  \BibitemOpen
  \bibfield  {author} {\bibinfo {author} {\bibfnamefont {Z.-W.}\ \bibnamefont {Liu}}\ and\ \bibinfo {author} {\bibfnamefont {A.}~\bibnamefont {Winter}},\ }\bibfield  {title} {\bibinfo {title} {Many-body quantum magic},\ }\href {https://doi.org/10.1103/PRXQuantum.3.020333} {\bibfield  {journal} {\bibinfo  {journal} {PRX Quantum}\ }\textbf {\bibinfo {volume} {3}},\ \bibinfo {pages} {020333} (\bibinfo {year} {2022})}\BibitemShut {NoStop}%
\bibitem [{\citenamefont {Bengtsson}\ and\ \citenamefont {\.{Z}yczkowski}(2017)}]{bengtsson2017geometryquantstates}%
  \BibitemOpen
  \bibfield  {author} {\bibinfo {author} {\bibfnamefont {I.}~\bibnamefont {Bengtsson}}\ and\ \bibinfo {author} {\bibfnamefont {K.}~\bibnamefont {\.{Z}yczkowski}},\ }\href@noop {} {\emph {\bibinfo {title} {Geometry of Quantum States: An Introduction to Quantum Entanglement}}},\ \bibinfo {edition} {2nd}\ ed.\ (\bibinfo  {publisher} {Cambridge University Press},\ \bibinfo {address} {Cambridge},\ \bibinfo {year} {2017})\BibitemShut {NoStop}%
\bibitem [{\citenamefont {Langhoff}\ and\ \citenamefont {Davidson}(1974)}]{DavidsonCorrection}%
  \BibitemOpen
  \bibfield  {author} {\bibinfo {author} {\bibfnamefont {S.~R.}\ \bibnamefont {Langhoff}}\ and\ \bibinfo {author} {\bibfnamefont {E.~R.}\ \bibnamefont {Davidson}},\ }\bibfield  {title} {\bibinfo {title} {Configuration interaction calculations on the nitrogen molecule},\ }\href {https://doi.org/10.1002/qua.560080106} {\bibfield  {journal} {\bibinfo  {journal} {Int. J. Quantum Chem.}\ }\textbf {\bibinfo {volume} {8}},\ \bibinfo {pages} {61} (\bibinfo {year} {1974})}\BibitemShut {NoStop}%
\bibitem [{\citenamefont {Wang}(2026)}]{symmer_hamiltonian_database}%
  \BibitemOpen
  \bibfield  {author} {\bibinfo {author} {\bibfnamefont {Q.}~\bibnamefont {Wang}},\ }\href {https://github.com/Kee-Wang/Symmer-Hamiltonian} {\bibinfo {title} {Symmer-{H}amiltonian: a database of molecular electronic {H}amiltonians in qubit operator form}} (\bibinfo {year} {2026}),\ \bibinfo {note} {{CC-BY-4.0} license, 190 species, 1{,}594 Hamiltonians}\BibitemShut {NoStop}%
\bibitem [{\citenamefont {Johnson}(2022)}]{CCCBDB}%
  \BibitemOpen
  \bibfield  {author} {\bibinfo {author} {\bibfnamefont {R.~D.}\ \bibnamefont {Johnson}, \bibfnamefont {III}},\ }\href {https://doi.org/10.18434/T47C7Z} {\bibinfo {title} {{NIST} computational chemistry comparison and benchmark database, {NIST} standard reference database number 101}} (\bibinfo {year} {2022}),\ \bibinfo {note} {release 22, \url{http://cccbdb.nist.gov/}}\BibitemShut {NoStop}%
\bibitem [{\citenamefont {PennyLaneAI}(2026)}]{PennyLaneAI_DatasetsSource}%
  \BibitemOpen
  \bibfield  {author} {\bibinfo {author} {\bibnamefont {PennyLaneAI}},\ }\href@noop {} {\bibinfo {title} {Datasetssource}},\ \bibinfo {howpublished} {\url{https://github.com/PennyLaneAI/DatasetsSource}} (\bibinfo {year} {2026})\BibitemShut {NoStop}%
\bibitem [{\citenamefont {Huang}\ \emph {et~al.}()\citenamefont {Huang}, \citenamefont {Li},\ and\ \citenamefont {Zhong}}]{huang2026fast-9ec}%
  \BibitemOpen
  \bibfield  {author} {\bibinfo {author} {\bibfnamefont {X.}~\bibnamefont {Huang}}, \bibinfo {author} {\bibfnamefont {H.-Z.}\ \bibnamefont {Li}},\ and\ \bibinfo {author} {\bibfnamefont {J.-X.}\ \bibnamefont {Zhong}},\ }\bibfield  {title} {\bibinfo {title} {A fast and exact approach for stabilizer {R}\'{e}nyi entropy via the {XOR}-{FWHT} algorithm},\ }\Eprint {https://arxiv.org/abs/2512.24685} {arXiv:2512.24685} \BibitemShut {NoStop}%
\bibitem [{\citenamefont {Xiao}\ and\ \citenamefont {Ryu}()}]{xiao2026exponentially-96a}%
  \BibitemOpen
  \bibfield  {author} {\bibinfo {author} {\bibfnamefont {Z.}~\bibnamefont {Xiao}}\ and\ \bibinfo {author} {\bibfnamefont {S.}~\bibnamefont {Ryu}},\ }\bibfield  {title} {\bibinfo {title} {Exponentially accelerated sampling of {P}auli strings for nonstabilizerness},\ }\Eprint {https://arxiv.org/abs/2601.00761} {arXiv:2601.00761} \BibitemShut {NoStop}%
\bibitem [{\citenamefont {Alterman}\ \emph {et~al.}(2026)\citenamefont {Alterman}, \citenamefont {Qian}, \citenamefont {Seibert},\ and\ \citenamefont {Wang}}]{magic-and-correlation-code}%
  \BibitemOpen
  \bibfield  {author} {\bibinfo {author} {\bibfnamefont {S.}~\bibnamefont {Alterman}}, \bibinfo {author} {\bibfnamefont {F.}~\bibnamefont {Qian}}, \bibinfo {author} {\bibfnamefont {B.}~\bibnamefont {Seibert}},\ and\ \bibinfo {author} {\bibfnamefont {Q.}~\bibnamefont {Wang}},\ }\href@noop {} {} (\bibinfo {year} {2026}),\ \bibinfo {note} {\url{https://github.com/samalterman/magic-and-correlation}}\BibitemShut {NoStop}%
\bibitem [{\citenamefont {Alterman}\ \emph {et~al.}()\citenamefont {Alterman}, \citenamefont {Qian}, \citenamefont {Seibert}, \citenamefont {Wang}, \citenamefont {Miyake},\ and\ \citenamefont {Love}}]{magic-and-correlation-data}%
  \BibitemOpen
  \bibfield  {author} {\bibinfo {author} {\bibfnamefont {S.}~\bibnamefont {Alterman}}, \bibinfo {author} {\bibfnamefont {F.}~\bibnamefont {Qian}}, \bibinfo {author} {\bibfnamefont {B.}~\bibnamefont {Seibert}}, \bibinfo {author} {\bibfnamefont {Q.}~\bibnamefont {Wang}}, \bibinfo {author} {\bibfnamefont {A.}~\bibnamefont {Miyake}},\ and\ \bibinfo {author} {\bibfnamefont {P.}~\bibnamefont {Love}},\ }\href {https://doi.org/10.5281/zenodo.20798316} {\bibinfo {title} {10.5281/zenodo.20798316}}\BibitemShut {NoStop}%
\bibitem [{\citenamefont {Yao}\ and\ \citenamefont {Li}(2025)}]{yao2025efficient}%
  \BibitemOpen
  \bibfield  {author} {\bibinfo {author} {\bibfnamefont {Q.}~\bibnamefont {Yao}}\ and\ \bibinfo {author} {\bibfnamefont {H.}~\bibnamefont {Li}},\ }\bibfield  {title} {\bibinfo {title} {Efficient excited-state calculations for molecules based on contextual subspace method and symmetry optimizations},\ }\href {https://doi.org/10.1088/1367-2630/ae0360} {\bibfield  {journal} {\bibinfo  {journal} {New J. Phys.}\ }\textbf {\bibinfo {volume} {27}},\ \bibinfo {pages} {094508} (\bibinfo {year} {2025})}\BibitemShut {NoStop}%
\bibitem [{\citenamefont {Weaving}\ \emph {et~al.}(2023{\natexlab{b}})\citenamefont {Weaving}, \citenamefont {Ralli}, \citenamefont {Kirby}, \citenamefont {Love}, \citenamefont {Succi},\ and\ \citenamefont {Coveney}}]{PhysRevResearch.5.043054}%
  \BibitemOpen
  \bibfield  {author} {\bibinfo {author} {\bibfnamefont {T.}~\bibnamefont {Weaving}}, \bibinfo {author} {\bibfnamefont {A.}~\bibnamefont {Ralli}}, \bibinfo {author} {\bibfnamefont {W.~M.}\ \bibnamefont {Kirby}}, \bibinfo {author} {\bibfnamefont {P.~J.}\ \bibnamefont {Love}}, \bibinfo {author} {\bibfnamefont {S.}~\bibnamefont {Succi}},\ and\ \bibinfo {author} {\bibfnamefont {P.~V.}\ \bibnamefont {Coveney}},\ }\bibfield  {title} {\bibinfo {title} {Benchmarking noisy intermediate scale quantum error mitigation strategies for ground state preparation of the {HCl} molecule},\ }\href {https://doi.org/10.1103/PhysRevResearch.5.043054} {\bibfield  {journal} {\bibinfo  {journal} {Phys. Rev. Research}\ }\textbf {\bibinfo {volume} {5}},\ \bibinfo {pages} {043054} (\bibinfo {year} {2023}{\natexlab{b}})}\BibitemShut {NoStop}%
\bibitem [{\citenamefont {Viscardi}\ \emph {et~al.}(2026)\citenamefont {Viscardi}, \citenamefont {Dalmonte}, \citenamefont {Hamma},\ and\ \citenamefont {Tirrito}}]{Viscardi2025}%
  \BibitemOpen
  \bibfield  {author} {\bibinfo {author} {\bibfnamefont {M.}~\bibnamefont {Viscardi}}, \bibinfo {author} {\bibfnamefont {M.}~\bibnamefont {Dalmonte}}, \bibinfo {author} {\bibfnamefont {A.}~\bibnamefont {Hamma}},\ and\ \bibinfo {author} {\bibfnamefont {E.}~\bibnamefont {Tirrito}},\ }\bibfield  {title} {\bibinfo {title} {{Interplay of entanglement structures and stabilizer entropy in spin models}},\ }\href {https://doi.org/10.21468/SciPostPhysCore.9.1.012} {\bibfield  {journal} {\bibinfo  {journal} {SciPost Phys. Core}\ }\textbf {\bibinfo {volume} {9}},\ \bibinfo {pages} {012} (\bibinfo {year} {2026})}\BibitemShut {NoStop}%
\bibitem [{\citenamefont {Munizzi}\ and\ \citenamefont {Schnitzer}()}]{munizzi2026magic}%
  \BibitemOpen
  \bibfield  {author} {\bibinfo {author} {\bibfnamefont {W.}~\bibnamefont {Munizzi}}\ and\ \bibinfo {author} {\bibfnamefont {H.~J.}\ \bibnamefont {Schnitzer}},\ }\bibfield  {title} {\bibinfo {title} {Magic and non-clifford gates in topological quantum field theory},\ }\Eprint {https://arxiv.org/abs/2604.14271} {arXiv:2604.14271} \BibitemShut {NoStop}%
\bibitem [{\citenamefont {Ravi}\ \emph {et~al.}(2022)\citenamefont {Ravi}, \citenamefont {Gokhale}, \citenamefont {Ding}, \citenamefont {Kirby}, \citenamefont {Smith}, \citenamefont {Baker}, \citenamefont {Love}, \citenamefont {Hoffmann}, \citenamefont {Brown},\ and\ \citenamefont {Chong}}]{cafqa}%
  \BibitemOpen
  \bibfield  {author} {\bibinfo {author} {\bibfnamefont {G.~S.}\ \bibnamefont {Ravi}}, \bibinfo {author} {\bibfnamefont {P.}~\bibnamefont {Gokhale}}, \bibinfo {author} {\bibfnamefont {Y.}~\bibnamefont {Ding}}, \bibinfo {author} {\bibfnamefont {W.}~\bibnamefont {Kirby}}, \bibinfo {author} {\bibfnamefont {K.}~\bibnamefont {Smith}}, \bibinfo {author} {\bibfnamefont {J.~M.}\ \bibnamefont {Baker}}, \bibinfo {author} {\bibfnamefont {P.~J.}\ \bibnamefont {Love}}, \bibinfo {author} {\bibfnamefont {H.}~\bibnamefont {Hoffmann}}, \bibinfo {author} {\bibfnamefont {K.~R.}\ \bibnamefont {Brown}},\ and\ \bibinfo {author} {\bibfnamefont {F.~T.}\ \bibnamefont {Chong}},\ }\bibfield  {title} {\bibinfo {title} {{CAFQA}: A classical simulation bootstrap for variational quantum algorithms},\ }in\ \href {https://doi.org/10.1145/3567955.3567958} {\emph {\bibinfo {booktitle} {Proceedings of the 28th ACM International Conference on Architectural Support for Programming Languages and Operating Systems, Volume 1}}},\ \bibinfo {series
  and number} {ASPLOS 2023}\ (\bibinfo  {publisher} {Association for Computing Machinery},\ \bibinfo {address} {New York, NY, USA},\ \bibinfo {year} {2022})\ pp.\ \bibinfo {pages} {15--29}\BibitemShut {NoStop}%
\bibitem [{\citenamefont {Wang}\ \emph {et~al.}()\citenamefont {Wang}, \citenamefont {Zhukas}, \citenamefont {Miao}, \citenamefont {Dalvi}, \citenamefont {Love}, \citenamefont {Monroe}, \citenamefont {Chong},\ and\ \citenamefont {Ravi}}]{keecafqa}%
  \BibitemOpen
  \bibfield  {author} {\bibinfo {author} {\bibfnamefont {Q.}~\bibnamefont {Wang}}, \bibinfo {author} {\bibfnamefont {L.}~\bibnamefont {Zhukas}}, \bibinfo {author} {\bibfnamefont {Q.}~\bibnamefont {Miao}}, \bibinfo {author} {\bibfnamefont {A.~S.}\ \bibnamefont {Dalvi}}, \bibinfo {author} {\bibfnamefont {P.~J.}\ \bibnamefont {Love}}, \bibinfo {author} {\bibfnamefont {C.}~\bibnamefont {Monroe}}, \bibinfo {author} {\bibfnamefont {F.~T.}\ \bibnamefont {Chong}},\ and\ \bibinfo {author} {\bibfnamefont {G.~S.}\ \bibnamefont {Ravi}},\ }\bibfield  {title} {\bibinfo {title} {Demonstration of a {CAFQA}-bootstrapped variational quantum eigensolver on a trapped-ion quantum computer},\ }\Eprint {https://arxiv.org/abs/2408.06482} {arXiv:2408.06482} \BibitemShut {NoStop}%
\bibitem [{\citenamefont {Ralli}\ \emph {et~al.}()\citenamefont {Ralli}, \citenamefont {Weaving}, \citenamefont {Bickley}, \citenamefont {Coveney},\ and\ \citenamefont {Love}}]{qascf}%
  \BibitemOpen
  \bibfield  {author} {\bibinfo {author} {\bibfnamefont {A.}~\bibnamefont {Ralli}}, \bibinfo {author} {\bibfnamefont {T.}~\bibnamefont {Weaving}}, \bibinfo {author} {\bibfnamefont {T.~M.}\ \bibnamefont {Bickley}}, \bibinfo {author} {\bibfnamefont {P.~V.}\ \bibnamefont {Coveney}},\ and\ \bibinfo {author} {\bibfnamefont {P.~J.}\ \bibnamefont {Love}},\ }\bibfield  {title} {\bibinfo {title} {Quantum-accelerated self-consistent field: {A} hybrid algorithm},\ }\Eprint {https://arxiv.org/abs/2606.20176} {arXiv:2606.20176} \BibitemShut {NoStop}%
\bibitem [{\citenamefont {Sun}\ \emph {et~al.}(2020)\citenamefont {Sun}, \citenamefont {Zhang}, \citenamefont {Banerjee}, \citenamefont {Bao}, \citenamefont {Barbry}, \citenamefont {Blunt}, \citenamefont {Bogdanov}, \citenamefont {Booth}, \citenamefont {Chen}, \citenamefont {Cui} \emph {et~al.}}]{sun2020recent}%
  \BibitemOpen
  \bibfield  {author} {\bibinfo {author} {\bibfnamefont {Q.}~\bibnamefont {Sun}}, \bibinfo {author} {\bibfnamefont {X.}~\bibnamefont {Zhang}}, \bibinfo {author} {\bibfnamefont {S.}~\bibnamefont {Banerjee}}, \bibinfo {author} {\bibfnamefont {P.}~\bibnamefont {Bao}}, \bibinfo {author} {\bibfnamefont {M.}~\bibnamefont {Barbry}}, \bibinfo {author} {\bibfnamefont {N.~S.}\ \bibnamefont {Blunt}}, \bibinfo {author} {\bibfnamefont {N.~A.}\ \bibnamefont {Bogdanov}}, \bibinfo {author} {\bibfnamefont {G.~H.}\ \bibnamefont {Booth}}, \bibinfo {author} {\bibfnamefont {J.}~\bibnamefont {Chen}}, \bibinfo {author} {\bibfnamefont {Z.-H.}\ \bibnamefont {Cui}}, \emph {et~al.},\ }\bibfield  {title} {\bibinfo {title} {Recent developments in the {PySCF} program package},\ }\href {https://doi.org/10.1063/5.0006074} {\bibfield  {journal} {\bibinfo  {journal} {J. Chem. Phys.}\ }\textbf {\bibinfo {volume} {153}},\ \bibinfo {pages} {024109} (\bibinfo {year} {2020})}\BibitemShut {NoStop}%
\bibitem [{\citenamefont {Wang}(2024)}]{symmerpyscf}%
  \BibitemOpen
  \bibfield  {author} {\bibinfo {author} {\bibfnamefont {Q.}~\bibnamefont {Wang}},\ }\href {https://github.com/Kee-Wang/symmer-pyscf} {\bibinfo {title} {symmer-pyscf: {PySCF} integration for the {Symmer} qubit operator framework}} (\bibinfo {year} {2024}),\ \bibinfo {note} {version 0.1.0}\BibitemShut {NoStop}%
\end{thebibliography}%

\end{document}